\newcommand{\ie}{{\it i.e. }}
\newcommand{\eg}{{\it e.g. }}
\newcommand{\planck}{{\sc Planck~}}
\newcommand{\map}{{$MAP$~}}
\newcommand{\lsim}{\, \lower2truept\hbox{${<\atop\hbox{\raise4truept\hbox{$\sim$}}}$}\,}
\newcommand{\gsim}{\, \lower2truept\hbox{${>\atop\hbox{\raise4truept\hbox{$\sim$}}}$}\,}
      [1999/08/25 v1.4a Il Nuovo Cimento]
\documentclass{cimento}
\usepackage{threeparttable}
\usepackage{psfig}
\usepackage{epsfig}

\usepackage{amsmath}
\usepackage{amsfonts}
\usepackage{amssymb}
\title{Anisotropies of the Cosmic Microwave Background}
\author{M.~Bersanelli\from{ins:x},  D.~Maino\from{ins:x} \atque A.~Mennella\from{ins:y}}
\instlist{
  \inst{ins:x} Universit\'a degli Studi di Milano, Via Celoria 16, I-20133, Milano, Italy
  \inst{ins:y} CNR-IASF (Sez. di Milano), Via Bassini 15, I-20133,
  Milano, Italy
}
\PACSes{
\PACSit{98.80}{Cosmology}}
\begin{document}

\maketitle

\begin{abstract}

We review the present status of Cosmic Microwave Background (CMB)
anisotropy observations and discuss the main related
astrophysical issues, instrumental effects and data analysis
techniques. We summarise the balloon-borne and ground-based
experiments that, after $COBE$-DMR, yielded detection or
significant upper limits to CMB fluctuations. A comparison of
subsets of combined data indicates that the acoustic features
observed today in the angular power spectrum are not dominated by
undetected systematics. Pushing the accuracy of CMB anisotropy
measurements to their ultimate limits represents one of the best
opportunities for cosmology to develop into a precision science
in the next decade. We discuss the forthcoming sub-orbital and
space programs, as well as future prospects of CMB observations.

\end{abstract}

\section{INTRODUCTION}
\label{intro}

The Cosmic Microwave Background (CMB) radiation has played
a central role in modern cosmology since the time of its
discovery by Penzias and Wilson in 1965 \cite{penwil65}.
The existence of a background of cold photons was predicted several
years before by Gamow, Alpher and Herman \cite{gamow48,alpher48}
following their assumption that primordial abundances were
produced during an early phase dominated by thermal radiation.
Traditionally, the CMB is considered one of the three
observational pillars supporting the cosmological scenario
of the Hot Big Bang, together with light elements primordial
abundances (see, e.g., \cite{light}) and the cosmic expansion \cite{hubble26}. 
In recent years, CMB measurements are
widely considered as the single most fruitful field in
observational cosmology, thanks to the richness
and precision of information it can provide directly from
the early universe.

In the first decade or so after the discovery, experiments and
theoretical work focussed on establishing the nature of the CMB
itself, leading to a compelling evidence of a cosmic origin of
the radiation. By the early 80's the Hot Big Bang prediction of a
highly isotropic background \cite{alpher48} with a nearly
Planckian spectrum was remarkably supported by observation. The
growing CMB community then shifted the interest to observations
of first-order deviations from the ``idealised'', unperturbed
scenario. Spectrum experiments searched for spectral distortions
\cite{danedezo77} capable, if detected, to set constraints on
cosmological parameters, such as the baryon density $\Omega_b$,
and on the thermal history of the universe \cite{burigana91}.
Measurements of the CMB angular distribution were carried out
with increasing accuracy to detect CMB anisotropy.

A new phase was opened up by the successful outcome of the $COBE$
mission in the early 90's. The FIRAS experiment \cite{fixsen96}
established that the CMB spectrum is planckian within limits as
tight as 0.03\% in the frequency range 60-600~GHz with a
temperature of $T_0 = 2.725\pm 0.002$~K. Coupled with sub-orbital
measurements at low frequencies
\cite{bersa94,deami91,bensa93,smoot87,siro91,ruben00} very
stringent constraints to distortion parameters were placed
($|\mu|<9\times 10^{-5}$ for ``chemical'' distortions,
$y<1.5\times 10^{-5}$ for Compton distortions and $|Y_{ff}|
<3\times 10^{-6}$ for free-free distortions) leading to tight
upper limits on energy injections in the early universe.

The FIRAS results illustrate the level of precision that
cosmology can seek based on observations of the CMB. The
possibility of ``precision cosmology" with the CMB comes from the
combination of three factors. First, measurements of the CMB can
be made with exquisite accuracy thanks to the extraordinary
progress achieved by microwave and sub-mm technology in recent years. 
Second, we have now
good evidence that astrophysical emission in the microwaves, that
adds to the cosmological signal, does not prevent in principle
the observation of the subtle intrinsic characteristics of the
CMB. And, third, the theoretical interpretation of CMB data is
relatively simple, since the features we observe in the microwave
background carry information directly from epochs when all the
processes were still in the linear regime.

In addition to the FIRAS results, the other major result of 
$COBE$ was the first unambiguous detection, by the DMR
instrument, of anisotropies at a level $\Delta T/T \approx
10^{-5}$ on large angular scales \cite{smoot92}. This
breakthrough immediately stimulated the realisation of many new
experiments aiming at measuring the CMB angular distribution with
increasing resolution and sensitivity. As we shall see in detail,
today more than 20 independent projects carried out with different
technologies and from a variety of observing sites have reported
anisotropy detection at similar $\Delta T/T$ levels over a wide
range of angular scales. At present, these are far from
the precision measurements achieved by FIRAS on the CMB frequency spectrum,
but the new generations of space-based anisotropy experiments is
designed to eventually reach FIRAS-like accuracy in the angular
power spectrum. It is important however to highlight a fundamental
difference: while FIRAS made a very accurate measurement of an
essentially null result (no spectral distortions detected), the
present anisotropy data already demonstrate the presence of
angular structure at ``non-zero'' levels. The statistical
properties of these fluctuations and their dependence upon
cosmological parameters can be easily computed and compared to
the observed maps, without the complications that arise in
non-linear processes. The details of the angular power spectrum are
at reach and can reveal a wealth of genuine new information on
the early universe.

In this work we present an overview of the present status of the
observations of CMB anisotropy covering the ``post-$COBE$ era''.
This is an extremely rapidly evolving field (e.g.
\cite{hudodel,scowhisil}) and new important results are published
almost every month. This makes a comprehensive review a very
hard, if not impossible, task. In particular, the NASA $MAP$
satellite, launched in June 2001, is carrying out its first survey as we
write. By the time this work is published, the first release of
the $MAP$ data will be shortly expected, and will have a major
impact in the field. Furthermore, the planned ESA mission {\sc
Planck} will have reached a more mature stage and some
information given here may turn out to be incomplete.

While a detailed discussion of CMB polarisation is out of the
scope of this review, it will be briefly mentioned and discussed,
when relevent, in connection with temperature anisotropy. The
paper is organised as follows. In Section~\ref{cmbanirev} we
briefly discuss the standard scenario for the origin of CMB
fluctuations and outline the scientific information they encode.
In Section~\ref{astrolimit} we address the astrophysical
limitations faced by observations, mainly represented by
confusion emission of galactic and extragalactic origin. We then
discuss observational issues typical of CMB anisotropy experiments
(Section~\ref{obsissue}), the most important systematic effects
that they have to fight to reach the desired precision
(Section~\ref{siste}), and some of the aspects related to the
analysis of CMB data (Section~\ref{analysis}), in particular for
what concerns the challenges posed by the large data sets
expected in the near future. Then in Section~\ref{experiments} we
attempt an overview of the anisotropy experiments carried out in
the past decade, and we give a synthesis of the observational
status. Finally, we outline the main features of the $MAP$ and
{\sc Planck} missions and future sub-orbital programs, and
discuss some prospects for the future of CMB studies.

\section{CMB ANISOTROPY}
\label{cmbanirev}

According to the standard Hot Big Bang cosmology, the cosmic
expansion started about 15 billion years ago from a phase
characterised by high density and temperature ($\rho \simeq
10^{25}$ g/cm$^{3}$, $T\simeq 10^2$~Gev $\sim 10^{15}$~K at
$t\simeq 10^{-8}$~sec) and the universe is expanding and cooling
down since that time. At primordial high temperatures, matter and
radiation were tightly coupled and behaved like a fluid. At $t
\sim 3\times 10^5$ years ($z_{rec} \simeq 10^3$) the temperature
dropped to $\sim 3000$~K and protons were able to capture
electrons to form neutral hydrogen and other light elements
($^3$He, $^4$He, $^7$Li). This ``recombination'' suddenly reduced
the opacity for Thomson scattering, setting the photons free, and,
since that time, the majority of them have interacted only
gravitationally with matter. The sphere surrounding us at
$z\simeq 1000$, which represents the position at which the CMB
photons seen today last interacted directly with matter, is
called the Last Scattering Surface (LSS).

One of the basic predictions of a cosmic origin of the CMB is that its temperature
be higly isotropic. A remarkable anisotropy, at a level of
$\Delta T / T \sim 10^{-3}$ on angular scale of 180$^{\circ}$, was detected in
the mid '70s \cite{smogoremull77} and interpreted as an effect of the motion of our local
frame with respect the rest frame of the CMB. This signal can be written as:
\begin{equation}
T_{\rm obs} = T_0 \left[1 + \frac{v}{c}{\rm cos}\theta +
\frac{1}{2} \left(\frac{v}{c}\right)^2{\rm cos}2\theta +
\mathcal{O}(v^3)\right], \label{dipole}
\end{equation}
where $\theta$ is the angle between the line of sight
and the direction of motion, and $v$ is the observer's velocity.
The dynamic quadrupole (third term in Eq.~(\ref{dipole})) is rather
small ($\sim 1$\% of the dipole) and it is quite below the
intrinsic CMB cosmic quadrupole. From the $COBE$ measurements of
the dipole it is possible to derive the earth velocity relative
to the CMB: $v_\oplus = 371\pm 1$~km s$^{-1}$ towards $(l,b) =
(264^\circ,48^\circ)$ \cite{fixsen96}. Correcting for the motion
of the earth and sun in the Milky Way, one can derive the
velocity of the Local Group relative to the CMB, about 600~km
s$^{-1}$.

Apart from the locally induced dipole anisotropy, the
CMB field is indeed observed to be highly isotropic. Soon after
the CMB discovery, however, it was realised that the presence of
density fluctuations at the last scattering epoch would
necessarily induce angular anisotropy in the CMB intensity. In
fact, small departures from the isotropic distribution should be
present to explain the structures (galaxies and galaxy clusters)
observed today. Therefore mapping CMB fluctuations provides an
image of the LSS whose statistical properties depend on the
physical process responsible for the formation of the primordial
inhomogeneities and on the cosmological parameters describing the
structure and evolution of the universe.

Because on large scales ($\gtrsim 300$ Mpc) the universe is highly homogeneous and isotropic,
the space-time metric is simply described by the Robertson-Walker \cite{robwal} metric:
\begin{equation}
ds^2 = c^2dt^2 - a^2(t)\left[\frac{dr^2}{1-kr^2} + r^2(d\theta^2 +
{\rm sin}^2\theta d\phi^2)\right]\, ,
\end{equation}
where $k = \{0, \pm 1\}$ is the curvature, and $a(t)$
is an adimensional scale-factor. The dynamical evolution of
$a(t)$ is specified by Einstein's equation of General Relativity
once a stress-energy tensor is provided. Considering the content
of the universe as a perfect fluid with energy-mass density
$\rho$, pressure $p$ and 4-velocity $u^\mu = dx^\mu /ds$, the
tensor reads:
\begin{equation}
T_{\mu \nu} = -pg_{\mu \nu} + (p + \rho c^2)u_\mu u_\nu\, ,
\end{equation}
where $g_{\mu \nu}$ is the metric tensor. This
expression is completely determined once the equation of state $p
= p(\rho)$ is given. With this stress-energy tensor into the
Einstein's equations, one obtains the Friedmann equations:
\begin{equation}
\frac{\ddot{a}}{a} = - \frac{4\pi G}{3}\left(\rho +
\frac{3p}{c^2}\right) + \frac{\Lambda c^2}{3}\, ,
\end{equation}

\begin{equation}
\left(\frac{\dot{a}}{a}\right)^2 + \frac{k c^2}{a^2} = \frac{8\pi
G}{3}\rho + \frac{\Lambda c^2}{3}\, , \label{friedII}
\end{equation}
where $\Lambda$ is the cosmological constant.
Eq.~(\ref{friedII}) provides a direct link between energy density
and geometry of the universe. In particular, if $\Lambda =0$
there is a critical density $\rho_c = (3H^2)/(8\pi G)$ for which
$k=0$, where $H(t) = \dot{a}/a$ is the Hubble parameter. Since
the flat universe has a particular, well defined value for the
critical density, it is usual to express the mass-energy density
as $\Omega \equiv \rho/\rho_c$. As we shall see, recent CMB
results provide strong evidence that $\Omega_0\equiv \rho_0 /
\rho_c \simeq 1$ \cite{debe01,hanany00}. In this case
Eq.~(\ref{friedII}) for $\Lambda = 0$ has very simple solutions:
$a(t) \propto t^{2/3}$ for pressure-less matter dominated
universe and $a(t)\propto t^{1/2}$ for a universe filled with
relativistic matter.

If the universe is dominated by vacuum energy, or if $\Lambda$ is
large, the scale-factor grows exponentially: $a(t) \propto {\rm
exp}(Ht)$ where $H=\sqrt{\Omega_\Lambda c^2/3}$ and
$\Omega_\Lambda$ is the vacuum energy density parameter. This
solution, called the de-Sitter expansion, is of particular
interest in the context of the inflationary paradigm.

\subsection{The inflationary scenario}
\label{inflation}

Although the Hot Big Bang model is successful, it leaves many
open issues which have to be addressed. The most relevant are:
the flatness problem (why is the density so close to the critical
value?); the horizon problem (why does the CMB have the same
temperature at high degree of accuracy on the whole sky if causal
regions on LSS have angular sizes of only few degrees?) and the
origin of the primordial density fluctuations (why is the universe
indeed so clumpy on small scales while it is homogeneous and
isotropic on very large scales?).

In particular great attention was, and is, devoted to the problem
of the origin of inhomogeneities. Early models were based on a
pure baryonic universe, but the level of CMB anisotropy expected
in this simple scenario was too high ($\sim 10^{-3}$) to match
observations \cite{silk67,silk68,peebles70}. Therefore theorists
started to consider a universe composed by a mixture of baryons
and various kinds of dark matter.

Many assumptions have to be made about the initial conditions for
gravitational instability: the geometry of the universe (\eg
flat), the statistical distribution of initial density
fluctuations (\eg Gaussian) and their power spectrum. These
assumptions are included in the inflationary paradigm for which
the very early universe ($t \sim 10^{-34}$ sec) underwent an
exponential expansion (see \cite{linde} for a review). Inflation
solves both the flatness and the horizon problem and specifies
initial conditions for structure formation making specific
prediction on the statistics of the CMB anisotropy as well as on
matter distribution.

A plausible scenario for driving such an expansion is needed and
a physical mechanism, able to generate primordial density
fluctuations is provided by a cosmological scalar field (it is
usual, in particle physics, to represent the zero-spin particles
with a scalar field \ie which is unchanged under coordinates
transformations). If we denote with $\phi$ this homogeneous scalar
field, its energy density and pressure can be written as:
\begin{eqnarray}
\rho_\phi & = & \frac{1}{2} \dot{\phi^2} + V(\phi)\,  \\
p_\phi    & = & \frac{1}{2} \dot{\phi^2} - V(\phi)\, \\
\nonumber \label{scalarfield}
\end{eqnarray}
where the first terms can be regarded as kinetic energy while the
second are potential-binding energy. The Friedmann equations thus become:
\begin{eqnarray}
H^2 & = & \frac{8\pi G}{3}\left[ \frac{1}{2} \dot{\phi^2} +
V(\phi)\right] \, , \\
\ddot{\phi} & + & 3H\dot{\phi} = -\frac{dV(\phi)}{d\phi}\, , \\
\nonumber \label{fried}
\end{eqnarray}

A standard way to solve the above equations is to consider
the scalar field initially displaced from the minimum of the potential
$V(\phi)$ and then slowly approaching the minimum value of $V$. 
In this so-called slow-roll approximation
$(\dot{\phi^2} \ll V, |\ddot{\phi}| \ll H\dot{\phi})$, neglecting terms of higher orders
($\ddot{\phi}$ and $\dot{\phi^2}$), one finally gets:
\begin{equation}
a(t) \sim {\rm exp}(Ht), \hspace{2cm} H(t) = \left( \frac{8\pi
G}{3} V\right)^{1/2} \sim {\rm const.}
\end{equation}

Therefore when $\phi$ is slow-roll approaching the minimum of $V$
a quasi-exponential expansion results. Once inflation is over the
$\phi$ field starts to oscillate around the minimum position and
the decay of these oscillations may lead to particle production
and radiation (``re-heating''). Quantum fluctuations present
during inflation are then ``stretched'' by the accelerated
expansion to become, eventually, density perturbations. These
models predict fluctuations that are Gaussian in origin with a
power law power spectrum $P(k) \propto k^{n_S}$ that is close to a
scale-invariant spectrum $n_S=1$ \cite{bond96}. Therefore
inflation offers a natural physical mechanism for the origin of
primordial density fluctuations, which leave their imprint as
spatial variations in the CMB temperature. Different processes
are responsible  of coupling the primordial density fluctuations
to the radiation, and their efficiency is a strong function of
the angular scale. The dependency of the CMB field on the angular
scale is conveniently described by its angular spectrum.

\subsection{The CMB angular power spectrum}
\label{cmbPS}

Let us expand the CMB anisotropy on the celestial sphere in spherical-harmonic series
\begin{equation}
\frac{\Delta T}{T}(\theta,\phi) =
  \sum_{\ell m} a_{\ell m} Y_{\ell m} (\theta,\phi) \,,
\end{equation}
where $\ell \sim 180^\circ/\theta$, and $a_{\ell m}$ represent
the multipole moments that should be characterised \cite{coles95} by
zero mean, $\langle a_{\ell m}\rangle=0$,
and non-zero variance $C_\ell\equiv\left\langle |a_{\ell m}|^2\right\rangle$
(the angle brackets indicate an average over all observers in the universe;
the absence of a preferred direction implies that $\langle |a_{\ell m}|^2\rangle$
should be independent of $m$). The set of $C_\ell$ is known as the angular power
spectrum and it represents the key theoretical prediction provided by cosmological
models.

If the CMB temperature fluctuations are Gaussian, as suggested by
inflation models, then the angular power spectrum completely
defines their statistical properties. The power spectrum is
related to the 2-point correlation function:
\begin{eqnarray}
C(\theta) &=& \left\langle\left(\frac{\Delta T}{T}(\mathbf{n}_1)
\cdot
\frac{\Delta T}{T}(\mathbf{n}_2)\right)\right\rangle \nonumber \\
          &=& \sum_\ell \frac{(2\ell + 1)}{4\pi} C_\ell
          P_\ell({\rm{cos}}\theta)\ ,
\end{eqnarray}
where $\mathbf{n}_1$ and $\mathbf{n}_2$ are two
unit vectors separated by an angle $\theta$,
and $P_\ell$ are the Legendre polynomial of order $\ell$.
For inflationary models, the $C_\ell$ can be computed accurately as a
function of the cosmological parameters \cite{hu95a, seljak94}. To predict CMB
anisotropies one has to solve the equations for the evolution
of all particle species present (see, e.g., 
\cite{peebles70, vittorio84, bond84}). Alternative theories 
\cite{allen97,ferreira95,turok96,gangui00}
predict the generation of CMB anisotropy with
non-Gaussian fluctuations and these can be tested by studying the
CMB field with higher order moments.

We will next summarise the main sources able to produce
anisotropy in the CMB field. As usual, we distinguish between
intrinsic and secondary anisotropy, depending on whether they
originate at the LSS or at later times.

\subsection{Intrinsic anisotropies}
\label{subsec:intrinsic_anisotropies}

On angular scales larger than the horizon at last scattering
($\theta \gsim 2^\circ \sqrt{\Omega_0}$) CMB anisotropies are
produced by the Sachs-Wolfe effect \cite{sacwol67}, which is due
to metric perturbations that produce a change in the photons'
frequency. To simply understand it, it is worth noting that, in a
Newtonian context, metric perturbations are related to
perturbations in the gravitational potential $\Phi$ that, in turn,
are produced by density perturbation $\delta \rho$. Photons
climbing out of a potential well will therefore suffer
gravitational redshift and time dilation: they are observed at
different times and at different values of the scale factor $a$
with respect to unperturbed photons. Considering adiabatic
perturbation and a matter dominated universe the gravitational
term is given by:
\begin{equation}
\frac{\Delta T}{T} = \delta \Phi\, ,
\end{equation}
while the time-delation term is:
\begin{equation}
\frac{\Delta T}{T} = - \frac{\delta a}{a} = - \frac{2}{3}\frac{\delta t}{t} = -\frac{2}{3}\delta \Phi\, .
\end{equation}
Therefore the final net effect will be:
\begin{equation}
\frac{\Delta T}{T} = \frac{1}{3} \delta \Phi\, .
\end{equation}

One could also take into account potential changes with time along
the photon path from the LSS and us. In this way the photon
alters its redshift as it travels leading to a temperature
perturbation:
\begin{equation}
\frac{\Delta T}{T} = 2 \int {\dot\Phi} d\ell\, . \label{sw}
\end{equation}
These fluctuations are generally
regarded as secondary anisotrpies and
will be discussed in the next section.

The Sachs-Wolfe effect is responsible for the features in the CMB
power spectrum at low $\ell$ since it dominates at scales
$\theta \gsim 2^\circ$. At these large scales there is no causal
connection affecting the initial perturbations: they reflect
directly the initial power spectrum of matter density
fluctuations. It is possible to show \cite{bondefsta87} that if
the initial matter power spectrum has the form $P(k) \propto
k^{n_S}$, then:
\begin{equation}
C_\ell = Q_{\rm rms-PS}^2 \frac{4\pi}{5} { \frac{
\Gamma\left(\ell + \frac{n_S-1}{2}\right)
\Gamma\left(\frac{9-n_S}{2}\right)} {\Gamma\left(\ell +
\frac{5-n_S}{2}\right)\Gamma\left(\frac{3+n_S}{2}\right) }\ , }
\end{equation}
where $Q_{\rm rms-PS}\equiv \sqrt{5C_2/4\pi}$ is the
quadrupole normalisation \cite{smoot96}. If $n_S=1$ (\ie the
so-called Harrison-Zel'dovich spectrum) then $C_\ell \propto
1/(\ell(\ell+1))$. It is therefore usual to plot
$\ell(\ell+1)C_\ell$ as function of $\ell$ since it is possible
to immediately recognise the plateau at small $\ell$ due to the
Sachs-Wolfe effect and directly link it to the initial spectral index.

\begin{figure}
\centerline{\psfig{figure=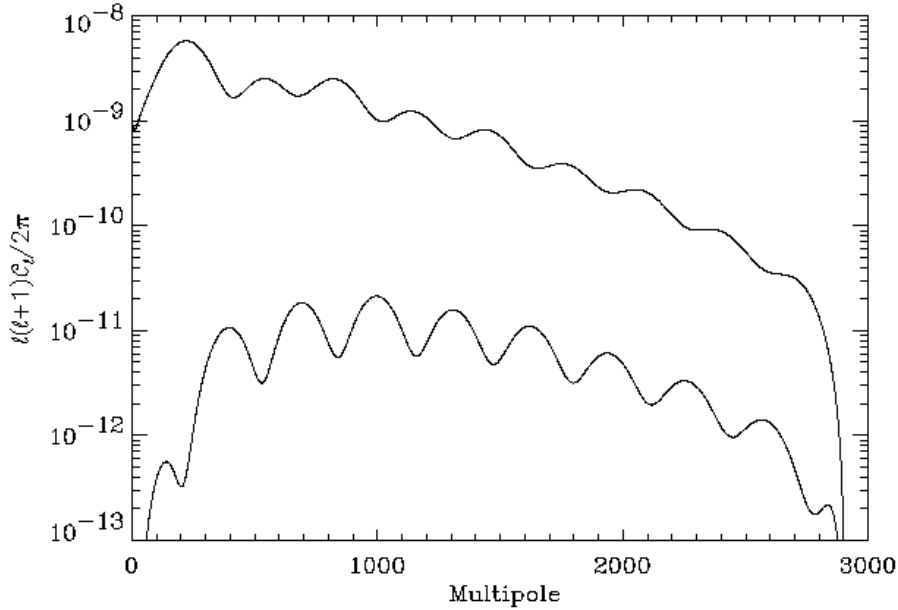,width=12cm}}
\caption{\small A typical angular power spectrum of temperature
(upper line) and polarisation (lower line) anisotropy for a cold
dark matter model with scale--invariant initial conditions. Only
scalar fluctuations are considered.} \label{pola}
\end{figure}

Anisotropies on scales $0.1^\circ\lsim\theta\lsim 2^\circ$ are
related to causal processes occurring in the photon-baryon fluid
until recombination. Photons and baryons are in fact tightly
coupled and behave like a single fluid. In the presence of
gravitational potential forced acoustic oscillations in the fluid
arise: they can be described by an harmonic oscillator where the
driving forces are due to gravity, inertia of baryons and
pressure from photons. Recombination is a nearly instantaneous
process and modes of acoustic oscillations with different
wavelength are ``frozen" at different phases of oscillation (see
Fig.~\ref{pola}). The first (so-called Doppler) peak at degree
scale in the power spectrum is therefore due to a wave that has a
density maximum just at the time of last scattering; the
secondary peaks at higher multipoles are higher harmonics of the
principal oscillations and have oscillated more than once.
Between the peaks, the valleys are filled (the power spectrum
does not go to zero) by velocity maxima which are 90$^\circ$ out
of phase with respect density maxima.

In the short but finite time taken for the universe to recombine
the photons can diffuse a certain distance. Anisotropies on
scales smaller than this mean free path will be erased by
diffusion, leading to the quasi-exponential damping
\cite{silk68,huwh97} in the spectrum at large $\ell$'s. This is
called ``Silk damping'' and becomes quite effective at $\ell\gsim
1000$, corresponding to angular scales $\theta\lsim10'$. Little
contribution from intrinsic CMB anisotropy is therefore expected
at arcmin scales, at least for the standard Cold Dark Matter (CDM)
models.

\subsection{Secondary Anisotropies}
\label{subsec:secondary_anisotropies}

Different processes can generate anisotropy in the path from the
LSS to the observer. Detailed observation of these effects
provides insight on the evolution of the universe after
recombination. A first effect is the Integrated Sachs-Wolfe
effect (ISW; see Eq.~(\ref{sw})) generated by time variations of the
gravitational potential in the photons' trajectory. These time
variations can be due to potential decay, gravitational waves
(tensor perturbations) or non-linear effects associated to
structure formation.

There are mainly three situations in which the potential time
derivative is not zero and they give rise to the Early ISW, the
Late ISW and the Rees-Sciama effect \cite{reessciama}
respectively. Soon after recombination the photon density is not
completely negligible and this causes $\Phi$ to decay producing
the Early ISW effect which peaks slightly to the left of the
first acustic peak. In open or $\Lambda$-dominated models, the
universe eventually becomes curvature or vacuum dominated
respectively. This produces a variation in $\Phi$ yielding the
Late ISW effect since curvature and vacuum are important at low
redshifts. This affects only very large angular scales. When
structures begin to form, entering a non-linear regime, the
approximation of constant $\Phi$ in time is no longer valid and
variations in $\Phi$ cause the Rees-Sciama effect, relevant for
the very small angular scales of the CMB power spectrum.

Another source of secondary anisotropy is gravitational lensing.
While the ISW does change photon energy but not their directions
from the potential gradient parallel to the photon path,
gravitational lensing alters photon directions leaving, to first
order, their energy unchanged. This is produced by the potential
gradient perpendicular to the photons path. The effect on CMB
angular power spectrum is a smoothing of the acoustic
oscillations at large and intermediate scales, while it adds
extra power at small angular scales \cite{zalda98} (ref. Martinez-Gonzalez).

All these are gravitational effects. Other processes producing
secondary anisotropy are related to local and global
re-ionisation of the universe. As for local re-ionisation, this
is usually located in galaxy clusters and is called
Sunyaev-Zel'dovich (S-Z) effect \cite{sunzel70}. This is the
result of the Compton scattering of the CMB photons by
non-relativistic electron gas within clusters of galaxies (see
\cite{repha95} for an excellent review).  This ``thermal''
effect, driven by the thermal motions of the electrons, results
in a systematic shift of photons from the Rayleigh-Jeans to the
Wien side of the frequency spectrum. With respect to the incident
radiation field, the change of the CMB intensity across a cluster
can be viewed as a net flux emanating from the cluster. The flux
is negative below the characteristic frequency, $\nu_0=217\,$GHz,
and positive above it. The change in the spectral intensity is
given by:
\begin{equation}
\Delta I = \frac{(2kT_0)^3}{(hc)^2}\frac{x^4e^x}{(e^x-1)^2}y
\left[x{\rm coth}\frac{x}{2} -4\right]\ ,
\end{equation}
where $x \equiv h\nu/kT_0$ is the dimension-less frequency and $y$ is the Comptonisation parameter:
\begin{equation}
y \equiv \frac{k\sigma_T}{m_e}\int dl T_e n_e\ ,
\end{equation}
where $\sigma_T$ is the Thomson scattering cross
section, $n_e$ and $T_e$ are the electron density and
temperature, respectively, and the integral is computed along the
line of sight.

An additional effect caused by scattering of CMB photons against
electrons in bulk motion is the so-called ``kinematic'' S-Z
effect. Its amplitude is proportional to  the line of sight
component of the peculiar velocity, $v_r$, and can then be used
to determine clusters peculiar velocities
\cite{sunzel80,rephlah91}.

If the universe becomes globally re-ionised, the effect on CMB
anisotropy is quite dramatic. The CMB is scattered by free
electrons and photons we see coming from a given direction may
instead be originated from a completely different direction on
the LSS. This leads to a damping of anisotropies on scales
smaller than the horizon size at the redshift of re-ionisation.
If the universe become re-ionised at redshift $z_r$ and remained
ionised up to now, the probability that a CMB photon never
experienced a scattering is $e^{-\tau}$ where $\tau$ is the
optical depth at re-ionisation. Since the power spectrum is the
square of fluctuations, this probability becomes $e^{-2\tau}$. In
a standard CDM universe $\tau$ becomes larger than 1, implying
that almost all photons are scattered at $z_r\sim 50$
corresponding to scales smaller than few degrees. Therefore the
larger $\ell$'s are suppressed by a factor $e^{-2\tau}$ while the
low $\ell$'s are left unaffected.

\begin{figure}[here]
\centerline{
    \psfig{figure=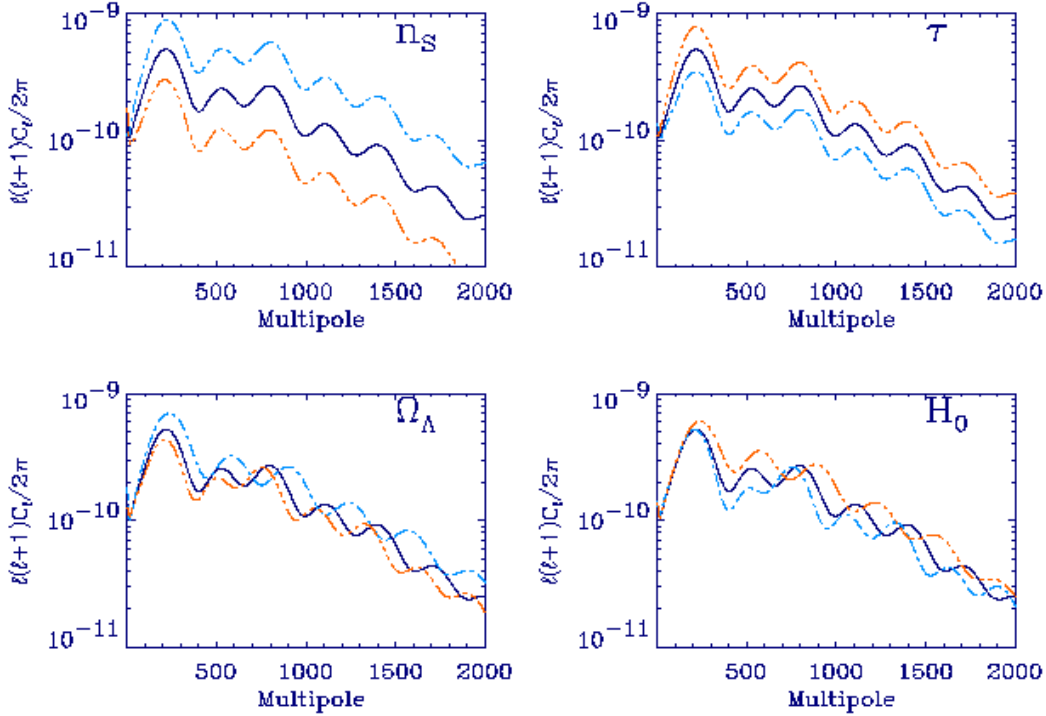,width=14cm}}
\caption{
Examples of the dependence of the shape of the angular power spectrum
$C_\ell$ on the value of the cosmological parameters. The
fiducial power spectrum (solid line) is a model with $\Omega_b=0.05$,
$\Omega_{\rm CDM} = 0.35$,
$\Omega_\Lambda=0.6$, $H_0=65$~km~s$^{-1}$~Mpc$^{-1}$,
$n_S=1.0$ and $\tau=0.3$.
In each panel, all the parameters are unchanged except for the one
indicated. The dash-dot-dot-dot line and the dash-dot line
refer to values:
$n_S = (0.8,1.2)$ (upper left);
$\tau = (0.0,0.5)$ (upper right);
$\Omega_\Lambda=(0.4,0.8)$ (lower left);
and $H_0=(40,80)$~km~s$^{-1}$~Mpc$^{-1}$ (lower right).}
\label{params}
\end{figure}

\subsection{Dependence on cosmological parameters}
\label{subsec:dependence_cosmological_parameters}

We have seen that a number of processes contribute to the
generation of CMB anisotropy field. Starting from such processes,
the shapes, heights and locations of peaks and troughs in the
angular power spectrum are predicted by all models which, like
CDM, are based on the inflationary scenario. Furthermore the
details of the acoustic features in the power spectrum depend
critically on the chosen cosmological parameters \cite{jungman96,
hu95}, which in turn can be accurately determined by a precise
measurement of the anisotropy pattern. In this section we briefly
discuss the sensitivity of the $C_\ell$'s on the value of some
fundamental parameters (see Fig.~\ref{params}).

\begin{itemize}

\item{\it Total density}: a decrease in $\Omega_0$
corresponds to a decrease in curvature and increase, although small,
in Late ISW
effect, with a corresponding shift of the power spectrum peaks
towards high multipoles. In particular it can be shown that the
angular scale of the first peak is $\ell_1\approx
200\sqrt{\Omega_0}$.

\item{\it Barion density}: an increase in the baryon
fraction $\Omega_b$ will increase odd peaks (compression phase of
the baryon-photon fluid) due to extra-gravity from baryons with
respect to the even peaks (rarefaction phase of the fluid
oscillation).

\item {\it Hubble constant}: a decrease
in $h$ ($h \equiv H_0/100$~km s$^{-1}$ Mpc$^{-1}$), maintaining
$\Omega_bh^2$ fixed at the nucleosynthesis value (typically
$\Omega_b h^2 = 0.019 \pm 0.002$, see \cite{burles98}),
corresponds to a delay in the epoch of matter radiation equality
and to a different expansion rate; this boosts the peaks and
slightly changes their location in $\ell$-space.

\item {\it Cosmological constant}: an increase in the
cosmological constant, $\Lambda$, in a flat space, again
corresponds to a delay in matter-radiation equality with a
boosting and shifting effect on the peaks. Furthermore the low
$\ell$'s are affected by the Late ISW effect.

\item {\it Spectral index}: increasing $n_S$  will raise
the angular spectrum at large $\ell$'s with respect to the low
$\ell$'s.

\item {\it Reionisation}: if the intergalactic medium was
re-ionised at $z \ll 1000$, then the power at $\ell> 100$
would be suppressed by a factor of $e^{-2\tau}$. Evidence
for early reionisation would set important constraints on
theories of galaxy formation and on the epoch at which the
first structures formed. A recent combined analysis \cite{tegzalham}
of CMB data and Large Scale Structure data together with Big Bang
nucleosynthesis priors yields $\tau \simeq 0.1$.


\item {\it Gravitational wave background}: gravitational waves
(tensor modes) generate additional CMB anisotropies, but only at
large angular scales. Inflationary models predict that the ratio
of tensor to scalar contribution to the quadrupole anisotropy
($r=C^{(T)}_2/C^{(S)}_2$) is related to the spectral index of
primordial tensor fluctuations, $n_T$.

\end{itemize}

As we shall see, present CMB anisotropy data already constrain
significantly some of these parameters. However, much more
accurate data are required to fully extract the cosmological
information encoded in the CMB power spectrum.

\subsection{CMB polarisation anisotropies}
\label{subsec:cmb_polarisation_anisotropies}

Theoretically, the degree of linear polarisation is directly
related to the quadrupole anisotropy in the photons when they
last scatter \cite{kaise83,kamion96,selzal97}. While the exact
scale dependence of polarisation depends on the mechanism for
producing the anisotropy, several general properties can be
identified. In particular, the polarisation power spectrum peaks
at angular scales smaller than the horizon at last scattering,
since the processes that produce it are causal (see
Fig.~\ref{pola}). Furthermore, the polarised fraction of the
temperature anisotropy is small, since only those photons that
last scattered in an optically thin region could have possessed a
quadrupole anisotropy (multiple scattering causes photon
trajectories to mix and hence erases anisotropy). This fraction,
which depends on the duration of last scattering, is expected to
be of ~5-10\% on a characteristic scale of tens of arcminutes.

CMB polarisation provides an important tool for reconstructing the
model of the fluctuations from the observed power spectrum. In
fact polarisation probes the epoch of last scattering directly,
unlike the temperature fluctuations which may evolve between LSS
and the present (see Sect.~\ref{subsec:secondary_anisotropies}).
This localisation in time is a very powerful tool for
reconstructing the sources of anisotropy. Moreover, different
sources of temperature anisotropies (scalar, vector and tensor)
give different patterns in the polarisation, both in its intrinsic
structure and in its correlation with the temperature
fluctuations themselves.

Finally, the CMB polarisation power spectrum provides information
complementary to the temperature power spectrum even for ordinary
(scalar or density) perturbations. This can be of use in breaking
degeneracies in the determination of the cosmological parameters,
thus constraining them even more accurately \cite{zasespe97}.


\section{ASTROPHYSICAL LIMITATIONS}
\label{astrolimit}

In a CMB experiment, the measured signal contains many different
contributions, beside the CMB, some of which are astrophysical
(both galactic and extra-galactic) in origin. These astrophysical
foreground emissions can be separated from the CMB up to a certain level of
accuracy by means of multi-frequency measurements,
although there is currently no emission component for which both
angular and frequency destributions are precisely determined.
Galactic and extragalactic microwave and sub-mm emission, 
while representing a challenge for CMB experiments,
yield by themselves interesting astrophysical information.

\subsection{Galactic synchrotron emission}
\label{subsec:galactic_synchrotron_emission}

Diffuse Galactic synchrotron emission originates from cosmic-ray
electrons accelerated in the Galactic magnetic field. Therefore
the intensity of this emission depends on the energy distribution
of electrons as well as on the structure of the Galactic magnetic
field (see \cite{smoot99} for a recent review). Synchrotron
radiation dominates at frequencies $\nu \lsim 10$~GHz which makes
it possible to obtain, in this frequency range, sky maps of genuine
synchrotron radiation. Large sky area surveys, however, suffer
significant uncertainties associated with calibration errors,
zero levels and scanning strategy artifacts. At frequencies below
1~GHz instrument gain and zero-level uncertainty are considerably
smaller than the observed signal and reliable sky maps have been
obtained at 408~MHz \cite{hasla82} over the whole sky, and at
1420~MHz \cite{reirei86} and 2326~MHz \cite{jona98} over large sky
fraction. The typical angular resolution of these surveys is
$0.85^\circ\div 0.3^\circ$.

The synchrotron brightness temperature is expected to scale with
frequency as a power law, $T_b \propto \nu^{-\beta}$, with a
spectral index $\beta$ which vary with frequency and position
according to the energy distribution of electrons and Galactic
magnetic field structure. The mean temperature brightness
spectral index between 38 and 1420~MHz is $\beta \simeq 2.7$ with
variations of at least 0.3 and the de-striped 408 and 1420~MHz
maps gave temperature spectral indices of 2.8 to 3.2 in the
northern galactic pole regions \cite{davies96}. Some evidence of 
a high frequency steepening of $\beta$ was
reported \cite{plata97} considering data taken from the
White Mountain site \cite{smoot85} with $18^\circ$ angular
resolution together with 408 and 1420~MHz maps to estimate the
spectral index between 1 and 10~GHz. A mean value of $\beta =
2.76\pm 0.11$ has been found.

As for the angular dependence of synchrotron emission there is
not a general agreement. Some authors \cite{tegefst96} suggested
that synchrotron angular power spectrum should scale as
$\ell^{-3}$ as observed for dust emission. However other authors
\cite{lase96} considering the region observed by the Tenerife
experiment (see Sect.~\ref{subsubsec:tenerife}) found a rather
low amplitude as well as a flatter angular dependence
($\ell^{-2}$).

Synchrotron radiation is expected to be polarised by a large
fraction (in principle up to $\sim$ 70\%). This could be a problem
in CMB polarisation measurement since the CMB polarised signal is
expected at the few $\mu$K level. Furthermore, the spectral
behaviour of synchrotron polarised emission as well as its angular
dependence are not known. Detailed studies exploiting the small
database of polarisation measurements available are currently
on-going \cite{bacci01}.

\subsection{Galactic free-free emission}
\label{subsec:free_free}

When a free electron is accelerated by the Coulomb field of ions,
free-free emission results. This occurs when hot electrons
($T_e\gsim 10^4$~K) interact with an ion, starting from an
un-bound state and ending, after interaction, in another un-bound
state. The physics of this process is well know. It is possible to
calculate the emissivity per volume along the line of sight of a
given distribution of electrons. When integrating the emissivity
we obtain the optical depth and then the brightness temperature
which scales as:
\begin{equation}
T_b^{ff} \sim 26\,\mu{\rm K}\,\,\, \nu^{-2.1} T_e^{-0.35}\ EM\ ,
\label{eq:free_free}
\end{equation}
where $T_e$ is the electron temperature and $EM \equiv
\int N_i N_e dl$ is the emission measure which depends on the
number of ions, $N_i$, and electrons, $N_e$, per unit volume. Due
to the harder spectrum with respect to synchrotron emission,
free-free should dominate the microwave sky at frequencies
$\nu \gsim 10$~GHz. However, the signals are extremely weak and
no sky survey of free-free emission free of other components 
is available.

Free-free emission arises from two distinct components, one
discrete and one diffuse. The former is clearly associated with
H{\sc ii} regions. These are regions of intense star formation
where hot electrons are present. They are mainly localised along
the galactic plane ($|b|<5^\circ$) with very few exceptions (\eg
the Orion Nebula). In a recent work \cite{pala99} a catalogue of
about $\sim 1200$ H{\sc ii} regions at 2.7 GHz has been produced.
Since free-free emission is expected not to be polarised, H{\sc ii} regions could
be used to separate instrumental polarised components in CMB
measurements at sub-degree angular resolution.

As for the diffuse component, we have to rely on tracers of the
ionised interstellar medium. Most of the available information at
intermediate and high latitudes comes from H$\alpha$ surveys
from which it is possible to derive the
free-free brightness temperature:
\begin{equation}
T_b^{ff} \sim 1.68\mu{\rm K} \langle g_{ff}\rangle T^{0.4\ {\rm
to}0.7} \lambda I_{H \alpha}\ ,
\end{equation}
where $\lambda$ is in cm, $I_{H\alpha}$ is the intensity
of the H$\alpha$ emission and $\langle g_{ff}\rangle$ is the Gaunt
factor. Therefore a measurement of H$\alpha$ emission directly
yields microwave free-free brightness temperature.

A correlation between free-free and dust emission was found
\cite{kogut96} comparing the $COBE$-DMR maps with the DIRBE maps.
Modelling the emission as combination of dust and radio
components, a radio spectral index $\beta_{\rm radio} \simeq
-2.1^{+0.6}_{-0.8}$ was determined in good agreement with the expected
index of free-free emission (Eq.~(\ref{eq:free_free})). This was
later confirmed \cite{leitch97,deolico97} down to 14~GHz with a
spectral index consistent with free-free emission at 95\%
confidence level. However the microwave emission reported in
\cite{leitch97,deolico97} is 5-10 times larger than derived
from H$\alpha$ measurements in the same regions \cite{gau96}.
This is quite interesting: microwave emission is correlated
with dust showing a spectral index consistent with free-free but
the H$\alpha$ emission cannot account for the overall emission
observed. Recently \cite{drailaza98} a possible explanation of
this ``anomalous'' emission was suggested, invoking the
contribution from electric and magnetic dipole from small
spinning dust grains. This emission should in fact peak between
10 and 50~GHz, and it would
explain the observed correlation with dust emission as well as
the lack of H$\alpha$ emission. More multi-frequency measurements
are required to clarify the issue.

As for the spatial behaviour of free-free emission \cite{kogut96}
at angular scales $>7^\circ$ the angular spectral index was
estimated $\simeq 3.0$, a well-determined value for dust and HI
at degree scales. This was confirmed to some extent
\cite{deolico97} for the North Celestial Pole (NCP) region in a
correlation of Saskatoon data (see
Sect.~\ref{subsubsec:saskatoon}) with the IRAS and DIRBE maps.

\subsection{Galactic dust}

Galactic dust emission originates from dust grains heated by
interstellar radiation: dust absorbs UV and optical photons and
re-emits in the far-IR. Dust emission typically dominates
at $\nu\gsim 100$~GHz and the total intensity
depends upon gas chemical composition, dust to gas ratio and
grain composition, structure and dimension.

The spectral shape of dust emission can be well modelled by a
modified blackbody emissivity law $I_\nu \propto \nu^\alpha
B_\nu(T_d)$ where $\alpha \simeq 2$ is the emissivity, $T_d$ is
the dust temperature and $B_\nu(T)$ is the planckian function.
Exploiting $COBE$-DIRBE data and focussing on high galactic
latitude regions, a dust temperature of 18~K has been found
\cite{kogut96}. More recently a dust emission map obtained by
joining together the $COBE$-DIRBE and the IRAS maps with IRAS
angular resolution ($\sim 6'$) has been produced \cite{schle97}.
A best fit model with two dust temperature (16 and 9.5~K) and two
distinct emissivities (2.7 and 1.7 respectively) was proposed based on
exploiting the $COBE$-FIRAS data in the range between 100 and
2100~GHz.

The analysis of dust contamination is further complicated by the
highly non-Gaussian nature of its Galactic distribution. The new
Berkeley-Durham dust map \cite{schle97} shows that its power
spectrum is not well described by a global $\ell^{-3}$ power law
everywhere at high Galactic latitude, as it was claimed
\cite{gautier92}; in some high galactic latitude patches the
power spectra are closer to $\ell^{-2.5}$, but with amplitudes
differing from patch to patch.

\subsection{Extragalactic sources}
\label{subsec:extragalactic_sources}

Beyond galactic foregrounds, another unavoidable
fundamental limitation for CMB measurements come from
extra-galactic sources that are expected to be a significant
challenge for future high-resolution CMB experiments.
The issue of small scale fluctuations due to
extra-galactic sources has been long discussed in literature
(e.g. \cite{france,tegefst96}). Due to the sharp rise in the dust
spectrum with increasing frequency ($\propto \nu^{3.5}$)
different populations of bright sources below and above 200~GHz
show up. Radio sources (``flat''-spectrum radio-galaxies, BL Lacs
objects, blazars and quasars) dominate at low frequencies while
dusty galaxies strongly contribute at high frequencies.

The large uncertainties on number of counts and spectra for radio
sources do not allow a comprehensive model of source counts at
$\sim 100$~GHz, although simple models \cite{toffo97,dane95} seem
to be remarkably successful, as they properly account for deep
counts at 8.44~GHz recently produced. The situation for
dusty galaxies is more complicated since evolutionary properties are
poorly known and source counts are strongly evolution sensitive.
The situation is however improving with the ISO-CAM, ISO-PHOT and
SCUBA data \cite{france97} although limited to small sky areas.
Number of counts for both radio and dusty galaxies has been
studied in the context of the {\sc Planck} mission
\cite{toffo97} showing that at 5$\sigma$ level several hundreds
of sources will be detected at each frequency channel.

Even assuming that the resolved sources can be completely removed,
we are left with the background of the un-resolved ones. A
Poisson distribution over the whole sky would produce
a simple white-noise power spectrum, with the same power in all
multipoles $\ell$, so that it dominates cosmic signal on
small scales. For radio point sources \cite{toffo97} the level
for the power spectrum is found to be:
\begin{equation}
\sqrt{\ell(\ell+1)C_\ell/2\pi} \simeq \frac {{\rm sinh}^2(\frac{\nu}{113.6})}
{(\nu/1.5)^{4.75 -0.185 {\rm log}(\nu/1.5)}} \ell\ ,
\end{equation}
where $C_\ell$ has units of K$^2$ and $\nu$ is expressed
in GHz. The point source contribution is well below the level of
CMB fluctuations in the range $50\div 200$~GHz at angular scales
larger than few arcminutes. Source clustering would add some
power on larger scales; however, its contribution is found to be
generally small in comparison with the Poisson term
\cite{toffo97}.

\begin{figure}
\centerline{ \psfig{figure=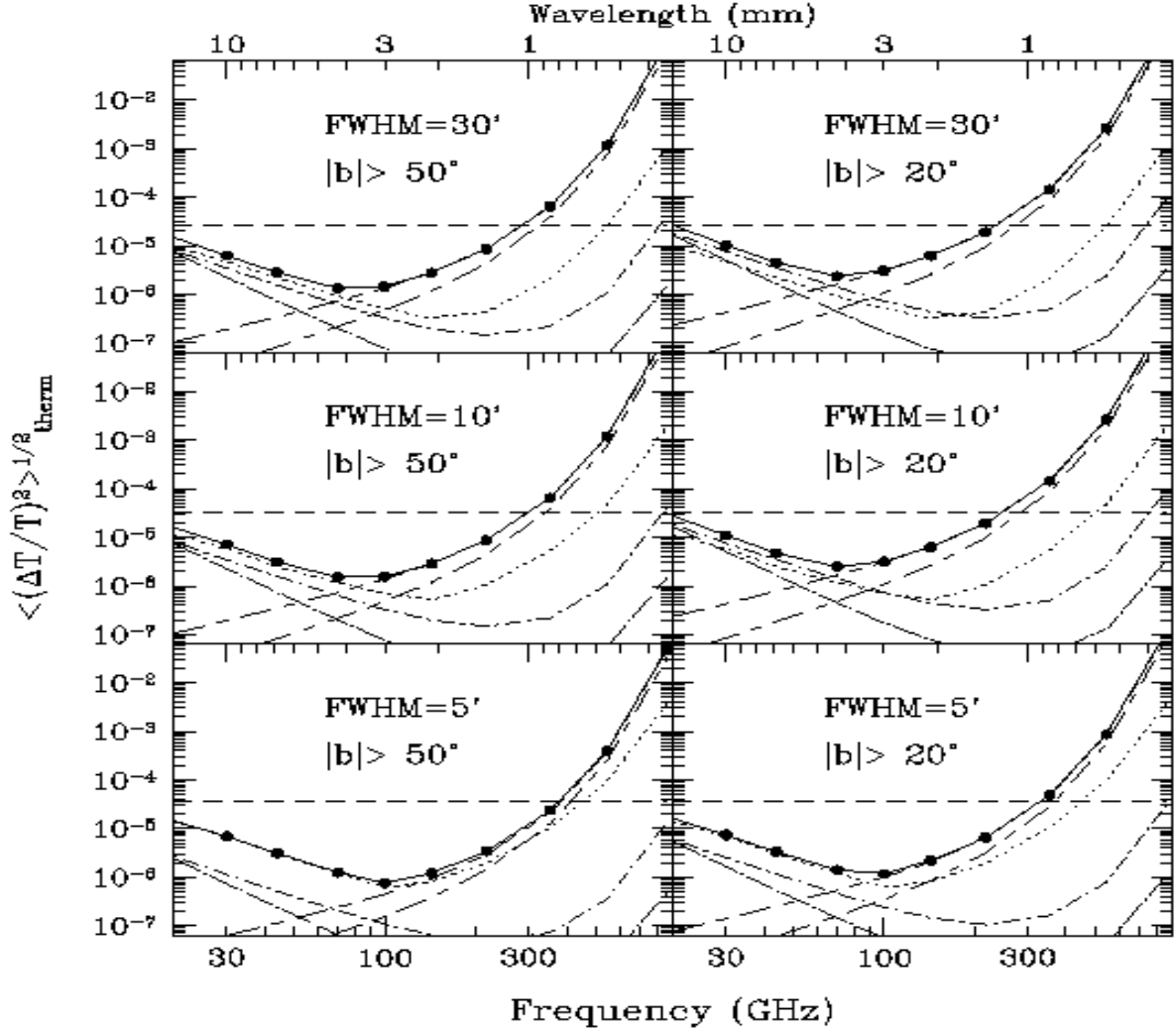,width=16cm,height=14cm}}
\caption{Temperature fluctuations for three different angular
resolutions and two different cuts in galactic emission, as a
function of frequency. The horizontal dashed line is the expected
level of CMB fluctuations from a standard CDM model. Dot-short
dashed, dot-long dashed and long-short dashed represent the mean
contribution from Galactic free-free, synchrotron and dust
emission respectively. Lower long/short dashed is a lower limit
for Galactic dust fluctuations. The dotted curves represent the
contribution from extra-galactic point sources fainter than 100
mJy \cite{toffo97}. The solid curve is the quadratic sum of all
the contributions and the filled circles are the selected {\sc
Planck} frequencies. From De Zotti \etal \cite{dezotti99}}.
\label{fore}
\end{figure}

Fig. \ref{fore} summarises the relative importance of galactic and
extra-galactic foreground fluctuations with respect to CMB
anisotropy as a function of frequency for different angular
scales and galactic cuts. In all cases, the combination of
foreground fluctuations in the range 70-100 GHz is at least an
order of magnitude below the typical CMB anisotropy. While this
shows that, with proper choice of frequency, foregrounds do not
represent a severe limitation to first-order statistical studies,
it is also clear that, to separate components at ~1\% level, a
careful multi-frequency approach of instrument design and data
analysis is mandatory.

\section{OBSERVATIONAL ISSUES}
\label{obsissue}

Following the $COBE$-DMR first detection \cite{smoot92}, 
observations of the CMB
anisotropy have experienced a time of explosion. At present, as
many as 70 papers report positive detection of $\Delta T/T$ at a
variety of angular scales and with different observing techniques.
Here we present some general issues relevant to CMB anisotropy
experiments, related to the observing strategy, design and
technical features which ultimately impact on the final precision
in the measurement.

\subsection{Cosmic and sample variance}
\label{subsec:cosmic_sample_variance}

The ``cosmic variance'' sets the ultimate limit on the accuracy
of our estimates of the power spectrum. In most theories the CMB
field is a single realisation of a stochastic process, and we
should not expect our observable universe to follow exactly the
average over the ensemble of possible realisations. This is
equivalent to say that the $a_{\ell m}$ coefficients are
independent identically distributed Gaussian random variables
(for a given $\ell$) and therefore the $C_\ell$ are a $\chi^2$
distribution with $2\ell +1$ degrees of freedom. The variance in
$C_\ell$ is then:
\begin{equation}
\frac{\Delta C_\ell}{C_\ell} =\sqrt{\frac{2}{2\ell+1}}\ ,
\end{equation}
which is quite important at low $\ell$ since a small
number of modes ($m = 2\ell +1$) is available.

Similarly, incomplete sky coverage degrades the accuracy on
$C_\ell$ since the observed region of the sky may not be
representative of the realisation as a whole. For experiments
which do not cover the full sky, this ``sample variance'' is
larger than the cosmic variance by a factor proportional to the
inverse of the fraction of the sky surveyed.

\subsection{Window functions and observing strategy}
\label{subsec:window_function}

The experiment window function $W_\ell$ is determined by the
instrument beam and sky scanning technique, and selects the
angular range at which the measurements are sensitive. Typical
beam switch experiments have $W_\ell$ peaking in a relatively
narrow range of the corresponding scales of the power spectrum
\cite{bond91}. Early measurements were designed to sparsely
sample the autocorrelation function at a fixed angular scale. On
the other hand, to obtain a model--free reconstruction of the
power spectrum one needs a function $W_\ell$ nearly constant over
a range of angular scales as wide as possible. This is the
characteristic of an imaging observation \cite{readhead92}, where
several different angular scales are simultaneously
probed. Interferometry and arrays of radiometers or bolometers
with proper sky scanning strategy, as we shall see, can both
obtain images of large areas of the sky with adequate sensitivity.

The sky power averaged over the region observed with a beam
characterised by a window function $W_\ell$ can be written as:
\begin{equation}
\left(\frac{\Delta T}{T}\right)_{rms}^2 = \sum_\ell \frac{2\ell +
1}{4\pi} C_\ell W_\ell \label{eq1}
\end{equation}
which is in fact the convolution of the sky power with
the function $W_\ell$. The most simplest window function
is that of a single Gaussian beam of width $\sigma_B$ (related to
beam Full Width Half Maximum by FWHM = $\sigma_B 2\sqrt{2{\rm
ln}2}$) for a full sky coverage:
\begin{equation}
W_\ell = {\rm exp}[-\ell(\ell+1)\sigma_B^2]\ .
\end{equation}

Note the high-$\ell$ cut-off due to finite angular
resolution, which occurs at $\ell_{\rm cut off} \cong 1/\sigma_B$.
If we consider an experiment that measures temperatures by
chopping between 2 or 3 beams in the sky, the window function has
additional terms:
\begin{equation}
{\rm exp}[\ell(\ell+1)\sigma_B^2]\ W_\ell =
\bigg\{\begin{array}{lr}
 2[1-P_\ell({\rm cos}\ \theta_c)] & {\rm 2-beam}\\
1/2[3-4P_\ell({\rm cos}\ \theta_c)+P_\ell({\rm cos}\ 2\theta_c)] & {\rm 3-beam} \\
       \end{array}
\end{equation}
where $P_\ell$ are the Legendre polynomial and
$\theta_c$ is the chopping angle. These observing strategies
introduce a low-$\ell$ cut-off. More generally $W_\ell$ is the
diagonal part of the window function matrix $W_{\ell\ell'}$ which
takes into account the coupling between different multipoles
$\ell$ due to the observational strategy and to the
non-orthogonality of the spherical harmonics on a limited region
of the sky (in general observed by a multiple beam switch
experiment).

The knowledge of the window function is extremely important when
comparing results from different experiments. It is usual to
represent power spectrum data in terms of $(\Delta T/T)_\ell^2
\equiv \ell (\ell + 1 )C_\ell /2\pi$ (band-power). Therefore from
the measured $(\Delta T/T)_{\rm rms}^2$ and using the expression
(from Eq.~(\ref{eq1})):
\begin{equation}
\left(\frac{\Delta T}{T}\right)_{\rm rms}^2 = \left(\frac{\Delta
T}{T}\right)^2_\ell \sum_\ell W_\ell \frac{2\ell
+1}{2\ell(\ell+1)}\ ,
\end{equation}
it is possible to obtain $(\Delta T/T)_\ell$.

\subsection{Instrument related variance}
\label{subsec:instrument_related_variance}

Detector sensitivity and angular resolution are key performance parameters.
Combining cosmic, sample and instrumental effects, the final
fractional error on the $C_\ell$'s can be written as follows
\cite{knox95}:
\begin{equation}
\frac{\delta C_\ell}{C_\ell} = \sqrt{\frac{2}{ {f_{\rm sky}(2\ell
+1)}}} \bigg[1+\frac{A\sigma^2}{N C_\ell W_\ell}\bigg]\, ,
\end{equation}
where $f_{\rm sky}$ is the fraction of the sky surveyed, $A$ is
the physical dimension of the surveyed area, $\sigma$ is the {\it rms} noise
per pixel and $N$ is the number of observed pixels.

We will discuss in some detailed the issue of sensitivity and how
technically this is addressed in Sect.~\ref{subsec:technology}. As
for the angular resolution, we note here that the FWHM of microwave
horns are limited to $>5^\circ$-$6^\circ$, which probes the power
spectrum only to $\ell \lesssim$ 20; as a consequence, to perform
sub-degree scale measurements either interferometry techniques or 
the use of a telescope is required. Interferometers
can achieve arcminute angular resolutions at cm wavelengths,
while telescopes need 
a primary reflector with an aperture $D\approx \theta_{\rm
FWHM} \lambda \approx$ 1 to 2 meters to reach $\sim$ 0.5$^\circ$.
Generally, Gregorian or
Cassegrain off-axis, clear-aperture optical systems are employed
to minimise diffraction from the sub-reflector and from the
support structure. Detailed shaping of the reflector surface may
be used to minimise beam aberration. We will discuss issue
related to optical effects in
Sect.~\ref{subsec:main_beam_distortions} and
\ref{subsec:straylight}.

\subsection{Atmosphere}
\label{subsec:atmosphere}

The radiation coming from the earth environment represents one of
the main problems for sub-orbital experiments, because it adds a
background which varies both in time and direction. The
uncertainty in the atmospheric contribution in turn affects the
capability of removing astrophysical foregrounds in the analysis.
While it has been proven that ground--based experiments can
produce high quality CMB anisotropy data, it is unlikely that
atmospheric noise can be rejected or removed to the levels
required to accurately subtract the galactic foregrounds on large
portions of the sky.

The atmosphere emits and absorbs radiation in a complicated way,
which depends on wavelength, pressure, temperature and chemical
composition, with O$_2$, O$_3$ and H$_2$O as the the most
important microwave-emitting molecules \cite{danese89}. 
Three spectral windows
($<$ 15~GHz, 30-40~GHz, 85-110~GHz) of the microwave atmospheric
spectrum are exploited by ground-based measurements usually
taking advantage of the optimal performance in this frequency
range of coherent receivers.

The impact of atmospheric radiation largely depends on the
variability of the water vapour content. For this reason
ground-based experiments are usually located in dry high altitude
sites. For example, the Izana site at Tenerife, Canary Islands
\cite{davies92,lasenby94} has a typical water vapour column of
$\sim 2$~mm and atmospheric antenna temperature $T_A\sim 4.7$~K
and $T_A\sim 8.7$~K at $28$ and $38$~GHz, respectively. The
Antartic Plateau is possibly the best ground-based site in terms
of atmospheric emission and stability \cite{meinhold91, meinhold93}. 
At the South Pole, at
30~GHz, the atmospheric emission amounts to $T_A\sim 4.6$~K, with
only 0.15~K contributed by water vapour (0.5~mm H$_2$O column),
and the atmospheric noise is at a level of $3\times 10^{-6}$
$T_A$. The impact of atmospheric noise on the data also strongly
depends on the instrument concept and scanning technique, and it
is particularly well suppressed in interferometer experiments. In
addition to water vapour fluctuations, pressure gradients in the
observed sky patches are likely to induce significant large-scale
variations of the O$_2$ emission, as direct measurements from the
South Pole site have shown \cite{bersanelli94}.

For balloon-borne experiments the overall signal from the
atmosphere is reduced by two to three orders of magnitude
depending on frequency. The reduced pressure yields a lower pressure
broadening of lines which are now clearly visible. The effects of
the atmosphere are strongly reduced (although not suppressed) so
that high--frequency measurements ($100 < \nu < 500$~GHz) can be
performed. Only for a space mission is the whole frequency range 
available with no limitations other than the unavoidable galactic
and extragalactic emission.

\subsection{Ground Radiation}
\label{ground_radiation}

Due to the large solid angle of the earth compared to the beam
angular resolution (say $\sim 0.5^{\circ}$), for a ground--based
experiment seeking final accuracies $\Delta T/T \sim$ $10^{-6}$,
the unwanted signal from the ground needs to be rejected by a
factor as high as 10$^{13}$ to fall below significance level. For
anisotropy experiments one is concerned only with the level of
the {\it variable} component of ground contamination from pixel
to pixel, which can be made, say, 1000 times smaller than the
total ground contribution; even in this assumption (quite
optimistic for large sky coverage experiments), the
requirement of $\sim -100$ dB rejection is still extremely tight,
and it becomes proportionally tighter moving to higher angular
resolutions. From balloon altitudes, since the distance of the
gondola from the earth is negligible compared to the earth
radius, the sidelobe and straylight rejection required is of the
same order as for ground-based experiments. A space mission from
a low--earth orbit would only marginally relax such extreme
requirement, since the earth would still cover about $1/4$ of the
total solid angle. In fact, microwave emission from the earth has
been a serious concern in the design and systematic error
analysis of the $COBE$--DMR experiment, even at the relatively
broad beams ($7^{\circ}$) of its antennas \cite{kogut92}. From
the $COBE$ 900 km circular orbit, the earth is a circular source
with angular diameter 122$^{\circ}$ and minimum temperature
285~K. Upper limits to the antenna temperature of the earth signal
contribution to the DMR 2--years data were at level 25--60 $\mu$K
\cite{bennett94}.

Clearly rejection of earth radiation is a major challenge to
balloon--borne or low--earth orbit experiments aiming to reach
sensitivities a factor $\sim$ 10 better than $COBE$--DMR with
beam areas smaller by a factor of $\sim$100 to $\sim$1000. Only
by moving the instruments to a far--earth orbit, is the earth's solid
angle greatly reduced, thus decreasing by the same factor the
required rejection. Orbits around the sun-earth libration point
L2 ($\sim 1,500,000$~km from earth) have been selected for both
$MAP$ and {\sc Planck}. The required rejection for earth
radiation becomes comparable to that required to suppress sun
radiation ($\simeq 10^{9}$), a level that can be obtained (and
tested) with careful, though conventional design of the optics
and shielding. We will come back to the more general issue of
off-axis straylight in Sect. \ref{subsec:straylight}.

\subsection{Calibration}
\label{subsec:calibration}

The radiation power collected by a generic CMB detector is
converted into a voltage $V$ that needs to be calibrated in
physical units. Calibration is typically in terms of antenna
temperature, $T_A \equiv P/k\Delta \nu$, proportional to the
power received $P$ per unit bandwidth $\Delta \nu$. For a linear
system $T_A = GV + T_{\rm offset}$, where $T_{\rm offset}$ is an
instrument (constant) offset term. By observing two sources of
known antenna temperatures $T_{A,1}$ and $T_{A,2}$, the
calibration constant $G$ is readily determined as $G
=(T_{A,1}-T_{A,2}) / (V_1 - V_2)$. Although in principle $G$ is a
constant characteristic of the receiver, in practice it undergoes
drifts and time fluctuations. Thus good calibration requires well
known, stable sources with adequate flux levels, to be observed
at time intervals short compared to the variation of $G$.

Controlled blackbody targets at liquid He or N temperatures and
suitable celestial sources are typically adopted. Jupiter, Mars,
Saturn and Venus provide signals at $\sim 10$ to $100$ mK in a
$\sim 10'$ beam at mm wavelengths, a level adequate for
calibration as well as for main beam mapping. Uranus and Neptune
are also detectable sources with lower signal level (few hundred
$\mu$K). The accuracy of planet calibration is limited to
$3$-$10\%$ by the uncertainty in their brightness temperature in
the microwaves (e.g. \cite{depater90}). Experiments at degree
scales also use the moon as a calibration source. Occasionally
strong radio sources are employed, such as Cas A, Carina Nebula,
Tau A. It may also be possible to use bright H{\sc ii} regions,
which however are often found in complex fields in the galactic
plane, and beam pattern effects must be carefully separated
\cite{pala99}.

For experiments covering significant sky areas with appropriate
scanning strategies and at frequencies below $\sim 300$ GHz, a
nearly ideal celestial calibrator is the CMB dipole: its
amplitude ($\Delta T_D \sim$ 3.37~mK) and distribution are
extremely well known \cite{kogut96b, fixsen96} and allow for
countinuous observation. Furthermore, the dipole has the same
frequency spectrum of the signal to be measured, i.e. the CMB
itself. Balloon experiments like Boomerang and Maxima (see 
Sect.~\ref{subsubsec:Boomerang} and \ref{subsubsec:Maxima})
achieved a few percent calibration accuracy using the dipole. Crosscheck
with an independent source is a desirable strategy. For example,
Boomerang also used an internal calibrator based on a germanium
lamp with a well known signal amplitude and temporal profile
\cite{masi99}. At frequencies $>300$ GHz the Galactic plane as
measured by FIRAS can be conveniently used.

The $MAP$ and {\sc Planck} missions will fully exploit the CMB
dipole for calibration. When high precision is required ($< 1\%$), both
instrumental and astrophysical effects must be carfully
considered in the process. In particular, foregrounds emission
from poorly known components add to the CMB dipole and can
introduce systematic calibration discrepancies. It has been shown
\cite{cappellini} that instroducing a frequency-dependent weight
function allows to optimise the procedure. As for the $COBE$--DMR
experiment \cite{bennett92b}, both missions can exploit the $\sim
0.3$~mK modulation of the CMB dipole in the 6-months period due
to the seasonal velocity of the earth around the sun
\cite{bersanelli96a}.

\subsection{Technology}
\label{subsec:technology}

The $\Delta T/T$ detection by $COBE$-DMR used coherent microwave
technology \cite{smoot90} in the frequency range 30-90~GHz; soon
after that, the FIRS balloon survey confirmed the detection of
cosmic structure with a bolometer receiver \cite{page90, meyer91}
operating at 170-680~GHz. Both radiometer and bolometric detector
technologies, after opening up the field of CMB anisotropy, have
rapidly and continuously evolved up to now, and both have
contributed to the dramatic progress in the past decade. The two
technologies are optimally employed roughly below and above $\sim
100$~GHz, respectively.

\subsubsection{\underline{Coherent receivers}}
\label{subsubsec:coherent_receivers}

Low system noise temperature $T_{\rm sys}$, large bandwidth
$\Delta\nu$, and good stability are the key features for high
performance radiometer. The minimum detectable temperature
variation (sensitivity) of a receiver is:
\begin{equation}
\Delta T(f) = k_R T_{\rm sys} \sqrt{ {1 \over \Delta\nu \tau} +
\left({\delta G(f) \over G}\right)^2 }\, ,
\label{eq:sensitivity}
\end{equation}
where $k_R \approx 1$ is a constant depending on the
radiometer scheme \cite{kraus66}, $\tau$ is integration time, and
$\delta G(f) / G$ represents the contribution from amplifier gain
and noise temperature fluctuations at post-detection sampling
frequency $f$. Cooling of the first stages of amplification
(typically to $\sim 100$ or $\sim 20$~K) is commonly used to
reduce $T_{\rm sys}$, and low-loss ($\sim 0.1 \div 0.5$ dB) wide band
($\delta \nu / \nu \sim$ $10\% \div 20\%$) front-end passive
components (such as feed-horns, transition, couplers, orthomode
transducers) are needed. Typically, the noise temperature of
current state-of-the-art low noise transistor amplifiers exhibit
a factor of 4--5 reduction going from 300~K to 100~K operating
temperature, and another factor $2 \div 2.5$ from 100~K to 20~K.
Amplifier fluctuations exhibit a characteristic power spectrum
$1/f^\gamma$ with $\gamma \approx 1$, i.e.
\begin{equation}
\Delta T(f) \simeq T_{\rm sys} \alpha \sqrt{1 + {f_{\rm k} \over f} }
\, ,\label{eq:sensitivity1}
\end{equation}
where $\alpha$ is a normalised amplitude of the noise
at high post-detection frequencies (i.e. in the pure white noise
limit), and $f_{\rm k}$, called knee-frequency, is the frequency at
which the white noise and $1/f$ components give equal
contributions to the power spectrum. In HEMT devices $1/f$ noise
is related to the presence of traps in the semiconductor
\cite{jarosik96}. As discussed in Sect.~\ref{subsec:1overf},
$1/f$ noise not only degrades the sensitivity, but it may also
introduce spurious correlations in the time ordered data and sky
maps. To minimise these effects, it is common to design the
instrument so that it takes differences between two nearly equal
signals, which may be the sky and a stable reference termination,
or two sky signals. The detected output of the radiometer can
then be synchronously demodulated to produce data stable on time
scales long compared to the switch period. A number of
differencing schemes have been adopted, including the classic
Dicke-switched scheme \cite{dicke46}, correlation designs
\cite{faris67}, and their combination with various beam switching
schemes \cite{partridge95}. Continuous-comparison designs have
been used at decimetre wavelengths \cite{haslam74, haslam81,
staggs96a}; more recently this technique was extended to higher
frequencies thanks to improved manufacturing and low-loss
performance of millimetre waveguide components
\cite{readpredmore85}. Beam switching strategies, in which the
beam is moved in the sky at a few Hz by a wobbling or rotating
reflector, have been extensively adopted, sometimes in
combination with total power receivers \cite{natoli01}.

In recent years, high electron mobility transistor (HEMT)
amplifiers have been widely and successfully employed both in
differential receivers and interferometers at frequencies up to
90~GHz.
These devices display a unique combination of features, including
very low noise performance, wide bandwidth, operability at
cryogenic temperatures and very low power consumption. The
current generation of HEMT amplifiers is based on indium
phosphide (AlInAs/GaInAs/InP heterostructure) which yields better
noise performances and lower power consumption than the more
traditional gallium arsenide technology (see, e.g.,
\cite{makunda87}). Recent progress in cryogenic HEMT noise
temperature has been quite dramatic. In the late 70's the
advances in gallium arsenide field-effect transistors, combined
with cryogenic cooling, made their noise performances competitive
with parametric amplifiers \cite{weinreb80}. In the past decade,
further major improvements have been obtained, leading in the mid
90's to state-of-the-art noise temperatures of 15~K at 40-50~GHz
and 50~K at 60-75~GHz \cite{pospieszalski94, pospieszalski95,
pospieszalski97}. Today, the best measured performances are
roughly a factor of 2 better (Sect.\ref{subsubsec:planck}).
Receivers based on SIS
(Superconductor--Insulator--Superconductor) junctions cooled to
4~K are capable of quantum limited detection, and are competitive
with HEMTs \cite{miller99} in the millimetre regime (e.g. in the
$\lambda \simeq 3$\,mm atmospheric window); however, they are
now strongly limited in bandwidth \cite{tucker79, kerr90}.

Corrugated feed horns are used to couple receivers with the sky
or optical system (see, e.g., \cite{xiaolei93}), and very low
sidelobe levels (down to $-90$\,dB \cite{toral89}) have been
demonstrated. Designs optimised for good primary illumination and
low edge taper include double profiled feeds \cite{gentili00,
villa02}, for which high performances have been demonstrated at high
frequencies \cite{bersanelli98}.

\subsubsection{\underline{Bolometric detectors}}
\label{subsubsec:bolometric_detectors}

Bolometers are thermal detectors in which incoming photons give
raise to phonons, and the temperature changes of the absorber is
measured by a thermometer, typically doped silicon or germanium
\cite{lamarre97}. The resistance $R$ of the thermometer changes
with its temperature, which can be measured by applying a
constant current. The typical application of bolometers in astrophysics
is in the sub-mm range (0.2$<\lambda<$3~mm), although they are also 
used in a variety of regimes, including X-ray.

A semi-empirical formula \cite{mather82} describes the bolometer
sensitivity in terms of noise equivalent power ($NEP$) as:
\begin{equation}
NEP^2 = {2kT_0 \over \pi} (f_0 C T_p) \left({9 \over A}+{25 \over A^2}\right) + 4kT_0 {36 W_0 \over A}\, ,
\label{eq:NEP}
\end{equation}
where $T_p$ is physical temperature, $f_0$ is frequency
response, $C$ is the material's heat capacity, $W_0$ is the
steady component of the optical power, and $A$ is given by:
\begin{equation}
A = -{d {\rm Log} R \over d {\rm Log} T} \approx 3.
\label{eq:A_bolometer}
\end{equation}

It is clear that reducing $T_p$, which controls both the first
and second terms in equation (\ref{eq:NEP}), one can in principle
reach extreme sensitivity. It can be shown that the critical
temperature below which the bolometer $NEP$ is limited by
unavoidable photon noise \cite{lamarre86} for a source with
blackbody temperature $T_S$ given by:
\begin{equation}
T_p < {1 \over 18} A T_S\, .
\label{eq:photon_noise}
\end{equation}

In the case of CMB observations, $T_S \approx 3$\,K, so that the critical
temperature is $T_p \approx 0.5$~K. Current state-of-the-art bolometer
technology, combined with the sophisticated cryogenic chains needed
to reach the extremely low temperatures (typically $0.3$~K to $0.1$~K)
required, allows to reach photon-noise-limited sensitivity.

Unlike coherent detectors, bolometers are sensitive to radiation
from any frequency and any direction. The absorber is normally
housed in a cavity which enhances the bolometer efficiency by
multiple reflection of the radiation. Coupling of the detector to
the telescope is achieved with either multimode concentrators,
such as Winston cones, or with single-mode corrugated feedhorns
(now preferred at least for frequencies $< 300$~GHz). To avoid
self-emission, the front-end temperatures must be kept as low as
possible. Frequency selection is achieved with high-efficiency
filters (typically rejecting out-of-band radiation at $10^{-9}$
level). These are cooled down enough to avoid stray emission
towards the bolometers, but at the same time they must be
compatible with the system cooling power.

A recent breakthrough in bolometer technology has been achieved
with the introduction of ``spider-web bolometers'', first
development at the JPL Center for Space Microelectronics
Technology \cite{bock96, lange96}, with silicon nitride
(Si$_4$N$_3$) micromesh. In this technique, a resistive grid is
used to absorb the radiation, instead of a continuous metallic
layer. These bolometers have better sensitivity due to the
reduction of the heat capacity $C$ of the absorber (see equation
(\ref{eq:NEP})). In addition, the web structure reduces the
detector mass by two orders of magnitude thus decreasing the
bolometer's sensitivity to microphonics. Also, its smaller
cross-section reduces dramatically the importance of cosmic rays
hits as well as of high frequency radiation. Noise equivalent
power of order $10^{-17}$\,W$\times$Hz$^{-1/2}$ have been
measured.

\section{INSTRUMENTAL SYSTEMATIC EFFECTS}
\label{siste}

Possibly the most challenging aspect in CMB experiments is the
need to reject and control all possible systematic effects. While
this is true for any experiment, the importance of systematic
effects is exacerbated in situations where the sought-for signal
is embedded in the noise, thus requiring long integration times.
When the goal is a mere ``detection'' of the CMB anisotropy one
can reasonably plan the rejection of spurious signals at $\sim
30\%$ of the CMB anisotropy, or $\sim$15-30~$\mu$K level. However,
in the present era of transition to precision measurements,
experimenters require control of systematics 10 times better, or
$\sim$1-3~$\mu$K level. High resolution CMB measurements at
sensitivity $\Delta T/T \approx 10^{-6}$ thus require an
environment and viewing field extremely free from local
contamination. Instruments must be designed with the aim to
control {\em in hardware} the level of systematic effects
introduced in the measured data: experience has demonstrated that
the more and the larger are systematic effects which have to be
removed in the data analysis, the less robust the final result
will be.

\subsection{Optical effects. Main beam distortions}
\label{subsec:main_beam_distortions}

\begin{figure}[here]
\begin{center}
\resizebox{6.cm}{!} {\includegraphics{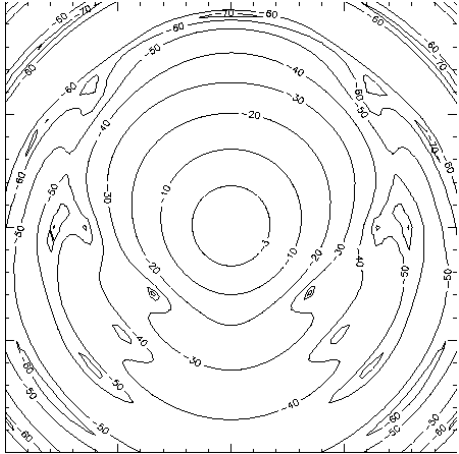}}
\end{center}
\caption{Typical simulated beam pattern at 100~GHz for the {\sc
Planck}-LFI instrument. Courtesy of F.~Villa.} \label{beam100}
\end{figure}

The observed signal is the result of the convolution of the
antenna beam pattern with the sky signal distribution (expressed
in antenna temperature):
\begin{equation}
T_{\rm A}(\theta_0,\phi_0) = \frac{ \int d\Omega
P_n(\theta-\theta_0,\phi-\phi_0)T_{\rm sky}(\theta,\phi)} {\int
d\Omega P_n(\theta,\phi)}\, , \label{convolution}
\end{equation}
where $T_{\rm sky}(\theta,\phi)$ is the sky temperature
in the direction $(\theta,\phi)$ and $P_n(\theta,\phi)$ is the
normalised antenna beam pattern. Ideally, the beam is a pure
symmetric Gaussian in shape. In practice non-idealities occur in
the $P_n(\theta,\phi)$ function. The effects on
$P_n(\theta,\phi)$ in its very central part are usually called
main beam distortions; the signal coming from secondary lobes is
instead referred to as stray-light signal. Fig.~\ref{beam100}
shows a typical simulated beam at 100~GHz for the {\sc
Planck}-LFI instrument: the beam pattern is quite symmetric down
to -10~$\div$ -20 dB while at lower level several structures
appear. The contour at -10 dB shows also a slight deviation from
a perfect circle.

Optical aberrations make the main beam (within few FWHM from the
beam centre) different from the ideal reference case of a pure
centrally symmetric Gaussian shape. Main beam distortions may
introduce a degradation of the angular resolution and of the
sensitivity per resolution element. These two effects can be seen
as orthogonal in the $\theta - \Delta T$ space of angular scales
and temperature anisotropy or, equivalently, in the $\ell -
C_\ell$ space \cite{burigana98}. The net effect is lowering the
maximum multipole $\ell$ probed and increasing the error on
$C_\ell$.

The need for multi-frequency, high sensitivity and high angular
resolution in present and future CMB experiments calls for a
multi-frequency focal plane arrays placed at the focus of an
optical system. As a consequence, the detectors located far from
the optical axis will suffer more beam distortion
effects. While sophisticated optical shaping can be optimised to
use a large focal area \cite{villa98, mandolesi00}, this effect
can be a limiting factor in the growth of array size.

\subsection{Optical effects. Side-lobe radiation}
\label{subsec:straylight}

\begin{figure}[here]
\centerline{
   \psfig{figure=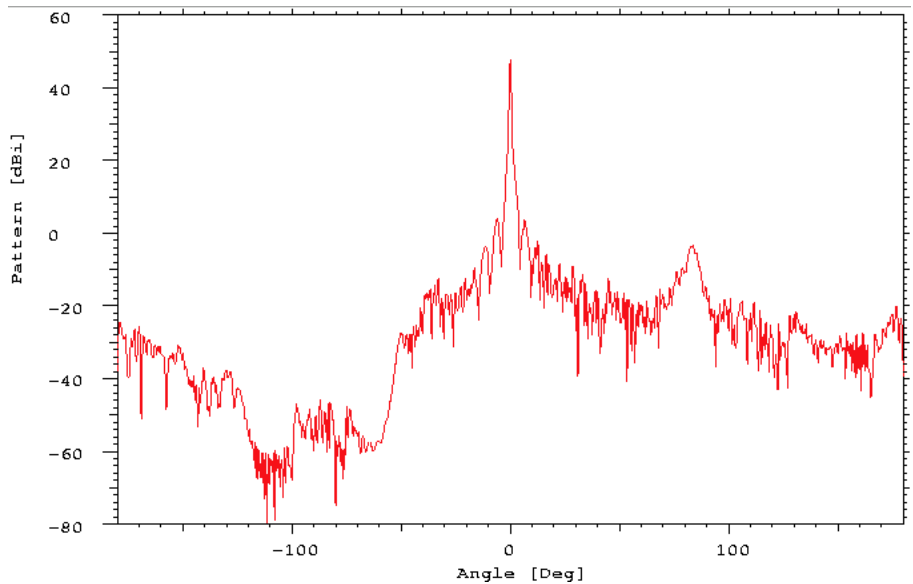,width=12cm}
} \caption{Cut of a simulated {\sc Planck}-LFI beam profile at
30~GHz. A part from the main beam, several other features are
present due to diffraction of the telescope, shield and
supporting structures. The integrated power of
the beam profile is dominated by the main beam ($\approx $ 98\%) and
the remaining small fraction is diluted on the full range of
angular scales. Courtesy of F.~Villa.} \label{stray}
\end{figure}

The radiation pattern at large angles from the main beam
(sidelobes) is due to diffraction and scattering effects from the
edge of the mirrors as well as from the nearby supporting
structures (see Fig.~\ref{stray}). In
Sect.~\ref{ground_radiation} we discussed the issue of sidelobe
effects from earth radiation. Of course the problem is much more
general. The source of stray-light may be both internal to the
instrument, e.g. emission from the telescope; and external, i.e.
from the surrounding environment, from small solid angle sources
like sun, moon, planets, or from diffuse sources like the
Galactic plane. The amplitude of the effect can be reduced by decreasing the edge
taper by design of the feed, at the cost of loss in angular
resolution. The level of stray-light depends on the orientation
of the beam profile with respect to external and internal sources.

\begin{figure}[here]
\centerline{
   \psfig{figure=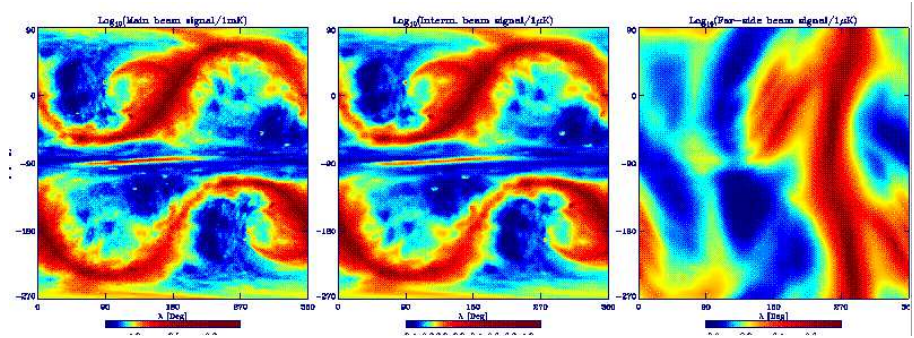,width=12cm}
} \caption{Synthetic view of the data stream from all scans of the
{\sc Planck}-LFI instrument at 30~GHz. The ecliptic coordinates
$\lambda$ and $\beta$ are the longitude and latitude of the
telescope axis \cite{mainophd}.} \label{mapstray}
\end{figure}

While the contamination from planetary bodies can be highly
reduced by choice of orbit and scanning strategy (see Sect.~\ref{subsec:ground}), 
for extended sky surveys little can be done to
avoid straylight from the Galactic plane. In the
frame-work of the {\sc Planck} mission, for example, a detailed
simulation has been carried out \cite{buri01} to address this
issue. In Fig.~\ref{mapstray} we report a map of the stray-light
signal for three distinct regions of the beam pattern: the main
beam (up to 1.2$^\circ$), an intermediate region (from 1.2 to 5
degrees) and the far pattern for $\theta > 5^\circ$. The
contribution from the intermediate part is quite similar to the
main beam signal and reproduces the galactic structures smoothed
and at a lower amplitude. This contamination is maximum around
the galactic plane, not useful to extract CMB
information. The far pattern shows a maximum value around few
$\mu$K with a completely different spatial pattern: this is the
galactic signal re-projected into the sky by the far beam pattern
coupled with the scanning strategy. Although the signal is
quite small and below the noise level, the stray-light from far
beam pattern contaminates high galactic latitude regions, most
valuable for CMB measurements. Furthermore the spatial pattern
introduces a non-Gaussian component which may be an
issue for statistical test on the Gaussianity distribution of the
CMB. Methods to remove the contaminations have been studied,
although more quantitative studies are needed. Note however that
this example refers to 30~GHz where galactic foregrounds are
relatively large (see Fig.~\ref{fore}): it is expected that the
impact of stray-light in cosmological channels (around 100~GHz)
will be lower due to the reduced amplitide of foregrounds.

\subsection{Pointing accuracy}
\label{subsec:pointing_accuracy}

The uncertainty in the pointing accuracy may be a serious source
of degradation, particularly for high resolution experiments. A
typical requirement is a 2$\sigma$ error of 1/10 of the beam FWHM.
Experiments from balloon often involve complicated motions both
of mirrors (chopping) and of the gondola making pointing
repeatibility a non-trivial issue. Satellite instruments
experiments are designed to observe the sky with a redundant
scanning strategy which involves periodic motion of the satellite
as a whole. Therefore the inertia tensor of the satellite has to
be known accurately in order to estimate the effects of
re-pointing of the satellite on its attitude stability. In
order to reduce oscillations induced by pointing maneuvers (both
in satellite and balloon-borne experiments) active oscillation
damping is sometimes employed.

The uncertainty in pointing direction, as well as the knowledge
of such uncertainty, has a direct impact on the effective angular
resolution of the experiments, even in the case in which the beam
shape is known with high accuracy, and this translates into a
restriction of the multipole range effectively probed. The
observed signal is a convolution of the sky signal with the beam
pattern (see Eq.~(\ref{convolution})) together with the statistical
distribution of the pointing uncertainty: the net result is a
convolution with a beam with a larger FWHM.

Beam shapes from complex optical system are difficult to measure
in the laboratory and are usually reconstructed during the
observation phase by means of bright point sources such as
planets or supernova remnants. The accuracy of beam
reconstruction is strongly dependent on the pointing accuracy.
Multi-feed experiments require also a ``geometrical'' calibration
of each feed. This means that feed positions with respect to a
fiducial reference point have to be determined. We have already
seen how off-axis beam patterns differ from a purely circular
symmetric Gaussian beam. One could expect to reconstruct the beam
shape from point sources and properly account for it. However,
pointing accuracy may be a limiting factor for the proper
knowledge of beam resolution and beam shape.

In order to solve for the exact pointing solution, one usually has
to collect ancillary information. This includes CCD-camera or
star-mapper systems to map the sky for bright known stars which
requires to know the geometrical configuration of the instrument
set-up (focal plane arrangement) and the relative orientation
between star-mapper and field of view which can be obtained
through observations of bright sources (planets like Jupiter).

\subsection{$1/f$-noise instabilities}
\label{subsec:1overf}

The presence of $1/f$ noise or of slow drifts in the data stream
is a well known problem in CMB experiments. These instabilities
typically originate from a combination of thermal fluctuations
and amplifier gain variations as well as variatons in the
emission of the atmosphere (Sect.~\ref{obsissue}). For HEMT based
microwave radiometers the $1/f$ noise is dominated by gain
fluctuations, while for bolometers instabilities may originate in
thermal fluctuations of the environment surrounding the detector.
The coupling between long-term instabilities and the observing
strategy may lead to artifacts in the final map showing up as
stripes in the scanning direction. These stripes can increase the
overall noise and alter the statistical
analysis of the CMB anisotropy.

To first order, $1/f$ noise can be parameterised by a single
parameter: the knee-frequency (see
Sect.~\ref{subsubsec:coherent_receivers}). Of course the
knee-frequency depends on different parameters and quantities
related to the specific instrument under consideration
\cite{seiffert02}. A constraint on $f_{\rm k}$ is that it has to be as
small as possible compared to the first redundancy frequency
$f_{\rm s}$ (linear slew, spin, etc) of the scanning scheme. It is
indeed demonstrated \cite{janssen96} that when $f_{\rm k}\gsim f_{\rm s}$ a
degradation in the final sensitivity will occur so that the
optimal solution would be to design instruments with $f_{\rm k} < f_{\rm s}$.
Where this is not possible one is left with the need of
removing the artifacts due to $1/f$ noise from the data. If the
knee-frequency is well known (and in principle through iterative
methods \cite{prunet} both the noise spectrum and $f_{\rm k}$ can be
derived from the data themselves) it is possible to correct the
data with a proper noise filter and remove the tail of the $1/f$
noise at low frequencies. In special cases another approach
\cite{buri97,maino99} can be applied: it makes no assumption on
the noise spectrum and takes advantage of a redundant observing
strategy in which repeated observations of the same pixel in the
sky can be used. This approach is usually referred to as
``destriping''. Even for relatively large values of the knee
frequency, the behaviour of $1/f$ noise can be well approximated
with a single additive baseline for each redundant data scan. The
complete set of intersections between scans is used to solve a
linear system where the unknowns are the scan baselines; once the
baselines are recovered, they are subtracted from the original
scan circle to obtain $1/f$ free data. This approach works
remarkably well showing a residual contamination $\lsim 4\%$ even
for $f_{\rm k} \simeq 7 f_{\rm s}$.

\begin{figure}[here]
\begin{center}
\resizebox{14cm}{!} {\includegraphics{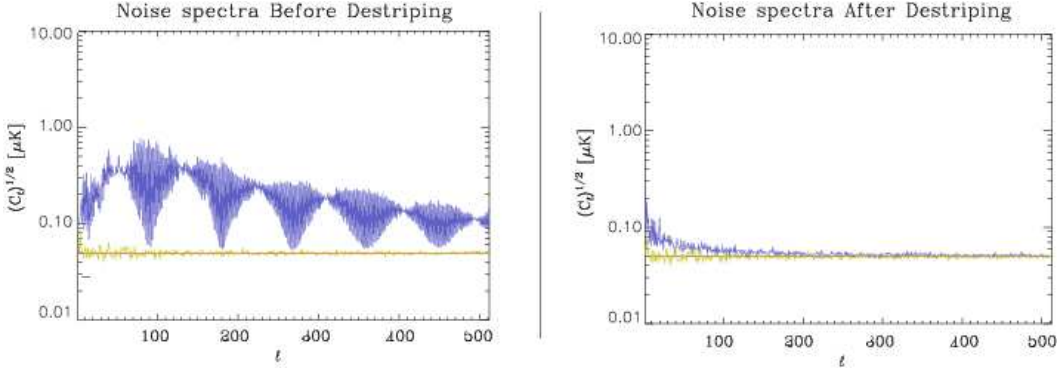}}
\end{center}
\caption{Noise angular power spectra before and after destriping
for $f_{\rm k} = 0.1$~Hz for a {\sc Planck}-LFI radiometer at 30~GHz.
The noise level after destriping is almost white and comparable
with the theoretical white noise level for multipole $\ell \gsim
100$ \cite{mainophd}.} \label{destri}
\end{figure}

This is shown in Fig.~\ref{destri} where the angular power
spectrum of a simulated {\sc Planck}-LFI noise map is plotted before
and after applying the destriping technique and compared to the
white noise level: a small residual noise is still present at low
multipoles, while for $\ell \gsim 100$ the noise spectrum is
almost indistinguishable from the theoretical white noise spectrum.

\subsection{Thermal and periodic fluctuations}

Another important class of systematic effects is represented by
periodic environment fluctuations ({\em e.g.} electrical and/or
thermal fluctuations) that may couple with the measured signal
and leave spurious signatures in the data. Temperature stability,
in particular, is one of the primary concerns that drives many
experimental issues, from the instrument thermal design to the
observation site and the scanning strategy. Even though a careful
choice of the observation site can optimise the ambient thermal
stability it is of primary importance that the whole system
(active cooling, receivers, and instrument configuration) be
properly designed in order to minimise the impact of internally
generated thermal fluctuations. The residual effect in the measured data stream can
be of the order of few mK peak-to-peak. Therefore it is important
to be able to estimate the impact of these spurious signals on the
final maps and to further reduce the effect by proper analysis of
the time ordered data (TOD).

Under quite general assumptions the peak-to-peak effect of
periodic oscillations on the reconstructed
sky map can be evaluated analytically \cite{mennella02}. If
we consider a periodic fluctuation, $\delta T$, of general
shape in the detected signal, we can expand it in Fourier series,
i.e.: $\delta T = \sum_{j=-\infty}^{+\infty}A_j \exp(i 2\pi f_j
t)$, where $f_j$ represents the different frequency components
in the fluctuation. For frequencies $f_j$ which are not synchronous with the
instrument characteristic scanning frequency, $f_{\rm s}$,
the measurement redundancy and the projection of the
TOD onto a map with a pixel size $\theta_{\rm pix}$
will damp the corresponding harmonic amplitude $A_j$ by a factor
proportional to $\sin(\pi f_j / f_{\rm s})$. For
frequencies synchronous with $f_{\rm s}$, instead, there will
be no damping and these signals will be practically
indistinguishable from the sky measurement. Therefore it is
critical that any spurious signal
which is synchronous with the characteristic scanning frequency
must be carefully controlled and kept at a negligible level {\em
in hardware}. For a spinning experiment (for example like \map and \planck) we
can estimate the final peak-to-peak effect of a generic signal
fluctuation $\delta T$ on the map as follows:
\begin{eqnarray}
\langle\delta T^{\rm p-p}\rangle_{\rm map} &\sim &
2\left[
    \frac{1}{N\times \theta_{\rm pix}/\theta_{\rm rep}}\left(
        \sum_{f_j < k\, f_{\rm s}}
        \left|\frac{A_j}{\sin(\pi f_j / f_{\rm s} )}\right| +
    \right.\right. \nonumber \\
&&\mbox{}
\label{eq:p2p_map_general}
\\
     & + & \left.\left.         \sum_{f_j < f_{\rm s},f_j \neq f_{\rm s}}
        \left|A_j\right|
    \right)+ \sum_{f_j = k\, f_{\rm s}} A_j\right],\nonumber
\end{eqnarray}
where $\theta_{\rm pix}$ is the pixel size, $\theta_{\rm
rep}$ is the {\em repointing angle} (i.e. the angle between two
consecutive scan circles in the sky), $N$ is the number of times
each sky pixel is sampled during in each scan circle and $f_{\rm s}$
represents the scan frequency. Note that
Eq.~(\ref{eq:p2p_map_general}) takes into account the damping
provided only by the measurement redundancy and by the scanning
strategy, without considering the possibility to detect and
partially remove these spurious signals from the TOD. Several
numerical strategies can be used to approach this issue and, as a
general rule, the removal efficiency is greater with ``slow''
fluctuations, i.e. with a frequency $f\ll f_{\rm s}$.

\begin{figure}[here]
\begin{center}
\resizebox{6cm}{!} {\includegraphics{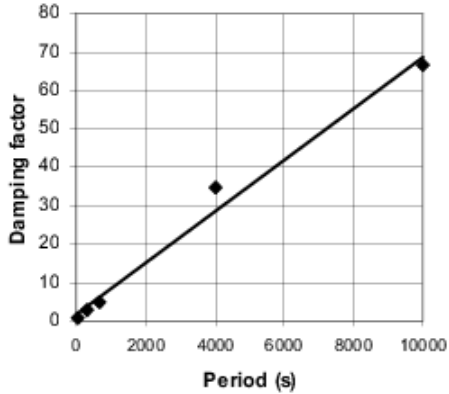}}
\end{center}
\caption{Destriping damping factor for periodic oscillations
versus the oscillation period. Dots indicate the values obtained
from maps while the solid line represents a linear interpolation.}
\label{fig:destriping_vs_Tf}
\end{figure}

An example is shown in Fig.~\ref{fig:destriping_vs_Tf} where we
plot the damping factor obtained by applying the same destriping
algorithm described in section~\ref{subsec:1overf} to sinusoidal
signal fluctuations with various periods; the damping factor is
defined as the ratio of the two values of the peak-to-peak effect
on the final map obtained before and after application of the
destriping algorithm to the data stream. The assumed scanning
strategy is representative of a \planck 30~GHz feed-horn.
The result clearly shows that the damping factor increases
roughly linearly with increasing values of the oscillation period.
If we denote with $F_j$ the damping factor obtained by applying a
certain algorithm to a periodic signal with frequency $\nu_j$
then we can write Eq.~(\ref{eq:p2p_map_general}) in the more
general form:
\begin{eqnarray}
\langle\delta T^{\rm p-p}\rangle_{\rm map} &\sim &
2\left[
    \frac{1}{N\times \theta_{\rm pix}/\theta_{\rm rep}}\left(
        \sum_{f_j < k\, f_{\rm s}}
        \left|\frac{A_j/F_j}{\sin(\pi f_j / f_{\rm s} )}\right| +
    \right.\right. \nonumber \\
&&\mbox{}\\
     & + & \left.\left.         \sum_{f_j < f_{\rm s},f_j \neq f_{\rm s}}
        \left|A_j/F_j\right|
    \right)+ \sum_{f_j = k\, f_{\rm s}} A_j\right].\nonumber
\label{eq:p2p_map_with_destriping}
\end{eqnarray}

\subsection{Summary}
\label{subsec:summary}

The effects discussed above, as well as the impact of atmospheric
and ground radiation (see Sect.~\ref{subsec:atmosphere}
and~\ref{ground_radiation}) and calibration (Sect.~\ref{subsec:calibration}), 
often represent the limiting factors
of anisotropy measurements and drive the design of the instrument
and data analysis. And the above list is far from complete.
Artifacts may be introduced in the data by a number of effects which
depend on the intrinsic details of the instrument, such as
variable offsets, detectors or amplifiers non-linearities,
time-changes of the optics emissivity and of the instrument
performance, electrical effects such as ground loops and power
instability, magnetic susceptibility, RF interferences. The data
processing itself (e.g. data quantisation, data compression,
pixelisation, co-adding of multiple scans from different
detectors) may be sources of spurious effects. As we shall see in
Sect.~\ref{experiments}, as the instrument sensitivity improves,
the instrument design, scanning strategy and data analysis are
more and more focussed on the control and removal of systematics.
Sometimes a design choice made to avoid a given effect may itself
introduce a secondary source of disturbance. As an example, many
sub--degree experiments use a moving (rotating or nutating)
subreflector to provide a beam--switching pattern to subtract
instrumental and atmospheric drifts (see e.g. \cite{davies92}).
In general, however, strategies that imply moving the
subreflector or other reflective parts may raise concern for
possible modulation of radiation from the earth (or from the
balloon) in the instrument sidelobes.

Each effect is characterised by a set of frequencies and
amplitudes, and couples in a peculiar way with the instrument
response and the observing strategy. While single effects have
been in some cases studied in great detail, of course in the real
world they are all hidden simultaneously in the data and may
couple with each other in non trivial ways \cite{burigana2000}. As
we shall see, experiments aiming at few $\mu$K sensitivities,
require a proportionally stringent rejection of systematic
effects. In addition, precision experiments call for a dedicated
effort to understand the impact of the various effects on
ancillary measurements such as calibration and beam
reconstruction.

\section{DATA ANALYSIS}
\label{analysis}

We now present the basic processes
involved in the data analysis of CMB anisotropy data which
eventually yield the $C_\ell$ coefficients and an estimation of the
cosmological parameters. The implementation of such
processes represents a computation challenge since current
experiments produce large amounts of data and much more are
expected from the forthcoming satellite missions. 
The data analysis processes could be regarded as a data
(lossless) compression: as an example, for the Boomerang
experiments \cite{debe00} it is possible to pass from the data
time stream of few$\times 10^6$ elements, to maps with
few$\times 10^4$ pixels, to power spectrum with coefficients
usually binned in few intervals ($\sim 10$) in $\ell$ space.

\subsection{Map-making}
\label{subsec:mmaking}
The first data compression involves passing from the TOD
to a map of the surveyed sky area. This
map gives a visual impression of what the data
look like, and helps in recognising possible systematic effects
and astrophysical contamination. In general the problem consists
of estimating $N_p$ parameters, the signal value in each pixel of
the final map, from a large data set of $N_t$ observations,
producing the best unbiased map. The TOD vector can be written as:
\begin{equation}
\mathbf{d} = A\mathbf{s} + \mathbf{n}\ ,
\end{equation}
where $\mathbf{s}$ is the pixelised sky signal, $\mathbf{n}$ is
the instrumental noise and $A$ is the pointing matrix with $N_t$
rows and $N_p$ columns. For a total power scanning experiment,
the pointing matrix has the simple form
\begin{equation}
A_{tp} = \bigg\{ \begin{array}{ll}
                    1 & {\rm if} (\theta_t,\phi_t) \in p \\
                    0 & {\rm otherwise} \\
                  \end{array}\, ,
\end{equation}
where $(\theta_t,\phi_t)$ are the coordinates at which the
instrument is pointing at the time $t$ falling into the pixel $p$.
The best unbiased map is obtained by minimising the quantity:
\begin{equation}
\chi^2 = (\mathbf{d}^t - \mathbf{s}^t A^t)N^{-1}(\mathbf{d} -
A\mathbf{s})\ ,
\end{equation}
where $N\equiv N_{tt'}=\langle n(t)n(t')\rangle$ is the noise covariance matrix. 
Detector noise properties can be derived directly from data either by means of 
iterative \cite{dore01} or non-iterative algorithm \cite{nato02}.
The generalised least squares solution gives:
\begin{equation}
\mathbf{\tilde{d}} = (A^t N^{-1} A)^{-1} A^t N^{-1}\mathbf{d}\ .
\end{equation}

Therefore in principle the solution is quite simple: one has to
compute the vector
$\mathbf{w} = A^t N^{-1}\mathbf{d}$ and the matrix $\Sigma^{-1} =
(A^t N^{-1} A)$ and then invert it to obtain
\begin{equation}
\mathbf{\tilde{d}} = (\Sigma^{-1})^{-1} \mathbf{w}\ .
\end{equation}
However, the process involves the inversion of a $N_p \times N_p$
sparse matrix which requires $\mathcal{O}(N^3_p)$ operations and
$\mathcal{O}(N^2_p)$ memory for storage. For $N_p$ in the range of
$10^4\div 10^5$ of current experiments, these scalings strongly
call for a multi-processor parallel machine. The situation is much more
dramatic for future satellite experiments with typical
number of pixel around $10^6$.

Several algorithms exist for sparse matrix inversion and the case
of Conjugate Gradients algorithms is quite interesting \cite{press96}.
This involves the products of a matrix with a vector, which is an
$\mathcal{O}(N_p^2)$ operation.
The method is iterative and in general converges in $N_p$
iterations unless a good preconditioner $\hat{\Sigma}$
(such that $\hat{\Sigma}\Sigma^{-1} -I \ll 1$) is found.
This is usually the diagonal part of the matrix $\Sigma^{-1}$:
\begin{equation}
\hat{\Sigma} = {\rm diag}(1/\Sigma_{ii}^{-1})\ .
\end{equation}

In this way the algorithm can converge in few tens of operations
requiring few minutes for the creation of a $N_p=10^4$ map on
single-CPU workstation. Recently \cite{nato01} this algorithm
has been implemented for the {\sc Planck} case: processing 
$N_t \sim 3\times 10^9$ elements procuding maps with
$N_p \sim 3\times 10^6$ pixels, takes about $10^3$ second
on 256 processors on an SP3 IBM parallel computer.

\subsection{Separating foregrounds from the CMB}
\label{subsec:separating_foregrounds}

As discussed in Sect.~\ref{astrolimit}, a key issue is the separation of
foreground emissions from the cosmological signal.
This takes advantage of the different frequency and spatial
behaviour of the foregrounds with respect to the CMB.
Several techniques have been developed for
this purpose such as the Wiener filter \cite{wiener49} and Maximum
Entropy Method (MEM) \cite{skilling89}.

Let us suppose $\mathbf{X}$ is a vector of $M$ observations
($M$ different frequencies) whose probability distribution
$P(\mathbf{X}|\mathbf{S})$ depends on the values of $N$
parameters $\mathbf{S}$ (the signal we want to separate). Let
$P(\mathbf{S})$ be the {\em prior} probability distribution of
$\mathbf{S}$ \ie our a-priori knowledge of the signals
$\mathbf{S}$. Given the data $\mathbf{X}$, the Bayes' theorem
states that the {\em posterior} probability of $\mathbf{S}$ is
given by:
\begin{equation}
P(\mathbf{S}|\mathbf{X}) = z \cdot
P(\mathbf{X}|\mathbf{S})P(\mathbf{S})\ ,
\end{equation}
where $z$ is a normalisation constant. Therefore one has to
construct an estimator $\hat{\mathbf{S}}$ of the true signal
maximising the posterior probability. The real problem is in the
choice of the appropriate prior distribution. Two different
approaches have been adopted: the entropic prior and the Gaussian prior.

The entropic prior, which is at the basis of the MEM method, has
the form:
\begin{equation}
P(\mathbf{S}) \propto {\rm exp}[\alpha H(\mathbf{S,m})]\,
\end{equation}
where $\mathbf{m}$ is a model vector to which $\mathbf{S}$
defaults in absence of data, $\alpha$ is a constant depending on
the scaling of the problem and $H$ is the cross entropy of
$\mathbf{S}$ and $\mathbf{m}$. The MEM method has been applied to CMB
map reconstruction \cite{hobson98} assuming the complete
knowledge of the frequency scaling of the various components. As
for the spatial distribution, two different cases have been considered:
the first assumes complete knowledge of the covariance matrix
of the signal and in this case the maps are reconstructed with
good accuracy; in the second case no assumption is made on the
covariance matrix. In this second case reconstruction is obtained
only in an iterative way and leads to less accurate results. The CMB power
spectrum is in general better recovered than those of the
foregrounds.

The Gaussian prior has the form:
\begin{equation}
P(\mathbf{S}) \propto {\rm exp}(-\mathbf{S}^t \mathbf{C}
\mathbf{S})\ ,
\end{equation}
where it is assumed that the emission of each component is a
Gaussian random field (which is not true in general). Furthermore,
a complete knowledge of both the noise ($\mathbf{N}$) and the
signal ($\mathbf{C}$) covariance matrices is assumed. This is the
Wiener filter which has been applied to data such as DMR
\cite{bunn94} and Saskatoon \cite{tegmark97}.

Recently a new method for foregrounds separation
has been proposed \cite{bacci00} based on neural network. This is
promising since it does not require any a-priori
knowledge of either frequency scaling or spatial
distribution of the components to be separated. The only requirement
is that at most one of the signal has to be Gaussian (the CMB) and
that the signals are statistically independent. Preliminary
results without instrumental noise are encouraging, yielding
output maps exactly reproducing the input ones, a part from a scaling constant.
On the same line,  a recent work \cite{maino02} implements a fast
Independent Component Analysis (ICA) technique applied to
simulated \planck observations. This approach takes into account
the instrumental noise, although it is assumed to be white and
uniformly distributed on the sky. Results are extremely good: the CMB
angular power spectrum is recovered to high precision (up to $\ell
\simeq 2000$) with the correct signal normalisation, as well as the expected CMB
frequency scaling. The algorithm is extremely fast (few minutes
to separate components for map with $\sim 10^6$ pixels).
One of the main limitations is that the algorithm it not able,
in the present version, to perform simultaneously beam deconvolution. This
implies that input maps have to be at the same angular resolution.

In recent years it has become clear that ``blind'' algorithms, such as those
exploiting ICA techniques, represent an optimal tool
to derive ``priors'' on the components to
be separated, making such assumptions less sensitive to
our incomplete
knowledge of astrophysical signals. These ``priors'' can then be
fed into more sophisticated algorithms, like MEM, for a full
complete analysis, inlcuding deconvolution.

\subsection{Power spectrum estimation}

Once we have compressed the TOD into a map and cleaned it
from astrophysical foregrounds and possible systematic effects,
we have to extract from it the cosmological information mostly embedded in
the $C_\ell$ coefficients. The problem of deriving the $C_\ell$'s scales as
$\mathcal{O}(N_p^3)$ and at present no general solution is available.

The problem is quite simple in principle. The set
of $C_\ell$, convolved with the window function $W_\ell$, are
obtained maximising the likelihood function:
\begin{equation}
\mathcal{L}(C_\ell|\mathbf{\tilde{d}}) = \frac { {\rm
exp}\left(-\frac{1}{2}\mathbf{\tilde{d}}^t \mathbf{C}^{-1}
\mathbf{\tilde{d}}\right)} {(2 \pi)^{N_p/2} ({\rm
det}\mathbf{C})^{1/2}}\ , \label{likelihood}
\end{equation}
where $\mathbf{C}$ is the map covariance matrix (signal plus
noise). The most likely power spectrum is found expanding the
log-likelihood function in Taylor series about the minimum. After
some computation the solution, found iteratively, is:
\begin{equation}
C_\ell^{n+1} = C_\ell^n - \frac{1}{2}
\sum_{\ell'}\mathbf{F}_{\ell\ell'}^{-1}\frac{\partial f}{\partial
C_{\ell'}}\bigg|_{C_{\ell'}^n}\ ,
\end{equation}
where $f = -2{\rm ln}\mathcal{L}$ and $\mathbf{F}_{\ell\ell'}$ is
the Fisher information matrix:
\begin{equation}
\mathbf{F}_{\ell \ell'} = - \bigg\langle
\frac{\partial^2}{\partial C_\ell \partial C_{\ell'}} {\rm ln}
\mathcal{L} \bigg\rangle\, .
\end{equation}

The system can be solved using the Newton-Raphson method which
usually converges in 3-4 iterations but the convergence is
quadratic only in a neighborhood of the root. Furthermore, as
stated above, the matrix inversion takes
$\mathcal{O}(N_p^3)$ operations, but in this case it is not easy to find a
good preconditioner as for the map-making problem. This is
because the $\mathbf{C}$ matrix depends both
on noise and signal which essentially belong to different spaces.
While the noise covariance matrix is diagonal in pixel space it is
dense in $Y_{\ell m}$ space and the reverse is true for the signal
covariance matrix. So a general solution is not at hand.
Furthermore any basis transformation that would make $\mathbf{S}$
and $\mathbf{N}$ diagonal or block-diagonal cost
$\mathcal{O}(N_p^3)$ operations and it is of no use. A good
preconditioner has been found by the $MAP$ team \cite{oh98} since
the $MAP$ scanning strategy generates noise patterns which are
nearly azimuthally symmetric. However this is not true for
balloon experiment and has to be carefully investigated for {\sc
Planck}. It is worth noting that the only case 
for which the maximum likelihood problem has been solved exactly, 
is $COBE$-DMR, for which the $\mathcal{O}(N_p^3)$ scaling was still treatable 
\cite{gorski94a, gorski94b, gorski96a}. 

A code called MADCAP has 
been developed \cite{borrill99} and has
been used to compute the maximum likelihood power spectrum for
the Boomerang \cite{debe00} and Maxima \cite{balbi00}
experiments. This works on a Cray-T3E parallel machine and
allows the computation of the power spectrum
in a reasonable time for a suitable number ($\approx$ 10)
of bins in $\ell$ space in the range 30-800. However, current
experiments require a solution able to handle much larger data sets.
Among
the new proposed approaches that dramatically reduce the number of
operations, we mention the Ring-Torus method \cite{wandelthansen01} and the
MASTER technique \cite{hivon02}. The first is extremely elegant
in its algebraic properties and works quite well for a limited
set of observing strategies, including those of $MAP$, as well as for
{\sc Planck}, if the spin-axis
is kept always on the ecliptic plane
 (see Sect.~\ref{subsubsec:planck}). In these cases exact maximum likelihood power spectrum
(Eq.~(\ref{likelihood})) is possible with $\mathcal{O}(N_p^2)$
operation since both signal and noise matrices are
block-diagonal in Fourier space.

The MASTER method is based on a direct spherical harmonic transformation
of the cleaned CMB map and allows to include specific features of
the observation (\eg survey geometry, scanning strategy, noise
properties) together with possible non-Gaussian and non-stationary
processes that can occur during the experiment life time. The
resulting power spectrum is corrected for characteristics of
the instrument and observing strategy as well as any
operation performed in the data analysis process, by means of
Monte Carlo simulations. The final power spectrum has been
demonstrated to be an unbiased estimator of the true CMB power
spectrum. This technique was applied to the Boomerang
data \cite{netter01} covering 1.8\% of the sky 
(57,000 pixels with $7'$ resolution). It is proved to
be extremely fast and easily parallelisable and can
be considered as a valuable tool for more extensive applications.
Recently applications of the MASTER approach have been implemented
using general least squared map-making algorithm \cite{balbi02} 
yielding similar results of the original MASTER method.

Other approaches to the problem involve 
computation of the 2-pt correlation function of the sky and then
invert it to obtain the angular power spectrum \cite{szapudi}. 
On the same line similar results have been obtained computing
the 2-pt correlation function of the CMB peak distribution \cite{kash}.

\subsection{Extracting cosmological parameters}

Once we have the CMB angular power spectrum, it is possible to
apply the usual likelihood approach to derive the cosmological
parameters $\mathbf{p} = (\Omega_b, \Omega_{\rm CDM}, \Lambda,
H_0, n_S, n_t, ...)$. Proceeding in a similar way as for the power
spectrum estimation, the Fisher information matrix for
cosmological parameters is given by:
\begin{eqnarray}
\mathbf{F}_{kk'} & = & - \bigg\langle \frac{\partial^2}{\partial
p_k \partial p_{k'}} {\rm ln} \mathcal{L} \bigg\rangle\\ \nonumber
 & = & -\sum_{\ell\ell'}
\frac{\partial C_\ell}{\partial p_k} \bigg\langle
\frac{\partial^2}{\partial C_\ell \partial C_{\ell'}} {\rm ln}
\mathcal{L} \bigg\rangle \frac{\partial C_{\ell'}}{\partial
p_{k'}}\\ \nonumber
 & = & \sum_{\ell\ell'}
\frac{\partial C_\ell}{\partial p_k} \mathbf{F}_{\ell\ell'}
\frac{\partial C_{\ell'}}{\partial p_{k'}}\\ \nonumber
\end{eqnarray}

The partial derivatives with respect to cosmological parameters can
be computed numerically using two-side finite differences of order
2-5\% of the parameter value using model spectra computed with fast
numerical codes \cite{selzal97}. However, in this case it is not
so easy to exploit the
Newton-Raphson method since the radius
of convergence is quite smaller and, in addition, the degeneracy
between various parameters creates features in the likelihood which
makes the Newton approach not optimal. One can instead proceed with a
simple $\chi^2$ fit:
\begin{equation}
\chi^2(\mathbf{p}) = \sum_{\ell \ell'} (C_\ell(\mathbf{p}) -
C_\ell^{\rm recov}) \mathbf{F}_{\ell\ell'} (C_{\ell'}(\mathbf{p})
- C_{\ell'}^{\rm recov})\ ,
\end{equation}
where $C_\ell(\mathbf{p})$ is the theoretically computed power
spectrum and $C_\ell^{\rm recov}$ is the recovered one. This
method has been tested by the $MAP$ team \cite{oh98} yielding
excellent results in the multipole range where the signal-to-noise
ratio is high enough ($\ell \lsim 600$).

It is clear that the final accuracy depends also on the accuracy
by which theoretical models are computed. Present fast algorithms
that integrate the Boltzmann equation to find the CMB angular
power spectrum, provide accuracy at the percent level, i.e. compatible
with the {\sc Planck} goal accuracy on the parameter estimation. It is
clearly of great interest
to develop codes that would be equally, or
even more, fast but with improved accuracy.

\section{ANISOTROPY EXPERIMENTS}
\label{experiments}

In this section we review the CMB anisotropy experiments carried
out from $COBE$-DMR up to now. Some of the key features of all experiments
leading to anisotropy detection are summarised in Table I-IV. We
first outline the DMR experiment, and then we review
ground-based and balloon-borne experiments. Finally we provide a synthesis of
the current status of power spectrum measurements.

\subsection{The $COBE$ Differential Microwave Radiometer}
\label{subsubsec:COBE}

\begin{figure}[here]
\begin{center}
\resizebox{8.cm}{!} {\includegraphics{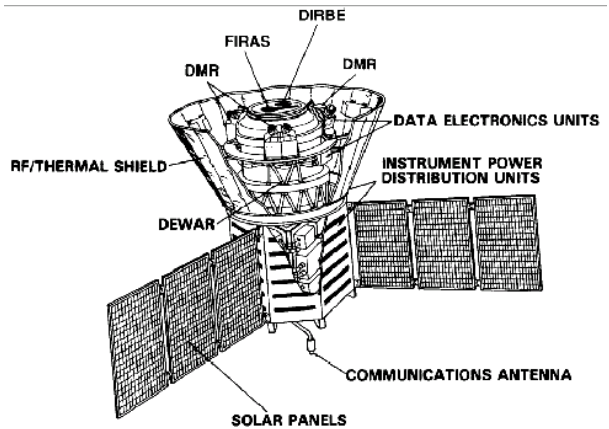}}
\end{center}
\caption{The Cosmic Background Explorer ($COBE$) satellite
showing the Differential Microwave Radiometers (DMRs), Data
Electronics Units, and Instrument Power Distribution Units as
mounted on the $COBE$ spacecraft. The spacecraft spins about its
axis of symmetry. (From \cite{smoot90} with permission.)}
\label{fig:cobe}
\end{figure}

On 18 November 1989, NASA's $COBE$ satellite (see Fig.~\ref{fig:cobe}) was
launched from Vandenberg Air Force Base into a near-polar circular orbit.
It was the starting point of one of the greatest achievements in the
history of CMB experiments, and indeed of cosmology at large: the
first unambiguous detection of structure in the CMB angular distribution.
The first detection was obtained on angular scales $> 7^\circ$ by the
Differential Microwave Radiometer (DMR) based on its first year of data
\cite{smoot92, bennett92a}. These early results were supported by detailed
calibration \cite{bennett92b} and systematic errors analysis \cite{kogut92}.
The first detection was soon confirmed by the positive cross-correlation
between $COBE$-DMR and sub-orbital observations \cite{ganga93,lineweaver95}.
Later, the results from DMR's first two years observations \cite{bennett94}
were shown to be consistent with those from the first year alone. The
DMR mission ended in December 1993, after 4 years of uninterrupted survey.
The definitive DMR analysis and results are reported in
\cite{bennett96,gorski96, hinshaw96a, hinshaw96b, wright96, kogut96a, 
kogut96b, banday96, banday97, fixsen97}.

The instrument \cite{smoot90} consisted of six Dicke-switched differential
radiometers, two independent systems at each observing band centred at 31.5,
53 and 90~GHz. Each radiometer measured the difference in power received from two
directions in the sky separated by 60$^\circ$, using a pair of corrugated horn
antennas at symmetric angles with the spin axis. The antenna beams were highly
symmetric with FWHM~$\simeq 7^\circ$. The receivers were superheterodyne systems 
switched at 100~Hz, using a single local
oscillator per frequency assembly.


The $COBE$ satellite spun at approximately 0.8~rpm and
precessed in a 900~km near-polar orbit, following the day-night
terminator. This allowed the instrument to always point away from
the earth and be perpendicular to the sun, thus minimising
straylight effects. The combination of the satellite spin (75~s),
orbit (103~min), and orbital precession ($\sim 1$~degree/day)
provided a redundant set of temperature differences spaced
$60^\circ$ apart, and allowed to cover the entire sky in a
six-month period. Systematic thermal and electrical modulations
occurred near the summer solstice, for about 60~days per year,
since the earth was not fully screened by the satellite shields
as the spacecraft approached the North Pole, while at the South
Pole eclipses occurred every orbit.


In-flight relative calibration was provided by solid state noise
sources injecting broad band power in the radiometers front end
every 2~hours. The moon signal provided further check to the
calibration, with an accuracy limited by the uncertainty in the
model of the moon antenna temperature and phases. The most accurate
absolute calibration was derived from the 30\,km\,s$^{-1}$
earth's motion around the solar system baricentre, which produces
a $\sim 0.3$~mK Doppler shift dipole of accurately known
amplitude (see Sect.~\ref{subsec:calibration}).


The careful analysis of potential and detected systematic effects
\cite{kogut92, kogut96b} included emission from earth and the
moon, the instrument's response to thermal instabilities, and the
effect of earth's magnetic field. The analysis showed that even
the largest detected effects did not contribute significantly to
the DMR maps. Data taken in worst-case situations for potential straylight
contamination and from thermal fluctuations were rejected
($\lesssim 10\%$ of the data), while the remaining data were
corrected using models of each effect. The estimated upper limit
to  residual systematic artifacts in the maps was $<6~\mu$K (95\%
C.L.), i.e. significantly less than the noise levels of sky signals.


With the full 4-year data, the typical signal-to-noise ratio of
the frequency-averaged map smoothed at $10^\circ$ was $\sim 2$,
which means that the final DMR CMB map does provide a visual
impression of the CMB structure in the sky. Monopole ($T_0 =
2.725 \pm 0.020$~K) and dipole 
($\Delta T_{\rm dip} = 3.353 \pm 0.024$~mK)
measurements from DMR were reported in \cite{kogut96b, fixsen96}.
The intrinsic quadrupole ($\ell = 2$) amplitude was found to be $\Delta T_2
= 10.0^{+3.8}_{-2.8}~\mu$K (68\% confidence level). The value
predicted by a power-law fit to the power spectrum yielded
$Q_{rms-PS} = 15.3~\mu$K \cite{kogut96a, hinshaw96b}. The observed
power spectrum at large angular scales, which was limited by
cosmic variance for $\ell < 20$, was found to be consistent with
a $n_S\simeq 1$ power law \cite{gorski96, hinshaw96b, wright96}.
Assuming $n_S = 1$, then the best-fit normalisation was found to
be $Q_{{\rm rms-PS};n=1} = 18 \pm 1.6$~$\mu$K. Finally, the
combined 31, 53, and 90~GHz map smoothed at $10^\circ$ yielded a
CMB rms of $29 \pm 1$~$\mu$K. The DMR full-sky map allowed also
testing for the Gaussianity of the temperature distribution
\cite{kogut96b}, which was found to be consistent with Gaussian
statistics.

These results set the stage for new major efforts: anisotropies
were now known to exist at amplitudes $\Delta T/T \sim$ $10^{-5}$,
and the race to unveil new features in the angular power spectrum
(especially with sub-degree scale observations) was started.

\subsection{Ground-based}
\label{subsec:ground}

High altitude, dry ground based sites have provided remarkable CMB
anisotropy results, in spite of the limitations of atmosphere and
ground emissions, although in small sky patches. 
Even if harsh and isolated, ground based
operations offer the enormous advantage of reachable instruments
and long integration times. The Python experiment, which operated
at the South Pole during the 1996-1997 austral summer, found that
the atmosphere could rarely be seen in the data; the experiment
was noise limited 75\% of the time \cite{lay00}. DASI data
suggest that during the austral winter, the efficiency is closer
to 90\%. The Polar environment during Antarctic winter is
probably the most favorable condition for a CMB ground based
experiment. In addition, the South Pole provides access to a
region of sky with some of the lowest dust column densities, as
inferred from the IRAS 100-$\mu$m map of the southern sky.

\subsubsection{\underline{UCSB South Pole}}
\label{subsubsec:south_pole}

Among the first experiments to systematically probe the sub-degree
portion of the power spectrum, this was a program by the
University of California at Santa Barbara which led to a series
of three measurement campaigns from the Amundsen-Scott South Pole
Station in the period 1988-1994. All of these experiments were
carried out with HEMT-based coherent receivers, cooled at
cryogenic ($T_{\rm phys} \simeq$ 4~K) temperatures. The observing
frequencies were in the atmospheric window around 30-40~GHz (Ka
and Q bands). The results from the 1988-1989 measurements are
described in \cite{meinhold91} and \cite{meinhold93} while the
results from the 1990-1991 measurements are detailed in
\cite{gaier92} and \cite{schuster93}.

The third campaign, during the austral summer 1993-1994, provided
the most accurate results of this effort \cite{gunderson95}. 
Each observation employed a 3 degree
peak-to-peak sinusoidal, single difference chop scanning
strategy. The observed sky was a $20^\circ \times 1^\circ$ strip.
Two sets of observations were performed: the first used a receiver
operating in three channels between 38 and 45~GHz (Q-band) with
FWHM varying from 1$^\circ$ to 1.15$^\circ$; the second
overlapped the first one, and used a receiver operating in four
channels between 26 and 36~GHz with FWHM varying from 1.5$^\circ$
to 1.7$^\circ$. Significant correlated structure was observed in
all channels for both observations leading to an amplitude
$\delta T_\ell \sim$ 30-35~$\mu$K at $\ell \simeq 60$. The spectrum
of the detected structure was proven to be compatible with CMB,
while a spectral analysis showed inconsistency (at $5 \sigma$
level) with a diffuse synchrotron or free-free emission. In
addition, the amplitude of the observed fluctuations was inconsistent with
20~K interstellar dust; however, the data were not able to
discriminate against contamination from flat or inverted spectrum
point sources.

\subsubsection{\underline{WD}}
\label{subsubsec:white_dish}

The ``White Dish'' experiment \cite{tucker93, peterson95} carried
out by a U.S. collaboration, was a ground-based bolometer detector
centred at 90~GHz, cooled at 100~mK by an adiabatic
demagnetisation cooler. The bolometer with 30\% bandpass had a
sensitivity $\sim 5$~mK$\times$Hz$^{-1/2}$. The instrument used an
on-axis Cassegrein telescope, with the secondary reflector
slightly displaced from the optical axis of the main reflector,
yielding a beam size of $12'$ \cite{tucker94, tucker93}. The
structure supporting the subreflector was designed to minimise
diffraction effects, and two radiation shields surrounded the
telescope during the observations. Deep side-lobes measurements
were performed showing a rejection $<-85$~dB at angles greater
than $50^\circ$.

The observations were made from the Amundsen-Scott South Pole
site in January 1993. Five interlocking circles in the sky were
scanned, with centres separated by $15'$ at constant elevation
angle, minimising spurious effects from the atmosphere, from the
ground as well as gravitational effects on the instrument.
Calibration was performed at the beginning and at the end of the
each 6-hour measurement run by placing warm absorber on the main
reflector and filling the beam with blackbody targets cooled with
liquid N$_2$ and O$_2$, yielding a calibration uncertainty of
$\sim 30\%$. A total of 150 hours data were collected, about 14\%
of which were used in the final analysis. No conclusive evidence
of anisotropy detection was obtained, but an upper limit was
found, $\Delta T/T < 2.3 \times 10^{-5}$ at 95\% confidence
level, corresponding to $\Delta T_\ell <$35~$\mu$K at $\ell\sim
60$.

\subsubsection{\underline{IAC-Bartol}}
\label{subsubsec:IAC_Bartol}

In the summer of 1994, a collaboration between the groups from
IAC and Bartol carried out observations from the Tenerife site at
angular scales $\sim 2^\circ$ \cite{femenia98}. Together with
Python and SuZIE, this was one of the few ground-based experiments
at frequencies $> 100$~GHz: four bands were used centred at 91,
142, 230, and 273~GHz. The instrument was based on bolometric
detectors in a 4-channel photometer cooled at approximately
0.33~K with $^3$He \cite{piccirillo91, piccirillo93}. An off-axis
Gregorian telescope was employed, with a 45-cm primary reflector
and a 28-cm secondary, leading to approximately equal beams with
FWHM$=2.03^\circ \pm 0.09^\circ$ at the four observing
frequencies. The primary reflector was sinusoidally wobbled, thus
moving the beam in the sky by about $\pm 2.6^\circ$ at 4~Hz. The
signal was then demodulated in software by evaluating the first
and second harmonics of the wobble frequency. Calibration was
performed with blackbody targets in the laboratory and against
the moon signal, leading to uncertainties at 5\% -- 25\% level,
depending on frequency. The drift-scan observations were centred
at $\delta = 40^\circ$, a region of the sky intensively observed
at 10-33~GHz by the Jodrell Bank - IAC experiment
\cite{hancock95, gutierrez00}, though at larger $\sim 5^\circ$
scales (see Sect.~\ref{subsubsec:tenerife}).

The main challenge was to disentangle the high atmospheric
contribution due to O$_2$, H$_2$O and O$_3$ emission lines at
millimetric wavelengths. The strong atmospheric signal showed up
as a correlated time-variable signature, which was removed by
relying on the 273~GHz channel as a monitor. Foreground Galactic
emission was excluded to significantly contribute based on
extrapolation of templates of Galactic emission and using
external data. The data demodulated at the two harmonics showed
some inconsistency which was attributed to residual atmospheric
effects. A two-component likelihood analysis gave an estimate of
the CMB fluctuation $\Delta T_\ell = 4.1^{+2.4}_{-2.2} \times
10^{-5}$~$\mu$K in the bandpower $\ell = 33^{+24}_{-13}$ and
$\Delta T_\ell = 2.0^{+1.0}_{-0.8} \times 10^{-5}$~$\mu$K in
$\ell = 53^{+22}_{-15}$. 

Recently a third campaign has been carried out with an improved angular resolution of
1.35$^\circ$ at all wavelengths \cite{romeo01}. These measurements provided
anisotropy detection in the multipole range $39 < \ell < 134$.
While this experiment showed that it is
possible to detect CMB anisotropy with ground-based measurements
at high frequency, it also clearly showed the very significant
limitation imposed by the atmospheric signal in this regime.

\subsubsection{\underline{Tenerife}}
\label{subsubsec:tenerife}

\begin{figure}[here]
\begin{center}
\resizebox{14.cm}{!} {\includegraphics{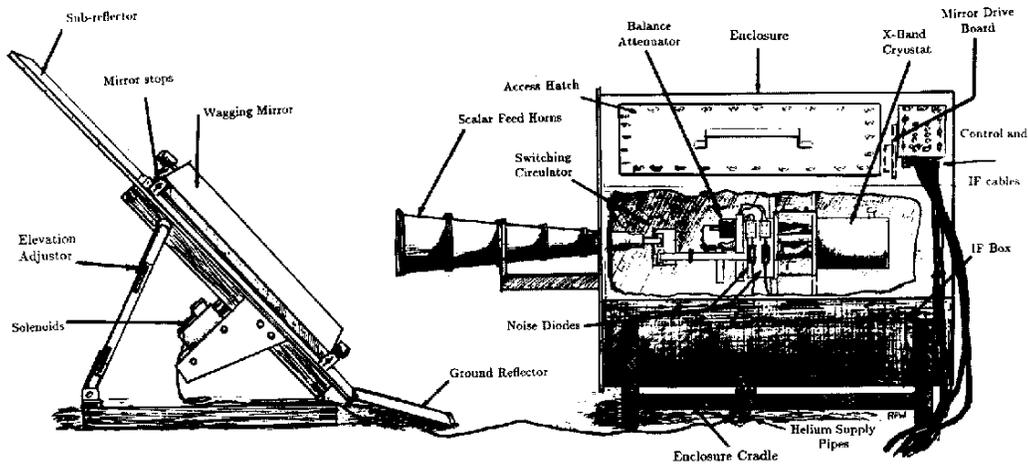}}
\end{center}
\caption{A sketch of the arrangement of the 10.45~GHz beam
switching system. A weatherproof enclosure houses the horn
support, the switching circulator and the cryogenically cooled
receiver system. The IF bands from the two independent receiver
systems are fed to a building 30~m distant, where they are
detected and recorded. The wagging mirror is operated by two
solenoids in a 14-s cycle; a guard mirror surrounds it.
Declination can be set by altering the mirror elevation. (From
\cite{davies92} with permission.)} \label{fig:tenerife}
\end{figure}

The Tenerife experiment employed three radiometers centred at
10.4, 14.9 and 33~GHz, designed and built at
Jodrell Bank (\cite{lasenby83}, \cite{watson92}, \cite{davies92},
\cite{davies96}), and operated in collaboration with the
Instituto de Astrofisica de Canarias (IAC) group in Tenerife.
Each instrument (see Fig.~\ref{fig:tenerife}) consisted of
dual-beam radiometer, with a pair of horns maintained fixed in
the horizontal direction, coupled to a tilted wagging secondary
flat reflector with a beamthrow of 8.1$^\circ$. This resulted in
a synthetic triple beam pattern in the sky, which helped
considerably in removing receiver long-term fluctuations as well
as the effect of atmospheric emission. At each frequency, two
independent Dicke-switched receivers cooled at $\sim$70~K with
liquid N$_2$ were used, thus improving sensitivity by a factor of
$\sqrt{2}$. The measurements were sensitive to angular scales of
$\sim 5^\circ$, or $\ell \sim$ 10-30, which correspond to the
Sachs-Wolfe plateau.

The observations were carried out at the Teide Observatory, Izana,
at an altitude of 2400 m. The good observing site (stable
atmospheric emission and typical water vapour content 2--3~mm
\cite{davies95}) and the advantageous observing strategy, helped
reducing data loss. Nonetheless, at 33~GHz, where the H$_2$O
contribution is highest, only about 20\% of the time was usable.

The data were taken by drift scans in right ascension and at a
fixed declination. By sampling at $2.5^\circ$ intervals in
declination with a $5.1^\circ$ FWHM, fully sampled
two-dimensional maps were constructed at each frequency
\cite{rebolo95}. A first set of observations scanned a sky strip
at declination $+40.0^\circ$ (a relatively low foreground region)
while subsequent observations covered adjacent sky strips,
between declination $+30^\circ$ and $+45^\circ$.

As discussed in Sect.~\ref{astrolimit}, in the 10-40~GHz range
the dominant foreground emission comes from diffuse Galactic
synchrotron and free-free emission. Between 10 and 33~GHz the
free-free component scales by a factor of $\sim 10$, while
synchrotron emission scales by as much as a factor $\sim 25$.
Thus the combination of the three instruments provided a good
spectral leverage, and in particular the 10.4~GHz channel was
used in the analysis to check against contamination from these
foregrounds.

The first clear detection of CMB structure by the Tenerife
collaboration \cite{hancock95} showed evidence of cosmic
anisotropy at a level $\sim 40$~$\mu$K. The Tenerife observations
provided the first ground-based detection of CMB anisotropy,
following the space-born DMR \cite{smoot92} and the balloon-borne
FIRS data \cite{meyer91}. It should also be mentioned that the
results from this experiment were the first in which the signal
to noise ratio per resolution element was sufficiently high to
actually ``image'' stable hot and cold spots in the CMB pattern.
The most recent analysis of the Tenerife effort
\cite{gutierrez99} covered observations over 5000 and 6500 square
degrees at 10 and 15~GHz respectively, centred around declination
$+35^\circ$. The ultimate receiver noise per resolution element
was 31 and 12~$\mu$K at 10 and 15~GHz respectively. The data
showed clear detection of structure at high Galactic latitude.
Including upper limits to possible galactic contamination
effects, the measured CMB signal was $\Delta T =
30^{+15}_{-11}$~$\mu$K. These values were highly
stable against the Galactic cut chosen and were found to be
compatible with the $COBE$ DMR data.
\subsubsection{\underline{Python}}
\label{subsubsec:python}

Five observing campaigns from the South Pole
Station were performed by the Python experiment,
lead by the group at Carnegie Mellon
University at Pittsburgh, each leading to
detection of CMB anisotropy at $\sim 1^\circ$ scales. The set up consisted of a 0.75~m
diameter off-axis parabolic telescope coupled either to a
bolometer system at 90~GHz or to a HEMT amplifier-based
radiometer at 40~GHz.

Observations from the first three seasons \cite{dragovan94,
ruhl95, platt97} were made at 90~GHz with an array of 4
corrugated horns feeding germanium bolometers cooled to 50~mK by
a $^3$He adiabatic demagnetisation refrigerator \cite{ruhl92}.
The four-point chop scan strategy yielded CMB detections at $\ell
\simeq 90$ and $\ell \simeq 170$. The beam was well approximated
by a Gaussian with FWHM $\sim 0.75^\circ$. Radiation from the sky
was first reflected by a chopping (100~Hz) flat reflector, and
then focussed into the cryostat by the primary mirror. Single-mode
waveguide filters were used to define the band. An additional
``blind'' bolometer was mounted on the cold stage as a monitor
against systematic effects. In addition, the overlap in the
regions observed in the three campaigns was exploited for
diagnostics including effects from atmospheric emission, pointing
and beam calibration, cosmic ray hits in the detectors,
radio-frequency interference, and sun radiation. The instrument
calibration was based on a combination of warm and cold targets
and the moon, leading to an accuracy of $\pm\sim$10\% limited by
systematic effects. The lack of spectral information prevented
direct discrimination of galactic foregrounds, although analysis
from external surveys \cite{dragovan93} indicated that galactic
contamination was likely to be small.

In the fourth and fifth Python campaigns \cite{kovac99},
\cite{coble99}, the telescope was equipped with a HEMT amplifier
radiometer, covering the 37-45~GHz range with a beam of about
1$^\circ$. The receiver consisted of two focal-plane feeds, each
with a single HEMT amplifier and a diplexer splitting the signal
at 41~GHz before detection, thus giving four data channels
\cite{kovac99}. Calibration was performed with liquid nitrogen,
liquid oxygen and ambient temperature external targets, leading
to $<15\%$ accuracy. In the most recent campaign \cite{coble99}
two regions were observed, sampling 598 square degrees in the
sky: the first included the fields of the previous four Python
observations, the second encompassed the region observed with the
Advanced Cosmic Microwave Explorer (ACME) telescope
\cite{gundersen95}. To search for potential foreground
contamination the data were cross-correlated with several
foreground templates  \cite{schlegel98, haslam82}
(Parkes-MIT-NRAO 4.8~GHz radio survey, PMN), and significant contamination from
known foregrounds was excluded.

The power spectrum results from Python were consistent with
$COBE$-DMR at low $\ell$, and indicated an increase at multipoles
$40 < \ell < 260$ from larger to smaller angular scales.

\subsubsection{\underline{Viper}}
\label{subsubsec:viper}

The Viper experiment, also led by the group at Carnegie Mellon
University at Pittsburg, was conceived as a follow up of Python
at higher angular resolution. The first results were obtained
with observations from the South Pole \cite{peterson00} at
40~GHz. At this frequency the Viper 2.15~m telescope yielded a
16~arcmin FWHM main beam, with an angular sensitivity in the
range $100 < \ell < 600$.

The rather sophisticated optics consisted of an aplanatic
Gregorian with the addition of a flat, electrically driven
chopping mirror (oscillating at 2.35~Hz), and of a fast hyperbolic
mirror redirecting the radiation to the radiometer
\cite{griffin98}. Extensive shielding around the primary and a
10-m baffle surrounding the whole instrument were employed to
minimise ground radiation pickup. The radiometer was a
two-channel receiver based on HEMT cryogenic ($\sim$20~K)
amplifiers, coupled to the telescope through corrugated feed
horns. The instrument measured the total power from 38 to 44~GHz,
in two sub-bands, and was calibrated using targets at ambient and
liquid-nitrogen temperatures with an estimated accuracy of $8\%$.
The main beam pattern ($0.26^\circ \pm 0.01^\circ$) and chopper
throw ($3.60^\circ \pm 0.01^\circ$) were calibrated using
observations of Venus.

Particular care was taken to remove offsets synchronous
with the chopping mirror modulation, e.g. due to synchronous
changes of illumination of the secondary mirror or to scattering by
snow grains on the optics.

The estimate of galactic foregrounds was based on the overlap with
the sky region observed by Python. Using IRAS data
\cite{beichman88} and the Parkes-MIT-NRAO survey data, an
upper limit $< 1$~$\mu$K was estimated from dust, synchrotron or
free-free emission \cite{coble99}. Due to the higher angular
resolution extra-galactic point sources were a more serious
concern compared to Python. Using PMN 4.8~GHz data the effect from
the brightest known point source (PMNJ2256-5158) was subtracted
at a level of $<2$~$\mu$K. Extrapolation of IRAS~100 $\mu$m
sources near the observed region lead to sub-$\mu$K limits. In
this and similar measurements, however, there remains the
possibility of contamination from a class of undetected objects with
spectra that closely mimic the CMB, such as dusty galaxies at
high $z$. High angular resolution sky surveys would be needed to
rule out this potential source of confusion. Interpreting the
power observed as due to CMB anisotropy, the results yielded
$\Delta T_\ell \approx 60$~$\mu$K, with a hint of the presence of
the first peak near $\ell \approx 200$.

After these measurements the Viper telescope was used to carry
out a survey of a complete sample of 14 luminous X-ray clusters
at $z<0.1$ \cite{romer99}, with the aim of measuring the
Sunyaev-Zel'dovich signal in the 45 - 350~GHz range.

\subsubsection{\underline{Saskatoon}}
\label{subsubsec:saskatoon}

This ground-based program, carried out by the Princeton group,
probed angular scales between 0.5$^\circ$ and 3$^\circ$, or
multipoles $60 < \ell < 360$, and produced one of the first
convincing evidences for the presence of the first peak in the power spectrum. 
The observations were made in the winters of 1993, 1994 and 1995 from
a site in Saskatoon, Saskatchewan, Canada. The instrument
\cite{wollack93, wollack96} used an off-axis parabolic reflector
coupled to cryogenically cooled total power receivers based on
low-noise HEMT amplifiers.
After the parabola, the beam was directed in a flat reflector
chopping in azimuth at 3.9~Hz, while the radiometer channels were
synchronously sampled at 250~Hz. Extended aluminum shields were
built around the instrument. The observations were made by two
independent radiometers, alternating each other at the focus of
the same telescope: one operated in the Ka-Band (26-36~GHz) and
another in the Q-Band (36-46~GHz). Each of these bands were
divided into 3 frequency ``sub-bands'' and 2 polarisations. The
beam FWHM of the Ka-Band and of the Q-Band were 1.42$^\circ$ and
1.04$^\circ$ degrees, respectively. To improve the sensitivity to
multiple angular scales, the beam was scanned back and forth
about the vertical axis, giving an effective beam pattern that
was synthesised in software. This technique also helped to remove
atmospheric noise and allowed to suppress the effect of ground
and solar radiation. For calibration and beam shape measurements,
Cassiopea-A was used as a source.

As mentioned, the Saskatoon data provided evidence for a raising
spectrum, going from $T_\ell = 49^{+8}_{-5}$~$\mu$K at $\ell =
87$, to $T_\ell = 85^{+10}_{-8}$~$\mu$K at $\ell = 237$. The
1993-1995 data were also used to construct a Wiener-filtered map
\cite{tegmark97} in a $15^\circ$ cap centred in the NCP, with an
angular resolution of $\sim 1^\circ$. The signal-to-noise ratio
in the map was of order 2. Some individual hot and cold spots
were significant at the 5 $\sigma$ level and were found to be
consistent for different observations. The spectral coverage of
the Saskatoon experiment was too low for a self-consistent
removal of foregrounds, so external information was used. No
significant correlation was found with point sources, with the
Haslam 408~MHz map, or with the Reich and Reich 1420~MHz map.
However, some correlation was found with the DIRBE 240, 140 and
100~$\mu$m maps, with an rms amplitude of the correlated
component $\sim 20\%$ of the CMB signal \cite{deOliveira97}.

\subsubsection{\underline{MAT/TOCO}}
\label{subsubsec:mat_toco}

As a continuation of the Saskatoon and QMAP experiments (see
Sect.~\ref{subsubsec:saskatoon} and \ref{subsubsec:QMAP}), the
MAT (Mobile Anisotropy Telescope) collaboration was established
between Princeton University and University of Pennsylvania to
extend the program of sub-degree measurements. The instrument
used the telescope successfully employed in two QMAP balloon
flights in 1996. The MAT instrument, sensitive to angular scales
from $\ell \simeq 40$ to $\ell \simeq 600$ had angular resolution
and frequency coverage chosen to overlap and complement the $MAP$
data (see Sect.~\ref{subsubsec:MAP}). In addition to channels at
31 and 42~GHz, overlapping $MAP$, MAT had a 144~GHz channel with a
0.2$^\circ$ degree beam. At this angular scale $MAP$ observes at
90~GHz. The 144~GHz channels are relatively immune from
extra-galactic sources and galactic foreground emission, and will
be useful to constrain potential contamination in $MAP$.

The receiver took advantage of both HEMT and SIS technologies,
both cooled with a mechanical refrigerator. Two Ka-band HEMT
channels (26-36~GHz) observed the same pixel on the sky in two
orthogonal polarisations, and four Q-band HEMT channels
(36-46~GHz) observed two orthogonal polarisations at each of two
independent pixels. The front-end HEMTs were cooled at 36~K while
the SIS mixers were operated at 4~K. An off-axis 85 cm paraboloid
was illuminated by the focal array of corrugated feeds with an
edge taper of about -25 dB, providing beam widths
$\sim$0.7$^\circ$ at Ka-Q bands and $\sim$0.2$^\circ$ in D band. A
large chopping flat reflector (1.8~m) was used to sweep the beam
across the sky, with a scanning strategy similar to that of
Saskatoon \cite{wollack96}.

The observations were carried out from Cerro Toco in northern
Chile from a site at an altitude of 5240 m with excellent
atmospheric characteristics. Two analyses have been published
covering the 1997 data at 30-40~GHz \cite{torbet99} and the 1998
data at 144~GHz \cite{miller99}. The observations in the two
campaigns concentrated on the same sky region, and the analysis
indicated that the detected structure is fully compatible with
the CMB spectral distribution. The results suggested that the
angular spectrum peaks at approximately 85~$\mu$K at
$\ell\sim$200, consistent with a $\Omega_0 \approx 1$ scenario.

Detailed design features and analysis of systematic effects was
described in \cite{miller02}, including microphonics, earth emission, atmospheric noise.
Systematic residual offsets at 2-8~mK level were observed. TOCO was calibrated using
Jupiter \cite{ulich81, griffin86}
providing a signal level in the range 15--350~mK, with an intrinsic calibration error of 5\%.

A quantitative analysis of foregrounds was carried out using the SFD dust map
\cite{schle97}, the Haslam map \cite{haslam82}, and the radio and IRAS source
compilations from WOMBAT\footnote{see WOMBAT home page at http://astron.berkeley.edu/wombat/}.
The dominant source in the Ka and Q bands was from the radio point sources as traced by the
4.85~GHz PMN catalog \cite{griffin93, gregory93, griffith94}. No significant contamination
from either dust or point sources was found at 144~GHz.

Recently \cite{miller02} a detailed analysis of the combined QMAP
and MAT/TOCO experiments was published. The MAT/TOCO experiment
provided a self-consistent data set with evidence of both the
rise and the fall of the first peak, with an estimated position
$\ell_{peak} = 216 \pm 14$.

\subsubsection{\underline{CAT}}
\label{subsubsec:CAT}

The Cosmic Anisotropy Telescope interferometer was developed by
the Cambridge group, and carried out observations from a site
near Cambridge. It is a three-element array that can operate at
any frequency between 13 and 17~GHz. The system has an observing
bandwidth of 500~MHz, a system temperature of $\sim 50$~K and it
simultaneously records data from orthogonal linear polarisations.
The baselines can be varied from 1 to 5~m, and for CMB
observations a synthesised beam of approximately 0.5$^\circ$ is
used. The three antennas have a diameter of 70~cm and the primary
beam has a FWHM of 2$^\circ$ at 15~GHz, and they are mounted on a
single turntable which can track in azimuth. To reduce the effect
of spill-over and artificial radio interference, the telescope is
surrounded by a 5-m high ground shield which, on the other hand,
limits observations to elevations $>30^\circ$.

The first set of CAT multifrequency observations made during 1994
and 1995 \cite{scott96, sullivan96} consisted of a $2^\circ\times
2^\circ$ area of the sky covered at frequencies of 15.5 and
16.5~GHz. Foreground radio sources brighter than 10~mJy were
removed using the Ryle Telescope. A comparison with earlier CAT
measurements at 13.5~GHz supported the evidence that the observed
structure at 16.5~GHz was to be attributed mainly to CMB
anisotropy. The overall broadband power at $330<\ell<680$, was
found to be $\Delta T_\ell/T = 2.0^{+0.4}_{-0.4} \times 10^{-5}$.

Further observations of a second $2^\circ\times 2^\circ$ field
\cite{baker99} after removal of discrete radio sources showed
detection of significant structure. A Bayesian analysis (see,
e.g., \cite{gull89}) suggested that in the small angular scale
range ($\ell=615$, $\theta \simeq 35'$), significant
contamination from Galactic emission was present. For these
multipoles, only an upper limit to CMB fluctuations was derived,
yielding $\Delta T_\ell/T < 2.0 \times 10^{-5}$. However, in the
lower half of the $\ell$ range the signal was found to be
dominated by CMB anisotropy. The average broad-band power centred
at $\ell = 422$ ($\theta \simeq 51'$) was estimated to be $\Delta
T_\ell/T = 2.1^{+0.4}_{-0.5} \times 10^{-5}$, a result fully
consistent with the detection \cite{scott96} obtained in a
different sky area. The importance of this result was to 
add further evidence for a downturn in the power spectrum at sub-degree scales.

\subsubsection{\underline{VLA}}
\label{subsubsec:VLA}

The Very Large Array of the National Radio Astronomy Observatory (NRAO) has
been used to set limits to CMB anisotropy at sub-arcminute angular resolution
since the early 80's \cite{fomalont84, knoke84, martin88, partridge88}.
In 1989 the VLA was equipped with low-noise receivers at 8.4~GHz, leading
to much better sensitivity to CMB fluctuations. VLA observations
\cite{fomalont93} established new upper limits in the angular range $0.17'$ to $1.33'$.

The most recent VLA observations at 8.4~GHz \cite{partridge97} were carried
out with a resolution of $6''$ in a $\sim 40$ arcmin$^2$ region, selected as
free from bright ($\sim 1$ mJy) radio sources, and which included one of
the survey areas of the HST Medium Deep Survey \cite{windhorst95}. The
data were collected from 1993 to 1995 in VLA compact configurations. As
a phase calibrator the source 1244$+$408 was regularly observed, and flux
(and polarisation) calibration was provided by 3C 286. The residual
variations in the gain of individual correlators were $< 3\%$ after
calibration and typically phase errors $<5^\circ$. A ``sum-image''
and a ``difference-image'' were constructed: the difference map was
used as a measure of the instrument noise in the corresponding sum map.
The correction for the effect of discrete radio sources included removal
of 46 bright ($>7$~$\mu$Jy) sources (most of which were listed in the catalog
of Kellerman et al \cite{kellerman86}), and estimating of the effect weak
radio sources with Monte Carlo analysis. Upper limits to CMB fluctuations
ranged between $\Delta T/T < 12.8 \times 10^{-5}$ at angular scales $0.17'$
and $\Delta T/T < 2.1 \times 10^{-5}$ at $1.33'$ scales (95\% confidence level).

\subsubsection{\underline{ATCA}}
\label{subsubsec:ATCA}

The Australia Telescope Compact Array (ATCA) is an interferometry
program carried out at the Australia Telescope National Facility,
CSIRO, to search for anisotropies on arcminutes scales. The
observing strategy uses full earth-rotation synthesis
observations in an ultra-compact 122~m array configuration, with
five 22-m antennas in an E-W line. Several precautions were
adopted to minimise systematic effects. For example, the array
phase centre was offset from the antenna pointing centre to keep
correlator offset errors well outside the primary beam; bandwidth
decorrelation was avoided by observing in multi-channel continuum
mode. The fields were selected to avoid strong discrete sources
and located at $\delta\sim -$50$^\circ$ to avoid shadowing and
maximise surface brightness sensitivity. The observations were
made at the highest available frequency, 8.7~GHz, to minimise
discrete source confusion. The three baselines between antennas
spaced 61~m apart were used to construct a model of the
foreground confusion with a strategy that minimised errors due to
sources variability and calibration. A first field was observed
in 1991 with a rms brightness sensitivity is 36~$\mu$K
\cite{subrahmanyan93}. The image showed no `excess' variance
above the telescope thermal noise and was used to place an upper
limit to anisotropy on arcminutes scales. New observations were
taken in 1994 and 1995 after the replacement of the 3~cm FET
amplifiers with HEMTs. After discrete sources subtraction, the
residual images were consistent with the expected thermal noise
(23~$\mu$Jy/beam), placing  a 95\% confidence upper limit of
$\delta T < 23.6$~$\mu$K. Polarisation limits have also been
obtained. Including more recent observations
\cite{subrahmanyan00} and assuming Gaussian distribution of the
CMB, upper limits are found of $\delta T_P < 11$~$\mu$K in
polarised intensity and $\delta T < 25$~$\mu$K in total intensity
for multipoles in the range $\ell =$~3350-6050. This provides
strong evidence for the turn over in the anisotropy spectral
power at large $\ell$.

\subsubsection{\underline{OVRO}}
\label{subsubsec:OVRO}

A long series of filled-aperture experiments have been carried
out at the Owens Valley Radio Observatory (OVRO), leading in the
late 90's to an unambiguous detection of CMB anisotropy at high
multipoles. Early measurements \cite{readhead89} were performed
at 20~GHz near the NCP using the 40-m telescope. These data
established an upper limit $\Delta T/T < 1.7 \times 10^{-5}$ at
$2'$ scales, improving previous limits at these scales by a
factor of 2. In the early 90's the ``RING40M'' experiment
\cite{myers93} surveyed 96 fields at $\delta\sim 88^\circ$
arranged in an interlocking scheme to improve detection of
systematics. These data were checked partially against
contamination from extragalactic sources using the VLA at low
frequencies. Detection of structure at arcmin-scales was reported
with amplitude $2.3 \times 10^{-5} < \Delta T/T < 4.5 \times
10^{-5}$, which however was attributed mainly to residual
foreground emission.

Recently Leitch et al \cite{leitch00} presented results of the
``RING5M'' observations on angular scales $7'$ to
$22'$ ($361 < \ell < 756$) at 31.7 and 14.5~GHz, using the OVRO
5.5-meter and 40-meter telescopes from 1993 to 1996. The 31.7~GHz
system coupled with the 5.5~m telescope provided an angular
resolution of $7.4'$. Both receivers used HEMT-based radiometers
and Dicke switching at 500 Hz between two feed horns separated by
$\sim 22'$ on the sky to minimise $1/f$ noise and atmospheric
instability. The 14.5~GHz feeds under-illuminated an 11~m patch
on the 40~m dish to match the 31.7~GHz beam. The 14.5~GHz
receiver was mounted in an off-axis configuration to reduce
ground spillover. The 5.5-meter telescope was illuminated from
the Cassegrain focus, so that the largest sidelobes of the feed
illumination pattern were directed to the sky.

Ancillary VLA observations at 8.5 and 15~GHz allowed removal of
discrete sources. After subtraction, the data showed significant
structure, attributed to a combination of a steep-spectrum
foreground component and to CMB anisotropy. The foreground
component was found to correlate with IRAS 100~$\mu$m dust
emission. The extracted CMB component was estimated to contribute
$\sim 88\%$ of the observed fluctuation at 31.7~GHz, yielding an
rms fluctuation amplitude of $82^{+12}_{-9}$~$\mu$K, which
correspond to $\Delta T_\ell = 56^{+8.5}_{-6.6}$~$\mu$K.

The RING OVRO results were in good agreement with CAT
\cite{scott96} at similar scales, and in conjunction with the
detections of medium scales and with the upper limits placed by
SuZIE \cite{church97} and VLA they supported evidence of a
decrease in the power spectrum at high multipoles.

\subsubsection{\underline{SuZIE}}
\label{subsubsec:SuZIE}

The Sunyaev-Zel'dovich Infrared Experiment (SuZIE) instrument and
scanning strategy were designed to detect the S-Z effect in
galaxy clusters \cite{wilbanks94} with minimal residual
systematics, but it proved effective in placing upper limits to
CMB primary anisotropy as well. SuZIE is  a 6-element bolometer
array \cite{holzapfel97} operated at the Caltech Submillimeter
Observatory on Mauna Kea. The array, cooled at 0.3~K, can be
operated at 142, 217 or 268~GHz, depending on the choice of
metal-mesh bandpass filters selected in front of the Winston
cones array. For CMB \cite{church97} it was operated at 142~GHz,
with angular resolution $1.7'$ and 11\% bandwidth. The measured
emission from the sky and from the warm Cassegrain optics was
36~K; to reduce systematic effects from spill over, only 8~m of
the 10.4~m primary aperture were used. In order to remove
common-mode atmospheric and telescope emission, electronic
differencing was performed between pixels in the same row of the
array.

The observations at Mauna Kea were carried out in 1994. Two sky
regions of size $\sim 36' \times 4'$ were selected as free from
known sources using IRAS catalogs and the NRAO 5~GHz Survey
\cite{becker91}, and were observed for 6-8 hour runs. Each pixel
was observed with both a dual-beam and a triple-beam chop, with a
typical sensitivity per pixel of $\sim 100$~$\mu$K in each chop.
To minimise systematic errors from position-dependent variations
in telescope spillover, the telescope position was kept fixed
relative to the earth and the source was observed while drifting
across the array as the sky rotated. The data were calibrated
using Uranus, through a convolution of the planet's millimetre
spectral model \cite{griffin93} with the instrument response. The
two-dimensional beam shape was measured using both Uranus and
Jupiter. The overall calibration uncertainty was estimated $\pm
8\%$, $1\sigma$.

Maximum-likelihood analysis was used assuming a Gaussian autocorrelation function for the
CMB, leading to an upper limit $\Delta T/T < 2.1 \times 10^{-5}$ (95\% confidence) for a coherence
angle of $1.1'$. The SuZIE upper limits are consistent with the results obtained at
similar angular scales with centimeter-wavelength observations (OVRO, VLA, ATCA), and provide
an important crosscheck from an independent wavelength regime affected by different potential
systematic effects and foreground contamination. In particular, one can expect the contribution
from point sources to be much smaller at SuZIE frequencies.

\subsubsection{\underline{JB-IAC}}
\label{subsubsec:JB-IAC}

The collaboration between the groups at Jodrell Bank and
Instituto de Astrofisica de Canarias lead to a measurement with
an interferometer centred at 33~GHz at degree angular scales. The
instrument \cite{melhuish99} used two horn reflector antennas
positioned to form a single E-W baseline, and it observed at
constant declination using the rotation of the earth so that no
moving parts are involved. The receivers used cryogenically
cooled, low noise, HEMT amplifiers, with bandwidth $\sim 10\%$.
In good weather conditions the system had an rms noise of 220
$\mu$K in a 2-minute integration. The measured main beam of the
interferometer (approximately a two-dimensional Gaussian with
$\sigma_{\rm RA} \times \sigma_{\rm Dec} \simeq 2.25^\circ \times
1.0^\circ$) yielded angular sensitivity in the range $\ell \simeq
109\pm19$.

Dicker et al \cite{dicker99} described the data taken in about 100 days of useful
observations from the Teide Observatory, Tenerife, in 1997-98 in a sky
strip at $\delta = +41^\circ$, taking advantage of the results of the 10~GHz beam-switching
Tenerife experiments in a largely overlapping sky region
\cite{hancock97, gutierrez00, lasenby96, davies98}. Galactic emission was shown to be
a factor $>10$ lower than the expected CMB fluctuations (see
Sect.~\ref{subsubsec:tenerife}). The source 3C84 was visible in the
data at a level consistent with its intensity as monitored in
the Metsahovi 22 and 37~GHz programme \cite{terasranta92}. All the other individual sources were
weaker than the measured signals in the raw data.
Relative calibration and drift removal was achieved by using a known periodical signal from
an internal source, stable to better than $\sim 4\%$ level in amplitude and $\sim 2^\circ$ in phase. The
internal signal was calibrated using the moon (which gave a signal $\sim$2--4~K in the beam), the
Crab Nebula as well as hot and cold loads.

The use of the MEM analysis technique (see 
Sect.~\ref{subsec:separating_foregrounds}) allowed to reduce the noise to
7~$\mu$K per $5^\circ$ RA beam. The detection of CMB anisotropy was unambiguous,
with $\Delta T_\ell = 43 \pm 12$~$\mu$K at $\ell = 109 \pm 19$. The main contribution
to the uncertainty was sampling error (21\%), which dominated over
the noise ($\sim 10\%$) and moon calibration (6.6\%).

\subsubsection{\underline{DASI}}
\label{subsubsec:DASI}

The Degree Angular Scale Interferometer, a compact 13-element 
interferometer sensitive to multipoles
$100 < \ell < 900$, was developed by a collaboration between the University of
Chicago and CalTech. The instrument \cite{halver01} used cryogenically cooled HEMT 
amplifiers in the spectral
window 26-36~GHz, in ten 1~GHz channels. The instrument was installed at the 
Amundsen-Scott South Pole
station during the 1999-2000 austral summer and carried out observations 
throughout the following austral winter.

The mechanism of the alt-azimuth mount telescope was
optimised for tracking
and pointing stability. The 13 antenna elements are arranged in a three-fold 
symmetric pattern on a rigid faceplate,
that can be rotated about its axis: this provides important diagnostic capabilities 
against systematic effects such
as cross-talk between antennas. For imaging purposes, the rotation also allows 
dense sampling of the u-v plane.

Each primary antenna consisted of a 20~cm diameter, wide
flare-angle corrugated horn with unobstructed aperture and
low sidelobes. To reduce cross-talk from correlated amplifier
noise, the receivers were equipped with front-end isolators, and
each antenna element was surrounded by a corrugated shroud. Each
element used a high density polyethylene lens, which allowed a
very compact horn design, while at the same time flattening the
aperture distribution. At Antarctic winter temperatures,
the lens contribution to the system temperature was $<2.5$~K. The
measured field of view of the interferometer was 3.4$^\circ$ with
first sidelobes below -20~dB. Exact antenna spacings on the
faceplate were optimised to yield uniform u-v
coverage over the accessible angular range. Each horn was coupled
to a cryogenically cooled 4-stage Indium Phosphide HEMT amplifier
operating at 26-36~GHz \cite{pospieszalski93, pospieszalski94}.
Typical system temperatures for DASI are of order 25~K, for an
rms sensitivity of $\sim$0.8~mK$\times$Hz$^{-1/2}$ in a 1~GHz band
on a single baseline. Besides yielding spectral information, the
ten 1-GHz channels allowed frequency synthesis mapping. The
effective snapshot resolution when all 10 bands were combined was
approximately $20'$. The effective sensitivity in $\ell$-space
was a nearly flat composite window function (including all
independent baselines at a single frequency) in the 100-800
multipole range.

Ground radiation, present at a level much greater than the CMB signal,
strongly constrained the DASI observing strategy. The observations strategy 
was a compromise between ground radiation rejection, observing time on each 
field, and sample variance on the CMB signal 
(see Sect.~\ref{subsec:cosmic_sample_variance}). Dedicated tests showed 
that the ground signal was stable over several days, and a suitable 
subtraction scheme was applied.

In its first season, DASI mapped CMB fluctuations in 32 fields, 
each 3.4$^\circ$ across. The results of this first campaign mapped 
the power spectrum in the range $100 < \ell < 900$ with high signal-to-noise. 
In the analysis \cite{leitch02} constraint matrices are used to discriminate 
systematic effects such as ground radiation, and for separating astrophysical 
foreground components. No evidence was found of significant contributions
from diffuse foregrounds, except for point sources. From the DASI data alone it was possible
to identify and remove point sources to approximately 0.25~Jy. A maximum likelihood temperature
spectral index $-0.1 \pm 0.2$ (at 1-$\sigma$ level) was found, thus consistent with CMB. The DASI
first peak in the power spectrum at $\ell \approx 200$ is evident and in good agreement
with results from TOCO, Boomerang and Maxima. In addition, the DASI data suggest the presence of
further peaks at $l \approx 550$ and at $\ell \approx 800$. The angular locations of these peaks is
consistent with the second and third harmonic peaks predicted by adiabatic inflationary cosmological models.

\subsubsection{\underline{CBI}}
\label{subsubsec:CBI}

The Cosmic Background Imager collaboration involved CalTech, the NRAO, the Canadian Institute for
Theoretical Astrophysics, the University of Chile and other international Institutions. The CBI
interferometer observations were sensitive to scales from $\sim 5'$ to $\sim 1^\circ$ (i.e.
from $\ell \simeq 300$ to $\ell \simeq 3000$). The 13-elements array operated in
ten 1~GHz-wide frequency bands from 26~GHz to 36~GHz using low-noise, broad-band
cryogenic HEMT amplifiers. The multi-frequency detection allowed to separate
synchrotron and free-free foregrounds, while unresolved extragalactic sources
were measured with the 40 meter OVRO telescope and subtracted from the data.

The feed-horns and the receivers were cooled at $\sim 6$~K,
reaching receiver noise temperatures of $\sim$15~K. A system of
quarter-wave and half-wave plates provided circular polarisation
and rejection of inter-antenna coupling effects. The
instantaneous field of view was $44'$ and its angular resolution
ranged from $4.5'$ to $10'$. A 9-hour observation with the CBI
yields an image covering a $2^\circ \times 2^\circ$ field with
rms noise $\Delta T/T \simeq 3 \times 10^{-6}$ on $10'$ pixels.
The CBI was located at an altitude of about 5000 meters near San
Pedro de Atacama, in northern Chile, where it was installed in
1999.

A classical Cassegrain optical design was adopted with shielded
0.9-m-diameter antenna elements \cite{padin00}. The elements can be
close-packed for maximum sensitivity, although this increases the
scattering effects between adjacent antennas, expected at level of
$\sim$15~$\mu$K. However scattered noise is not completely
correlated with the receiver noise and it reaches the correlator
with some delay. In addition the mounting structure allowed to rotate
the planar array about its axis, so that signals from the sky and
instrumental systematics can in principle be separated out. The
elevation range of the mount was restricted to $> 40^\circ$ to reduce
ground pickup, evaluated below $<3$~$\mu$K.  To minimise
scattering from the secondary support structure, some of its critical
parts were made with expanded polystyrene foam, essentially transparent
at $\sim 1$~cm wavelength.

Preliminary results \cite{cartwright01, padin01} showed a sharp
decrease in $C_\ell$ in the range $\ell = 400 - 1500$. An
estimate of flat band-powers yielded $\delta T_\ell =
58.7^{+7.7}_{-6.3}$~$\mu$K in $\ell = 603_{-166}^{+180}$ and
$\delta T_\ell = 29.7^{+4.8}_{-4.2}$~$\mu$K in $ \ell =
1190_{-224}^{+261}$. Images from the visibility data by the usual
synthesis-imaging procedures were also obtained \cite{padin01}.

Recently, a spectacular set of observations have been released by
the CBI team \cite{pearson02, sievers02, myers02, bond02}. Images
and power spectra were presented for three pairs of sky fields,
each of size $145' \times 165'$ (total sky area $\sim 40$
deg$^2$), using overlapping pointings (mosaicing). The power
spectrum results at $\ell < 2000$ were in agreement with
other experiments (in particular Boomerang, Maxima and DASI), but
the CBI multipole range extended up to $\ell \approx 3000$,
covering for the first time structures with mass-scales
corresponding to clusters of galaxies ($10^{14}$ to $10^{17}$
M$_\odot$). The spectrum, reconstructed from the interferometer
data \cite{myers02} suggests detection of the first 4 peaks
(and possibly of the fifth), and showed damping of power at $\ell
\sim$ 2000, consistent with simple inflation models. An excess
power was observed at high $\ell$, suggestive of a
Sunyaev-Zeldovich effect from clusters of
galaxies \cite{bond02}. These CBI data improved
constraints on the main cosmological parameters \cite{sievers02}
and provided further strong support for cosmological models
dominated by cold dark matter and dark energy, and with a
scale-invariant spectrum of primordial density fluctuations.
Further observations made in 2001 are being analysed by the CBI
team, and are expected to further improve the sensitivity and the
resolution of power spectrum estimates.

\subsection{Balloon experiments}
\label{subsec:balloon}

Observations of the CMB from high altitude balloons have a long
history \cite{weiss80, partridge95}, although only recently have they
become sophisticated enough to produce interlocking scan
strategies and accurate pointing reconstruction (see, e.g.,
\cite{staren00, debe00, lee01, miller02}. While the advantage
over ground-based experiments of the reduction of atmospheric
emission (typically by a factor of $10^3$) is great, in
conventional flights the available observing time is only about
10-12~hours. In recent years, long-duration balloon flights
(10-15 days) have been successfully flown, and in the future
ultra-long duration flights (over 100~days) may become possible,
thus competing with space experiments.

\subsubsection{\underline{FIRS}}
\label{subsubsec:FIRS}

The Far Infrared Survey balloon-borne program was started at MIT
in 1982 and produced high sensitivity maps of the sky
\cite{page90, meyer91} published soon after the DMR first
detection. The payload was hanging 600~m below the balloon at an
altitude of 35~km and the observations covered about 25\% of the
sky in a single flight. The instrument observed at 170, 290, 500
and 680~GHz with the two lower channels dedicated to CMB
fluctuations and the other two used as monitors for interstellar
dust and atmospheric fluctuations. The detectors were  monolithic
silicon bolometers cooled at 0.24~K.

The instrument angular resolution was about 3.8$^\circ$ through a
single horn. The sky signal was measured by comparing it to that
of a stable, thermally controlled reference load at a frequency
of 4.5~Hz with a cryogenic chopper. A secondary modulation was
provided by the rotation (8$^\circ$ per sec) of the gondola,
which observed at zenith angle of 45$^\circ$. The sensitivity at
170~GHz was $\sim 0.5$~mK$\times$Hz$^{-1/2}$, leading to detection
of CMB fluctuations in about 6 hours. The CMB dipole was used to
calibrate the 170 and 290~GHz channels, while at higher
frequencies an ambient temperature mylar target was used. Thermal
drifts in the main instrument optics \cite{page94}, and cosmic
rays hitting the detectors were among the main spurious effects
of concern, and both needed to be subtracted in the data analysis.

The IRAS 100~$\mu$m map was convoluted to the FIRS beam to
evaluate the galactic component. Little correlated emission was
found at $|b| > 15^\circ$. Combining different analysis
approaches on the FIRS data yielded an estimated CMB fluctuation
at a level $\Delta T = 19 \pm 5$~$\mu$K \cite{page95}. FIRS was
the first sub-orbital experiment to detect evidence of CMB
anisotropy. Limits on the power spectrum spectral index $n_S$
were also placed using the FIRS data \cite{ganga94}. A
cross­correlation analysis between the FIRS 170~GHz partial sky
and the full-sky $COBE$ DMR data \cite{ganga93} showed strong
correlation between the data sets, confirming that the source of
the observed structure is consistent with CMB anisotropy. The
power of this cross correlation analysis was enhanced by the
largely different observation technique, frequency ranges, data
reduction methods, and type of potential systematic effects
involved in the FIRS and DMR measurement.

\subsubsection{\underline{ARGO}}
\label{subsubsec:ARGO}

This degree-scale balloon experiment was carried out by the group at Universita\` di Roma, 
la Sapienza. The instrument \cite{debernardis94}
consisted of a 1.2~meter Cassegrain on-axis telescope with a wobbling secondary, 
coupled to bolometric detectors cooled at 0.3~K.
Four frequency bands were used at 150, 250, 375, and 600~GHz with $\sim 20\%$ bandwidths. 
The measured beam size was $52'$
(FWHM) modulated by a sinusoidal $1.8^\circ$ beamthrow.

The experiment was flown in 1993 from the balloon facility of the
Italian Space Agency in Trapani-Milo. The $^3$He evaporator
temperature during flight stabilised at 0.28~K with a stability
$<1$~mK/hr \cite{palumbo94}. A-posteriori pointing reconstruction
was based on a CCD camera with accuracy of few arcmin. In-flight
calibration was obtained by scanning a blackbody strip-target
partially filling the antenna beams. The in-flight sensitivity
was measured to be consistent with ground testing with a $NEP$ of
2 to 16 $\times 10^{-15}$~W$\times$Hz$^{-1/2}$ from 150 to
600~GHz. A scan nearby the moon was used to test sidelobe
response at $-50$ dB level at angles $\sim 10^\circ$ off axis,
although ground spillover was of particular concern due to the
optical setup \cite{debernardis94}.

A first set of observations were directed towards a low foregrounds region 
in Hercules \cite{debernardis94} yielding
a set of 63 independent sky differences. Correlation between the 600~GHz 
channel and the 100~$\mu$m IRAS map yielded an
estimate of the residual contribution of dust emission at 150 and 250~GHz 
of $<20\%$ and $<3\%$ respectively. These data
led to a detection of CMB anisotropy $\Delta T_\ell \simeq 39$~$\mu$K. 
A second subset of data was obtained with
observations at lower Galactic latitudes in Aries and Taurus, which exhibit 
significant cirrus component particularly in the high
frequency channels. Masi et al \cite{masi96} presented an analysis of separation 
of the galactic component yielding a statistically significant detection of CMB anisotropy 
$\Delta T_\ell \simeq 47 \pm 7$~$\mu$K rms (95\% confidence level).

\subsubsection{\underline{BAM}}
\label{subsubsec:BAM_B}

The Balloon-borne Anisotropy Measurement was a medium angular
scales experiment flown in 1995 \cite{halpern95, tucker97a,
tucker97}. The instrument used a cryogenic differential Fourier
transform spectrometer coupled to a 1.65 meter, prime-focus,
off-axis telescope and it was sensitive to angular scales from
$0.7^\circ$ to a few degrees. Data were obtained in 5 spectral
channels centred at frequencies in the range 110~GHz to 250~GHz.

The BAM spectrometer was previously used for absolute measurements of the CMB spectrum from a sounding rocket
\cite{gush90, gush92}. The optical elements were operated at $\sim 2$~K, while the bolometric detectors were cooled at 0.26~K.
Collimators defining the two inputs viewed the same portion of the primary to avoid systematic effects due
to thermal gradients across mirror. In addition, to minimise the risk of synchronous spurious signals, no
warm moving reflectors were used. The only moving optical element was the mirror assembly in the cryostat
whose rapid scanning was used to vary the optical length difference in the interferometer. This produced
interferograms at the bolometers whose amplitudes were proportional to the brightness difference between the
two spectrometer inputs. Spectra were obtained a posteriori by Fourier-transforming the interferograms with
respect to optical delay.

BAM was launched from the U.S. National Scientific Balloon Facility in Palestine,
Texas on 7 July 1995. Shortly before the gondola reached the floating altitude of
41.4~km, scans of Jupiter were performed for photometric calibration, which yielded an
accuracy of 20\%, and beam reconstruction. Unfortunately, due to the failure of a
memory chip in the pointing system, reliable commanding of the gondola was restored only for
the last 30 minutes of observation. In spite of these difficulties and the short integration
time, single difference measurements were obtained on 10 fields which produced a statistically
significant detection of anisotropy
$\Delta T_\ell = 55.6^{+29.6}_{-15.2}$~$\mu$K at $\ell \sim$ 75
(90\% confidence level).

\subsubsection{\underline{MAX}}
\label{subsubsec:MAX_B}

The Millimeter-wave Anisotropy experiment was carried out by a collaboration
between groups at U.C. Berkeley and U.C. Santa Barbara. The instrument was flown
five times between 1989 and 1994 to measure half-degree scale anisotropy. The
first flight \cite{fisher92} provided instrument performance tests, and in the
four successive flights observations of nine different sky regions were performed
with seven positive anisotropy detections
\cite{aslop92, gundersen93, meinhold93a, clapp94, devlin94, tanaka96}.

The instrument was a multi-band bolometric receiver mounted on an
attitude-controlled 1-meter Gregorian off-axis telescope
\cite{fisher92, aslop92, meinhold93b}. A single-pixel bolometer
in the focal plane received radiation from a system of dichroic
mesh filters, which split the beam in the different frequency
bands. Several aspects of the instrument were modified during the
development of the program, including frequency bands and
bolometer operating temperature. In the first 3 flights the MAX
composite bolometers at 180, 270 and 360~GHz (with $\sim$
30\%-40\% bandwidths) were cooled at 280~mK by a $^3$He
refrigerator; in the fourth and fifth flights, the instrument
included an additional single-mode channel centred at 105~GHz,
and was operated at 85~mK by an adiabatic demagnetisation
refrigerator. The sensitivity improved significantly during the
duration of the program, up to a factor of 10. The instrument was
calibrated by observing planets (Venus and Jupiter) during flight
and, in addition, by using a partially reflecting ($\sim 0.1\%$)
membrane as a periodic calibration standard. To fight systematic
errors, different scanning strategies were adopted, including a
double frequency scheme based on secondary reflector modulation
(5.4~Hz, as an optimal trade off between bolometer constant and
1/$f$ noise), and the gondola azimuthal scan (0.0075~Hz, to
reject atmospheric noise and bolometer temperature fluctuations).
Residual systematic effects due to earth or moon sidelobe
contamination, balloon straylight, atmospheric fluctuations were
analysed and a ``blind'' bolometer was included in the focal
instrument as a check against radio frequency interferences.

The spectral coverage at high frequency was sufficient to
discriminate the measured fluctuations against dust contribution.
During the third and fifth flights \cite{meinhold93, lim96} a sky
region near $\mu$-Pegasi was observed, which was found to be
dominated by a dust component as indicated by the observed
correlation with IRAS 100~$\mu$m map: only upper limits to CMB
anisotropy could be placed for these observations. Extrapolation
of the 408 MHz map \cite{haslam82} showed that synchrotron or
free-free emission could contribute only marginally (typically
less than $\sim 10\%$) to the observed fluctuations. No radio
sources available in catalogs were bright enough to contribute
significantly.

The MAX results were among the first ones suggesting a band power at degree-scales
higher than that of $COBE$-DMR. A complete analysis of the fourth and fifth MAX
flights detections \cite{tanaka96} yields five data points at multipoles
$78 < \ell < 263$ with $\Delta T_\ell$ in the range 33~$\mu$K -- 54~$\mu$K,
with typical uncertainties $\sim 30\%$, including $\sim$ 10\% calibration error.

\subsubsection{\underline{MSAM}}
\label{subsubsec:MSAM}

The Medium Scale Anisotropy Measurement was a balloon-borne bolometric instrument which
observed a region of about 10~deg$^2$ with half-degree resolution. The MSAM gondola was launched
three times in the period 1992-95 from Palestine, Texas, and each flight detected a clear CMB
anisotropy signature. A second version of the MSAM instrument (MSAM II) with complementary
frequency coverage has been launched in 1997. The first
two flights \cite{cheng94, cheng96} observed overlapping fields on the sky and demonstrated
the repeatability of the measurement \cite{inman97, knox98} while the third flight \cite{cheng97}
extended the sky coverage thus improving the coverage of the angular power spectrum.

The instrument \cite{fixsen96a} was a multi-channel bolometric detector cooled to 0.24~K at the
focus of a 1.4 m off-axis Cassegrain telescope. The beam had a FWHM of $28'$, and was moved
by a three-position chopping secondary by $\pm 40'$ in the sky. The instrument design and
support electronics was essentially the same as the one used for FIRS \cite{page94} with four
channels in the range 170 -- 678~GHz. Improvements in the gondola between the first and second
flight allowed  reduction of straylight contamination \cite{fixsen96a}. MSAM used an actively
pointed gondola yielding a final pointing accuracy of $2.5'$, small compared to the beam size
($28'$) and samples ($14'$).
The observing scheme was a slow azimuth scan of a region crossing the meridian above the NCP
covering an area sufficiently large to probe multipoles down to $\ell \sim 40$.
Jupiter was used to calibrate the signal and to map the beam
response in flight. Relative calibration accuracy between different flights was
estimated $\sim 4\%$, while the absolute uncertainty was $\sim 11\%$ \cite{inman97}.
The beam pattern uncertainties were dominated by pointing errors during raster scans of Jupiter.

The data analysis \cite{cheng94, cheng96, kowitt97} included
removal of spikes caused by cosmic rays, spurious electrical
pickup or telemetry dropouts (total loss 10\% to 30\% of the raw
data). An overall analysis of the three flights yielded
anisotropy detection in three band power estimates centred at
$\ell = 34$, $\ell = 101$ and $\ell = 407$.

\subsubsection{\underline{QMAP}}
 \label{subsubsec:QMAP}

The HEMT-based QMAP instrument, developed by the Princeton group,
was flown twice in 1996 to perform degree-scale observations at
31 and 42~GHz. The instrument and telescope essentially used the
same configurations as in the MAT/TOCO experiment (already
described in Sect.~\ref{subsubsec:mat_toco}), but without the
D-band SIS channel. Also, QMAP used liquid cryogens while TOCO
used a mechanical refrigerator to cool both the HEMTs and the SIS
mixers. Four of the six QMAP HEMT amplifiers were the same used
for the Saskatoon measurement, and were later used in the two
TOCO observing campaigns. Some evidence of degradation in the HEMT
performance in time \cite{miller02} above that expected from the
different operating temperatures \cite{pospieszalski89} was
observed.

Although QMAP and TOCO used very similar instruments, their
observational approaches were highly different: TOCO was designed
to measure the angular power spectrum, while QMAP was designed to
make a true map of the sky. The QMAP chopper swept horizontally at
4.7~Hz and the gondola wobbled in azimuth with a period of 100~s
about a meridian containing the NCP. This gondola motion,
combined with the earth rotation, produced a highly interlocking
scan pattern that helped separation of instrumental
effects from the celestial signal. The QMAP data
were pixelised on the sky and the analysis was performed
at the map level. Because of the short duration of the flights,
the offsets were found to be stable. In the mapmaking analysis,
the offsets were projected out of the CMB data using a technique
described in \cite{deoliveiracosta98}.

The instrument beams, with FWHM $\sim 0.8^\circ$, were mapped
using Cas-A and Saturn. Consistency checks were possible also
using the 31 and 43~GHz TOCO beams mapped with Jupiter. Small
variations in the beams from year to year were noted and
attributed to slight changes in the optical alignments.
Photometric calibration of QMAP was performed against Cas-A,
(unresolved at the $\sim 0.8^\circ$ beam size) using flux
data from \cite{baars77, chini84, mezger86}.

Definite detection of anisotropy power were reported both for the
first flight ($> 15\sigma$ in the range $40 < \ell < 140$
\cite{devlin98}), and for the second flight ($> 20\sigma$ in the
range $40 < \ell < 200$
\cite{herbig98}). Crosschecks searching for systematic effects
ensured that the signal was dominated by a component fixed on the
sky (thus not of instrumental origin), and foreground analysis
showed that the fluctuation was consistent with a CMB spectrum.
The second flight produced maps covering 83~deg$^2$
with resolution $42'$--$54'$ in Ka and Q band. After the first
release of the QMAP data, Mason et al. \cite{mason99} reported
new measurements Cas A in Ka band leading to an increase in the
temperature scale of the QMAP data \cite{miller02} by 6.6\%, and
reduced calibration uncertainty.

Based on a combination of the QMAP and Saskatoon data (known as
``QMASK'') Xu et al. \cite{xu02} have studied the power spectrum
on angular scales 1-6 degrees. The relatively large sky area of
the QMASK map (648 deg$^2$) allows to reduce the limitations of
sample-variance and to place significant constraints in the $30<
\ell <200$ range, bridging between Boomerang and $COBE$-DMR. Park
et al \cite{park01} carried out an analysis of the Gaussianity of
CMB anisotropy. The QMAP and Saskatoon maps were found to be
consistent with Gaussianity, while the QMASK map showed a mildly
non-Gaussian genus curve which did not appear to be explained by
known foreground contamination. However, one has to be extremely
cautious about the propagation of subtle systematic effects in
CMB Gaussianity analysis or other statistical analysis pushing at
the limit of instrumental sensitivity.

\subsubsection{\underline{Boomerang}}
 \label{subsubsec:Boomerang}

Boomerang (Balloon Observations of Millimetric Extragalactic
Radiation and Geophysics) was designed for a series of ``Long
Duration Balloon'' (LDB) flights around Antarctica. The program,
still on-going, was carried out by Italian and US teams at
Universit\'a di Roma La Sapienza and Caltech. After a test flight
in 1997 \cite{mauskopf00, melchiorri00}, the payload was launched
for its first LDB flight in December 1998 and landed roughly 10.5
days later, with 257 hours of data acquired at an altitude of
39~km. The instrument used a 1.2~m off-axis, aluminum parabolic
mirror, coupled to an array of bolometric detectors at 90, 150,
240 and 410~GHz, with sets of 2, 6, 3 and 4 channels
respectively. A large dewar (65~l of liquid He and 75~l of liquid
N) was able to continuously pre-cool the detectors to $\sim$4~K
for $\sim$12~days \cite{masi98}. A $^3$He refrigerator
\cite{masi99a} was mounted inside the dewar cooling the detectors
to 0.28~K. The bolometers reached typical sensitivities of 0.15 --
0.2 mK$\times$Hz$^{-1/2}$. The observation strategy consisted of
scanning back and forth the telescope in azimuth at an angular
velocity which could be set to 1 or 2 deg/sec. Both angular
velocities were used at different phases of the measurements in
order to decouple potential systematic effects synchronous with
the scan period. In addition, the elevation of the field of view
was changed every observing day, covering azimuth ranges of
$40^\circ$, $45^\circ$ and $50^\circ$.

\begin{figure}[here]
\begin{center} 
\resizebox{12.cm}{!}
{\includegraphics{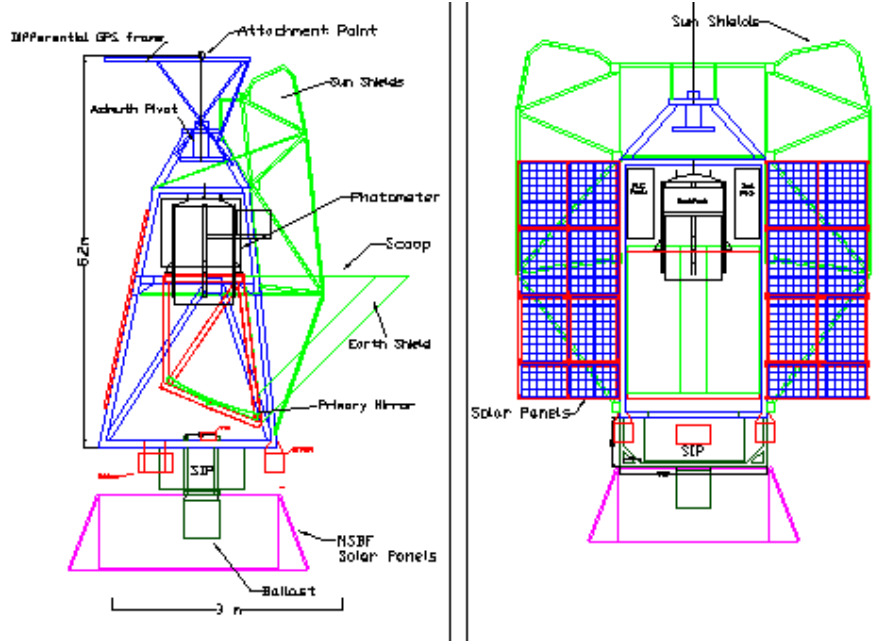}} 
\caption{The Boomerang Payload
in the LDB configuration. (From \cite{debe99} with permission.)}
\end{center}
\label{fig:Boomerang} \end{figure}

The first results \cite{debe00} were based on the data of a
single detector at 150~GHz, while a more detailed analysis
covering four 150~GHz detectors and a larger data set was later
presented \cite{netterfield02} including a refined estimate of
beam shape reconstruction and calibration. The most significant
residual systematic uncertainty at large multipoles came from the
poor knowledge of the effective beam sizes, which in turn was
affected by pointing uncertainties (of order of 2.5$'$).
Extragalactic sources as well as some bright galactic H{\sc II} regions
were used to map the beams ($\sim 18'$ at 90~GHz, $\sim 9.5'$ at
150~GHz, $\sim 14'$ at 240~GHz, and $\sim 12'$ at 410~GHz). To
map the near sidelobes, azimuthal near-field cuts were measured
down to $-20$ dB, but discrepancies of order 10\% between the
modelled and measured beams were found for the multi-mode Winston
concentrators. The window function was generated based on
physical optics models of the far field patterns. The observed
sky patch was chosen to avoid the Galactic plane and to include
some selected point sources. Calibration at 150~GHz was obtained
using the CMB dipole. The second analysis \cite{netter01} lead to
a 10\% correction in the power spectrum calibration with respect
to the first one \cite{debe00}.

A powerful test against systematic effects was to difference the
map made from data acquired in the first half of the flight with a
scan velocity of 2$^\circ$/sec from that of the second half of
the flight at 1$^\circ$/second. Such difference map is sensitive
to solar and ground contamination. One of the 150~GHz channels
was found to be contaminated and was not used in the analysis.
The same test was applied to the combination of the 150~GHz
channels, leading to a residual at $\ell < 400$ of $\sim
7$~$\mu$K, possibly due to atmospheric contamination.

The Boomerang results lead to the first robust and self-consistent
measurement of the first acoustic peak and probed the power
spectrum in the range $75 < \ell < 1025$: it thus represented a
major breakthrough in the field. Fits to Boomerang data to
estimate cosmological parameters \cite{lange01, netterfield02}
indicate that with the prior assumption $0.45 < h < 0.90$ the
values of both $\Omega_0$ and $n_S$ were strongly constrained to
be close to unity, and supported values of $\Omega_b$ consistent
with those suggested by cosmological nucleosynthesis
\cite{burles98}. Taking into account large scale structure
observations and supernovae Type 1a light curves, the constraints
were even more powerful. The best estimate for the Hubble
parameter was $h = 0.68 \pm 0.09$, in excellent agreement with the
results of the Hubble Space Telescope key project
\cite{freedman01} of $h = 0.72 \pm 0.08$, and strong evidence for
the existence of dark matter and dark energy was found with
$\Omega_\Lambda \approx 0.7$ and $\Omega_m \approx 0.3$.

\subsubsection{\underline{Maxima}}
 \label{subsubsec:Maxima}

A follow-up of the MAX project, the Maxima collaboration involves
U.C. Berkeley, the University of Minnesota/Twin Cities, CalTech,
University of Rome, and the IROE-CNR in Florence. The experiment
\cite{lee98} was a balloon-borne bolometric receiver sensitive to
multipoles $80 < \ell < 800$. The receiver housed 16 bolometric
channels (compared to the single photometer of MAX) with
frequency bands centred at 150, 240, and 410~GHz. The bolometers
were cooled to 100~mK, by an adiabatic demagnetisation
refrigerator, and a pumped $^3$He stage provided an intermediate
temperature of 300~mK. The focal assembly was coupled to a 1.3~m
diameter off-axis Gregorian telescope, yielding a typical angular
resolution of $\sim 10'$. The warm secondary mirror used in the
MAX experiment was replaced with cold secondary and tertiary
mirrors, which allowed to insert a cold baffle in the optical
path. To reduce the effects of temperature gradients on the
optics, the large primary mirror, rather than the secondary, was
modulated such that the optical beam on the primary remained
stationary.

The first flight in its full configuration \cite{hanany00} was
launched in August 1998, and produced one of the most important
data sets together with the Boomerang results. A 124~deg$^2$ sky
region (0.3\% of the sky, 3200 independent pixels) was imaged
during the 7-hour flight in a direction near Draco with low
galactic dust contamination. The data were calibrated using
in-flight measurements of the CMB dipole. A map of the CMB
anisotropy obtained by combining 150 and 240~GHz data, without
foreground subtraction, yielded angular information in the range
$36 < \ell < 785$. The spectrum showed clear evidence of a peak
with an amplitude of $78 \pm 6$~$\mu$K at $\ell \sim 220$ and an
amplitude varying between $\sim 40$ and $\sim 50$~$\mu$K for $400
< \ell < 785$.

Recently \cite{lee01} the analysis of the first flight was
extended to smaller angular scales, up to $\ell = 1235$, using
data from three 150~GHz channels in the central 60~deg$^2$ of the
map. The results of the new analysis, which used improved
resolution in the map making, were consistent with the findings of
\cite{balbi00} for $\ell < 785$. For higher multipoles, evidence
was found of excess power at $\ell \sim 860$ over the average
level at $411 < \ell < 785$, suggesting the presence of a third
acoustic peak \cite{stompor01}. At the same time the observed
power spectrum was found not consistent \cite{stompor01} with a
number of the non-standard (but still inflation-based) models
that have been proposed to improve the quality of fits to the
overall CMB data set. Combined with the $COBE$-DMR data, the
Maxima map gave best fit values for a number of cosmological
parameters; in particular, for the baryon density, $\Omega_b
h^2\simeq 0.0325{\pm 0.0125}$, and the total density, $\Omega_0 =
0.9{+0.18\atop -0.16}$ (95\% confidence level). Gaussianity tests
were also performed \cite{wu01} leading to upper limits to
possible deviations at scales between 10$^\circ$ and 5$^\circ$.

\subsection{Summary and discussion of up-to-date CMB measurements}
\label{summary_and_discussion}

Tables \ref{tab:results_1} though \ref{tab:results_4}
contain information about various experimental features
and results of those measurements leading to anisotropy detection.
\vfill

\begin{center}
\begin{threeparttable}
\small
\caption{Summary of CMB experiments and detection reported in
Fig.~\ref{fig:summary_plot_a}}\label{tab:results_1}
\begin{tabular*}{13.6cm}{lcccccrr}
\hline
\            &                                    &                         &            &               &            &                         &                       \\
\ Name       & Type \tnote{a} /Tech.\tnote{b}     & $\theta$,$f_{\rm sky}$  & $\nu$ \tnote{c}      & $T_{\rm det}$ & Calibr.    &  $\ell\pm\Delta \ell$   & $\delta T_\ell$        \\
\            & Altitude                           &                         & [GHz]      & [K]           &            &                         & [$\mu$K]               \\
\            &                                    &                         &            &               &            &                         &                       \\
\hline
\            &                                    &                         &            &               &            &                         &                       \\
\ $DMR$      & S/R                                & $7^\circ$               & 31.5       & 290\tnote{d}  & 0.7\%      &   2.1$^{+0.4}_{-0.1}$   &  8.5$^{+16.0}_{-8.5}$ \\
\            & 900~km                             & 100\%                   & 53         & 140\tnote{e}  & S/C motion &   3.1$^{+0.6}_{-0.6}$   & 28.0$_{-10.3}^{+7.5}$ \\
\            &                                    &                         & 90         &               & Moon       &   4.1$^{+0.7}_{-0.7}$   & 34.0$^{+6.0}_{-7.2}$  \\
\            &                                    &                         &            &               & Internal   &   5.6$^{+1.0}_{-0.9}$   & 25.1$^{+5.3}_{-6.6}$  \\
\            &                                    &                         &            &               &            &   8.0$^{+1.3}_{-1.2}$   & 29.4$^{+3.6}_{-4.1}$  \\
\            &                                    &                         &            &               &            &   10.9$^{+1.3}_{-1.2}$  & 27.7$_{-4.5}^{+3.9}$  \\
\            &                                    &                         &            &               &            &   14.3$^{+1.3}_{-1.6}$  & 26.1$_{-5.2}^{+4.4}$  \\
\            &                                    &                         &            &               &            &   19.4$^{+2.7}_{-2.8}$  & 33.0$_{-5.4}^{+4.6}$  \\
\            &                                    &                         &            &               &            &                         &                       \\
\ Tenerife   & G/R                                & $5.1^\circ$             & 10.4       & 70            & ...       &   20.0$^{+10.0}_{-8.0}$ & 32.5$^{+10.1}_{-8.5}$  \\
\            & 2400~m                             &  15\%                   & 14.9       &               & Moon        &                         &               \\
\            &                                    &                         & 33         &               & Sun           &                         &                 \\
\             &                                    &                         &            &               & Gal. plane           &                         &                 \\
\            &                                    &                         &            &               &            &                         &                          \\
\ FIRS       & B/B                                &3.8$^\circ$              & 170        & 0.24          & ...         &   11.0$^{+17.0}_{-9.0}$ & 29.4$^{+7.8}_{-7.7}$  \\
\            & 35~km                              & 25\%                    & 290        &               & Dipole     &   3.1$^{+0.6}_{-0.6}$   & 28.0$_{-10.3}^{+7.5}$ \\
\            &                                    &                         & 500        &               & Internal   &   4.1$^{+0.7}_{-0.7}$   & 34.0$^{+6.0}_{-7.2}$   \\
\            &                                    &                         & 680        &               &            &                         &                        \\
             &                                    &                         &            &               &            &                         &                \\
\ IACB       & G/B                                & 2$^\circ$               & 91         & 0.33          & 5-25\%\tnote{f} &  33.0$^{+26.0}_{-16.0}$ & 111.9$^{+65.4}_{-60.1}$\\
\            & 2400~m                             & 0.5\%                      & 142        &               & Moon       &  53.0$^{+26.0}_{-19.0}$ & 54.6$^{+27.2}_{-21.9}$ \\
             &                                    &                         & 230        &               &            &                         &                \\
             &                                    &                         & 273        &               &            &                         &                \\
\ IAC        &  G/B                               & 1.35$^\circ$            & 96-270     &  0.33         & 10\%         &  39.0$^{+38.0}_{-24.0}$ & 34.0$^{+8.0}_{-6.0}$   \\
\            &  2400~m                            & 2.4\%                     &            &             & Moon         &  61.0$^{+28.0}_{-22.0}$ & 40.0$^{+7.0}_{-6.0}$   \\
             &                                    &                         &            &               &            &  81.0$^{+27.0}_{-20.0}$ & 41.0$^{+8.0}_{-8.0}$   \\
             &                                    &                         &            &               &            &  99.0$^{+24.0}_{-18.0}$ & 50.0$^{+10.0}_{-9.0}$  \\
             &                                    &                         &            &               &            &  116.0$^{+23.0}_{-14.0}$& 46.0$^{+10.0}_{-9.0}$  \\
             &                                    &                         &            &               &            &  134.0$^{+20.0}_{-22.0}$& 56.0$^{+11.0}_{-10.0}$ \\
             &                                    &                         &            &               &            &                         &                \\
\ SP         & G/R                                & 1.5$^\circ$             & 38-45      & 4             & 10\%       &  61.0$^{+41.0}_{-31.0}$ & 30.2$^{+8.9}_{-5.5}$   \\
\            & 2800~m                             &                         &            &               & Moon       &  61.0$^{+41.0}_{-31.0}$ & 36.3$^{+13.6}_{-6.1}$  \\
\            &                                    &                         &            &               & Internal   &
                &                        \\

             &                                    &                         &            &               &            &                         &                    \\
\ Saskatoon  & G/R                                & 1.04$^\circ$            & 26-46      & 20            & 13\%       &  87.0$^{+44.0}_{-35.0}$ & 49.0$^{+8.0}_{-5.0}$   \\
             & 480~m                              & 1.42$^\circ$            &            &               & Cas-A      & 166.0$^{+39.0}_{-48.0}$ & 69.0$^{+7.0}_{-6.0}$   \\
             &                                    & 0.5\%                   &            &               &            & 237.0$^{+36.0}_{-48.0}$ & 85.0$^{+10.0}_{-8.0}$  \\
             &                                    &                         &            &               &            & 286.0$^{+33.0}_{-44.0}$ & 86.0$^{+12.0}_{-10.0}$ \\
             &                                    &                         &            &               &            & 349.0$^{+51.0}_{-46.0}$ & 69.0$^{+19.0}_{-28.0}$ \\
             &                                    &                         &            &               &            &                         &                \\
\ JB-IAC     & G/R-I                              & 1$^\circ$               & 33         & 20            & 6.6\%      &     109$^{+19.0}_{-19.0}$  & 43.0$^{+12.0}_{-12.0}$ \\
\            & 2800~m                             & 0.8\%                      &            &               & Moon       &                            &                        \\
\            &                                    &                         &            &               & Tau-A      &                            &                        \\
             &                                    &                         &            &               &            &                            &                        \\
\hline
\end{tabular*}
\begin{tablenotes}
    \small
    \item[a] S = Space, G = Ground, B = Balloon
    \item[b] R = Radiometer, B = Bolometer, I = Interferometer
    \item[c] Where a frequency range is indicated (\eg 26-46) then the receiver operated in a wide
            frequency band. See text for further detail.
    \item[d] 31.5~GHz radiometers.
    \item[e] 53 and 90~GHz radiometers
    \item[f] Depending on frequency
\end{tablenotes}
\end{threeparttable}
\end{center}

\begin{center}
\begin{threeparttable}
\small
\caption{Summary of CMB experiments and detection reported in
Fig.~\ref{fig:summary_plot_a}. Continued}\label{tab:results_2}
\begin{tabular*}{13.6cm}{lcccccrr}
\hline
\            &            &                         &            &               &            &                         &                       \\
\ Name       & Type/Tech. & $\theta$,$f_{\rm sky}$\tnote{a}  & $\nu$      & $T_{\rm det}$ & Calibr.    &  $\ell\pm\Delta \ell$     & $\delta T_\ell$        \\
\            &            &                         & [GHz]      & [K]           &            &                           & [$\mu$K]               \\
\            &            &                         &            &               &            &                         &                       \\
\hline
\            &            &                         &            &               &            &                         &                       \\
\ QMAP       & B/R        & 54$'$                   & 31         & 2.3            & 10-13\%    &     80.0$^{+41.0}_{-41.0}$ & 47.0$^{+6.0}_{-7.0}$   \\
\            & 30~km       & 42$'$                   & 42         &               & Cas-A      &   110.0$^{+65.0}_{-63.0}$  & 52.0$^{+5.0}_{-5.0}$   \\
\            &            & 0.2\%                   &            &               &            &   126.0$^{+54.0}_{-54.0}$  & 59.0$^{+6.0}_{-7.0}$   \\
             &            &                         &            &               &            &                            &                        \\
\ ARGO       & B/B        & 52$'$                   & 150        & 0.28          & 10\%       &  95.0$^{78.0}_{-44.0}$  & 39.1$^{+8.7}_{-8.7}$   \\
\            & 35~km      & (65)                    & 250        &               & Internal   &  95.0$^{78.0}_{-44.0}$  & 46.8$^{+9.5}_{-12.1}$  \\
\            &            &                         & 375        &               &            &  95.0$^{78.0}_{-44.0}$  & 46.8$^{+9.5}_{-12.1}$  \\
\            &            &                         & 600        &               &            &  95.0$^{78.0}_{-44.0}$  & 46.8$^{+9.5}_{-12.1}$  \\
             &            &                         &            &               &            &                         &                \\
\ Python     & G/R-B      & 45$'$-1$^\circ$         & 40         & 11\tnote{b}   & 15-20\%    &  88.0$^{+17.0}_{-39.0}$ & 60.0$^{+15.0}_{-13.0}$ \\
\            & 2800~m     & 0.3\%                   &    90      & 0.5\tnote{c}  & Internal   &  170.0$^{+69.0}_{-50.0}$& 66.0$^{+17.0}_{-16.0}$ \\
             &            &                         &            &               &            &                         &                \\
\ Python V   & G/R        &  $\sim 1^\circ$         & 37-45      &               &  12-15\%   & 44.0$^{+25.0}_{-15.0}$  & 22.0$^{+4.0}_{-5.0}$   \\
             & 2800~m     &  1.44\%                  &            &               &  Internal  & 75.0$^{+15.0}_{-15.0}$  & 24.0$^{+6.0}_{-7.0}$   \\
             &            &                         &            &               &            & 106.0$^{+15.0}_{-15.0}$ & 34.0$^{+7.0}_{-9.0}$   \\
             &            &                         &            &               &            & 137.0$^{+15.0}_{-15.0}$ & 50.0$^{+9.0}_{-23.0}$  \\
             &            &                         &            &               &            & 168.0$^{+15.0}_{-15.0}$ & 61.0$^{+13.0}_{-17.0}$ \\
             &            &                         &            &               &            & 199.0$^{+15.0}_{-15.0}$ & 77.0$^{+20.0}_{-28.0}$ \\
             &            &                         &            &               &            &                         &                \\
\ BAM        & B/B        & 42$'$                   & 110-250    & 0.26          & 20\%       &  74.0$^{+82.0}_{-47.0}$ & 55.6$^{+29.6}_{-15.2}$ \\
             & 41.4~km    & (10)                    &            &               & Jupiter    &                         &                \\
             &            &                         &            &               &            &                         &                \\
\ TOCO97     & G/R        & 42$'$\tnote{c}          & 31         & 36 \tnote{d}  & 5\%        & 63.0$^{+18.0}_{-18.0}$    & 40.0$^{+10.0}_{-9.0}$  \\
\            & 5240~m     &                         & 42         &               & Jupiter    &  86.0$^{+16.0}_{-22.0}$   & 45.0$^{+7.0}_{-6.0}$   \\
\            &            & 15$'$                   & 144        & 4             &            & 114.0$^{+20.0}_{-24.0}$   & 70.0$^{+6.0}_{-6.0}$   \\
\            &            & 1.45\%                  &            &               &            & 158.0$^{+22.0}_{-23.0}$   & 89.0$^{+7.0}_{-7.0}$   \\
\            &            &                         &            &               &            & 199.0$^{+38.0}_{-29.0}$   & 85.0$^{+8.0}_{-8.0}$   \\
             &            &                         &            &               &            &                           &                        \\
\ TOCO98     &         &                         &            &               &            & 128.0$^{+26.0}_{-33.0}$   & 55.0$^{+18.0}_{-17.0}$ \\
\            &            &                         &            &               &            &  152.0$^{+26.0}_{-38.0}$  & 82.0$^{+11.0}_{-11.0}$ \\
\            &            &                         &            &               &            &  226.0$^{+37.0}_{-56.0}$  & 83.0$^{+7.0}_{-8.0}$   \\
\            &            &                         &            &               &            &  306$^{+44.0}_{-59.0}$    & 70.0$^{+10.0}_{-11.0}$ \\
             &            &                         &            &               &            &                           &                        \\
\ MAX        & B/B        & 30$'$                   & 105        & 0.28 \tnote{e}& 10\%        &139.0$^{+108.0}_{-67.0}$ & 49.4$^{+7.8}_{-7.8}$   \\
             & 36~km      & (5)                     & 180        & 0.085         & Venus      &                         &                \\
             &            &                         & 270        &               & Jupiter    &                         &                \\
             &            &                         & 360        &               & Internal   &                         &                \\
             &            &                         &            &               &            &                         &                \\
\hline
\end{tabular*}
\begin{tablenotes}
    \small
    \item[a] Number in parenthesis indicates number of pixels.
    \item[b] 11~K for 40~GHz radiometer, 0.5~K for 90~GHz bolometer
    \item[c] 42$'$ at Ka and Q bands, 15$'$ at D band.
    \item[d] Front-end HEMTS at 36~K, SIS mixers at 4~K.
    \item[e] 0.28~K first and second flight, 0.085~K third and fourth flight.
\end{tablenotes}
\end{threeparttable}
\end{center}

\begin{center}
\begin{threeparttable}
\small
\caption{Summary of CMB experiments and detection reported in
Fig.~\ref{fig:summary_plot_a}. Continued}\label{tab:results_3}
\begin{tabular*}{13.6cm}{lccccccrr}
\hline
             &            &                         &            &               &            &                           &                        \\
\ Name       & Type/Tech. & $\theta$,$f_{\rm sky}$  & $\nu$      & $T_{\rm det}$ & Calibr.    &  $\ell\pm\Delta \ell$      & $\delta T_\ell$        \\
\            &            &                         & [GHz]      & [K]           &            &                            & [$\mu$K]               \\
             &            &                         &            &               &            &                           &                        \\
\hline
             &            &                         &            &               &            &                           &                        \\
\ CAT        & G/R-I      & 30$'$                   & 13-17      & 17            & 10\%        & 397.0$^{+84.0}_{-65.0}$ & 50.8$^{+15.4}_{-15.4}$ \\
             & Sea-level  & 0.02\%                  &            &               & Cas-A         & 615.0$^{+102.0}_{-72.0}$ &49.0$^{+19.1}_{-13.6}$ \\
             &            &                         &            &               &            & 391.0$^{+84.0}_{-65.0}$ & 54.0$^{+9.5}_{-6.4}$   \\
             &            &                         &            &               &            & 615.0$^{+102.0}_{-72.0}$ &43.6$^{+13.6}_{-13.1}$ \\
             &            &                         &            &               &            &                         &                \\
\ MSAM       & B/B        & 28$'$                   & 170        & 0.24          & 11\%       & 84.0$^{+46.0}_{-45.0}$    & 35.0$^{+15.0}_{-11.0}$ \\
\            & 39.5~km       & 0.02\%                  & 180        &               & Jupiter    & 201.0$^{+82.0}_{-70.0}$   & 49.0$^{+10.0}_{-8.0}$  \\
\            &            &                         & 415        &               &            & 407.0$^{+46.0}_{-123.0}$  & 47.0$^{+7.0}_{-6.0}$   \\
\            &            &                         & 678        &               &            &                           &                        \\
             &            &                         &            &               &            &                           &                        \\
\ DASI       & G/R-I      & 20$'$                   & 26-36      & 20            & 7\%       & 118.0$^{+49.0}_{-14.0}$   & 61.4$^{+6.3}_{-7.1}$   \\
\            & 2800~m     & 0.9\%                   &            &               & Internal       &   203.0$^{+52.0}_{-30.0}$ & 72.7$^{+4.4}_{-3.9}$   \\
\            &            &                         &            &               &            &  289.0$^{+53.0}_{-28.0}$  & 60.5$^{+2.7}_{-2.9}$   \\
\            &            &                         &            &               &            &  377.0$^{+41.0}_{-35.0}$  & 40.6$^{+2.4}_{-2.5}$   \\
\            &            &                         &            &               &            &  465.0$^{+35.0}_{-47.0}$  & 43.4$^{+2.5}_{-2.6}$   \\
\            &            &                         &            &               &            &  553.0$^{+41.0}_{-47.0}$  & 53.3$^{+2.6}_{-2.8}$   \\
\            &            &                         &            &               &            &   641.0$^{+35.0}_{-41.0}$ & 40.8$^{+3.2}_{-3.6}$   \\
\            &            &                         &            &               &            &  725.0$^{+32.0}_{-49.0}$  & 44.8$^{+3.8}_{-4.1}$   \\
\            &            &                         &            &               &            &   837.0$^{+27.0}_{-74.0}$ & 48.2$^{+4.4}_{-4.9}$   \\
             &            &                         &            &               &            &                           &                        \\
\ Viper      & G/R        & 16$'$                   & 38         & 20            & 8\%        & 108.0$^{+121.0}_{-78.0}$  & 61.6$^{+31.1}_{-21.3}$ \\
\            & 2800~m     & ...                      & 45         &               & Internal   & 173.0$^{+114.0}_{-101.0}$ & 77.6$^{+26.8}_{-19.1}$ \\
\            &            &                         &            &               &            & 237.0$^{+99.0}_{-111.0}$  & 66.0$^{+24.4}_{-17.2}$ \\
\            &            &                         &            &               &            & 263.0$^{+185.0}_{-113.0}$ & 80.4$^{+18.0}_{-14.2}$ \\
\            &            &                         &            &               &            & 422.0$^{+182.0}_{-131.0}$ & 30.6$^{+13.6}_{-13.2}$ \\
\            &            &                         &            &               &            & 589.0$^{+207.0}_{-141.0}$ & 65.8$^{+25.7}_{-24.9}$ \\
             &            &                         &            &               &            &                           &                        \\
\ Maxima     & B/B        & 10$'$                   & 150        & 0.1           & 4\%       &     77.0$^{+33.0}_{-41.0}$ & 45.0$_{-6.0}^{+7.0}$   \\
\            & 38.4~km       & 0.3\%                & 240        &               & Dipole     &   147.0$^{+38.0}_{-36.0}$  & 54.0$_{-5.0}^{+6.0}$   \\
\            &            &                         & 410        &               &            &   222.0$^{+38.0}_{-74.0}$  & 78.0$_{-6.0}^{+6.0}$   \\
\            &            &                         &            &               &            &   294.0$^{+41.0}_{-33.0}$  & 62.0$_{-5.0}^{+5.0}$   \\
\            &            &                         &            &               &            &   381.0$^{+29.0}_{-45.0}$  & 48.0$_{-5.0}^{+6.0}$   \\
\            &            &                         &            &               &            &   449.0$^{+36.0}_{-38.0}$  & 38.0$_{-4.0}^{+5.0}$   \\
\            &            &                         &            &               &            &   523.0$^{+37.0}_{-37.0}$  & 44.0$_{-5.0}^{+5.0}$   \\
\            &            &                         &            &               &            &   597.0$^{+38.0}_{-36.0}$  & 43.0$_{-6.0}^{+6.0}$   \\
\            &            &                         &            &               &            &   671.0$^{+39.0}_{-35.0}$  & 46.0$_{-6.0}^{+6.0}$   \\
\            &            &                         &            &               &            &   746.0$^{+39.0}_{-35.0}$  & 47.0$_{-8.0}^{+8.0}$   \\
\            &            &                         &            &               &            &   856.0$^{+79.0}_{-70.0}$  & 56.0$^{+7.0}_{-7.0}$   \\
\            &            &                         &            &               &            &   1004.0$^{+81.0}_{-68.0}$ & 33.0$_{-22.0}^{+13.0}$ \\
\            &            &                         &            &               &            &   1147.0$^{+88.0}_{-61.0}$ & 15.0$_{-15.0}^{+29.0}$ \\
             &            &                         &            &               &            &                            &                        \\
\hline
\end{tabular*}
\end{threeparttable}
\end{center}

\begin{center}
\begin{threeparttable}
\small
\caption{Summary of CMB experiments and detection reported in
Fig.~\ref{fig:summary_plot_a}. Continued}\label{tab:results_4}
\begin{tabular*}{13cm}{lccccccrr}
\hline
             &            &                         &            &               &            &                            &                        \\
\ Name       & Type/Tech. & $\theta$,$f_{\rm sky}$  & $\nu$      & $T_{\rm det}$ & Calibr.    &  $\ell\pm\Delta \ell$      & $\delta T_\ell$        \\
\            &            &                         & [GHz]      & [K]           &            &                            & [$\mu$K]               \\
             &            &                         &            &               &            &                            &                        \\
\hline
             &            &                         &            &               &            &                            &                        \\
\ Boomerang  & B/B        & 10$'$                   & 90         & 0.28          & 10\%       &    101.0$^{+24.0}_{-25.0}$ & 59.3$^{+4.5}_{-4.9}$   \\
\            & 29~km      & 1.8\%                   & 150        &               & Dipole     &   151.0$^{+24.0}_{-25.0}$  & 68.5$^{+3.9}_{-4.2}$   \\
\            &            &                         & 240        &               &            &   201.0$^{+24.0}_{-25.0}$  & 74.7$^{+3.6}_{-3.7}$   \\
\            &            &                         & 410        &               &            &   251.0$^{+24.0}_{-25.0}$  & 75.5$^{+3.1}_{-3.3}$   \\
\            &            &                         &            &               &            &   301.0$^{+24.0}_{-25.0}$  & 62.3$^{+2.5}_{-2.5}$   \\
\            &            &                         &            &               &            &   351.0$^{+24.0}_{-25.0}$  & 50.9$^{+2.0}_{-2.1}$   \\
\            &            &                         &            &               &            &   401.0$^{+24.0}_{-25.0}$  & 42.9$^{+1.7}_{-1.8}$   \\
\            &            &                         &            &               &            &   451.0$^{+24.0}_{-25.0}$  & 45.5$^{+1.7}_{-1.8}$   \\
\            &            &                         &            &               &            &   501.0$^{+24.0}_{-25.0}$  & 47.6$^{+1.8}_{-1.9}$   \\
\            &            &                         &            &               &            &   551.0$^{+24.0}_{-25.0}$  & 47.8$^{+2.1}_{-1.9}$   \\
\            &            &                         &            &               &            &   601.0$^{+24.0}_{-25.0}$  & 45.3$^{+2.0}_{-2.0}$   \\
\            &            &                         &            &               &            &   651.0$^{+24.0}_{-25.0}$  & 44.0$^{+2.1}_{-2.3}$   \\
\            &            &                         &            &               &            &   701.0$^{+24.0}_{-25.0}$  & 42.7$^{+2.4}_{-2.5}$   \\
\            &            &                         &            &               &            &   751.0$^{+24.0}_{-25.0}$  & 37.9$^{+2.9}_{-3.1}$   \\
\            &            &                         &            &               &            &   801.0$^{+24.0}_{-25.0}$  & 43.8$^{+3.1}_{-3.4}$   \\
\            &            &                         &            &               &            &   851.0$^{+24.0}_{-25.0}$  & 47.3$^{+3.7}_{-3.9}$   \\
\            &            &                         &            &               &            &   901.0$^{+24.0}_{-25.0}$  & 41.8$^{+4.9}_{-5.4}$   \\
\            &            &                         &            &               &            &   951.0$^{+24.0}_{-25.0}$  & 31.3$^{+7.3}_{-9.4}$   \\
\            &            &                         &            &               &            &   1001.0$^{+24.0}_{-25.0}$ & 22.4$^{11.4}_{-22.4}$  \\
             &            &                         &            &               &            &                            &                        \\
\ CBI        & G/R-I      & 5$'$                    & 26-36      & 6             & 5\%        & 603$^{+180.0}_{-166}$     &  58.7$_{-6.3}^{+7.7}$  \\
\            & 5080~m     & 0.1\%                   &            &               & Jupiter    &  1190$^{+261.0}_{-224}$   &   29.7$_{-4.2}^{+4.8}$ \\
             &            &                         &            &               &  Tau-A     &                           &                        \\
\ CBI-2      & G/R-I      & 5$'$                    & 26-36      & 6             & 5\%        & 200$^{+100.0}_{-200.0}$   & 72.4$^{+13.7}_{-16.9}$ \\
\            & 5080~m     & 0.1\%                   &            &               & Jupiter    &   407$^{+93.0}_{-107.0}$  & 44.7$^{+5.0}_{-5.7}$   \\
\            &            &                         &            &               &  Tau-A     &  605$^{+95.0}_{-105.0}$   & 45.5$^{+3.9}_{-4.4}$   \\
\            &            &                         &            &               &            &   801$^{+99.0}_{-101.0}$  & 50.3$^{+3.8}_{-4.1}$   \\
\            &            &                         &            &               &            &  1002$^{+98.0}_{-102.0}$  & 29.3$^{+3.9}_{-4.4}$   \\
\            &            &                         &            &               &            &  1197$^{+103.0}_{-97.0}$  & 35.4$^{+3.8}_{-4.2}$   \\
\            &            &                         &            &               &            &  1395$^{+105.0}_{-95.0}$  & 21.6$^{+5.4}_{-7.4}$   \\
\            &            &                         &            &               &            &  1597$^{+103.0}_{-97.0}$  & 26.7$^{+5.5}_{-6.9}$   \\
             &            &                         &            &               &            &                           &                        \\
\ OVRO       & G/R        & 22$'$-7.4$'$\tnote{a}   & 14.5       & ...            & 4.3\%        & 537.0$^{+267.0}_{-205.0}$ & 56.0$^{+8.5}_{-6.6}$ \\
             & 1200~m     &                         & 31.7       &               & Jupiter      &                         &              \\
             &            &                         &            &               & Cas-A           &                         &               \\
\hline
\end{tabular*}
\begin{tablenotes}
    \small
    \item[a] Depending on frequency.
\end{tablenotes}
\end{threeparttable}
\end{center}

\newpage

\begin{figure}[here]
\begin{center}
\resizebox{12cm}{!} {\includegraphics{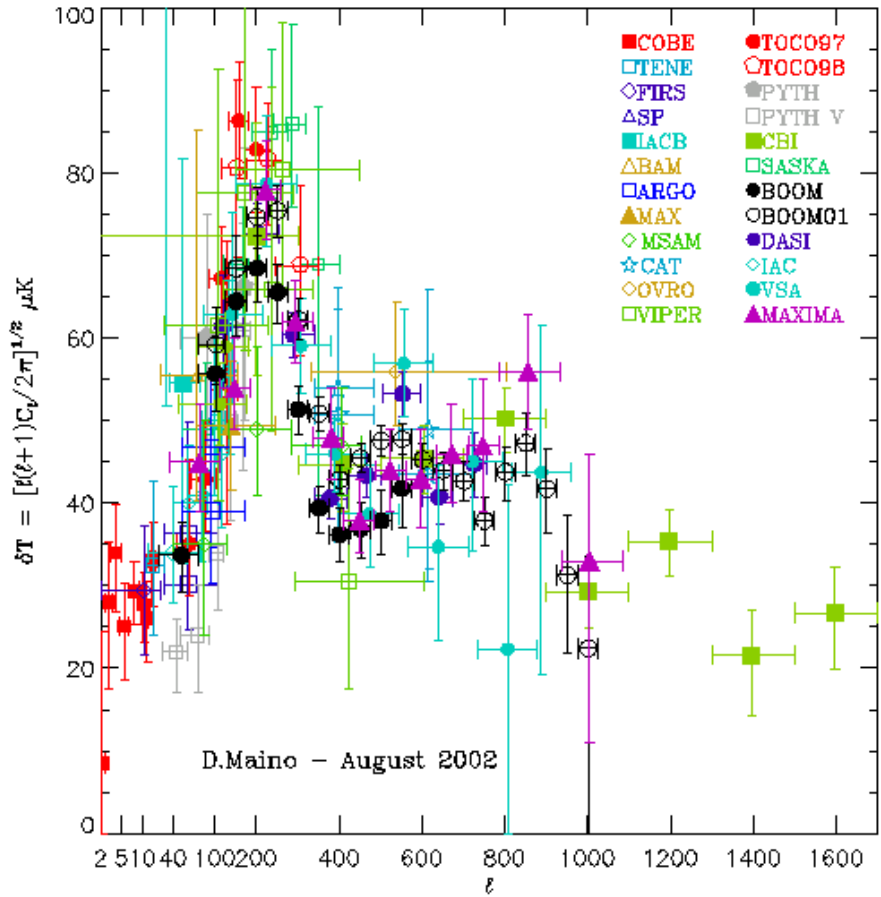}}
\end{center}
\caption{The full set of CMB anisotropy measurements, as described in Sect.~\ref{experiments}.}
\label{fig:summary_plot_a}
\end{figure}

All the power spectrum measurements are plotted
in Fig.~\ref{fig:summary_plot_a}. To obtain a clearer view of the observational status and
to check for self-consistency of the detections, we have
combined different results in a limited number of $\Delta \ell$ intervals.
We
considered separately ground-based and balloon-borne results,
which are characterised by widely different potential systematic errors.
In fact, in addition to the different observing environments,
these two categories typically represent independent instrumental approaches and
contaminated by different astrophysical foregrounds.
As it emerges from the discussion of the various experiments in the previous
sections, almost all balloon experiments (with the exception of QMAP) exploited,
in the relatively short flight-time available, the
excellent sensitivity of bolometric detectors at sub-mm wavelengths;
while most ground-based instruments (besides IAC, IAC-B and the 90~GHz
channel of Python) employed HEMT-based receivers, which are
best suited for ground observations in the low-frequency
($<100$ GHz) atmospheric windows (see Sections~\ref{subsec:atmosphere}
and~\ref{subsec:technology}).
As a consequence of the frequency split,
different foreground components (see Section~\ref{astrolimit}) dominate
the two subsets of measurements.

To combine the subsets of data, we followed
the technique described in \cite{wang00}.
Let $\mathbf{y}$ be a vector of CMB detections (expressed in terms of $\delta T^2$). The
assumed data model is
\begin{equation}
\mathbf{y} = \mathbf{Wx} + \mathbf{n}\, ,
\end{equation}
where $\mathbf{x}$ is the ``true'' power spectrum up to a fiducial high multipole $\ell_{\rm max}$
(we assume here $\ell_{\rm max} = 2000$), $\mathbf{W}$ is the window function for each
detection. The instrument error
is assumed to be a random variable with zero mean ($\langle \mathbf{n} \rangle = 0$) with a
variance given by $\mathbf{N} \equiv \langle \mathbf{nn}^t \rangle$, that is the sum of four
different sources of error:
\begin{equation}
\mathbf{N} = \mathbf{N}^{\rm meas} + \mathbf{N}^{\rm scal} + \mathbf{N}^{\rm ical} + \mathbf{N}^{\rm beam}\, ,
\end{equation}
where the first term is the quoted error in the measured $\delta T^2$, the other terms account for the
uncertainty in the calibration source ($\mathbf{N}^{\rm scal}$), for the calibration uncertainty specific
of each single experiment ($\mathbf{N}^{\rm ical}$) and for the beam uncertainty.
In general all these errors will depend upon the actual power spectrum $\mathbf{x}$. We however make
the approximation that relative errors are small and therefore in this limit the matrix $\mathbf{N}$ is
independent of $\mathbf{x}$.
If $\mathbf{W}$, $\mathbf{N}$ and $\mathbf{y}$ are known,
it is straightforward to invert the problem
and obtain the true power spectrum. Note that, from a mathematical point of view,
this is identical to the map-making problem (see Sect.~\ref{subsec:mmaking}).
In this case, however, the dimensions of the problem are such that it can be easily treated numerically.
Therefore the best, minimum-variance estimator $\mathbf{\hat{x}}$ of the underlying power spectrum
$\mathbf{x}$ is given by:
\begin{equation}
\mathbf{\hat{x}} \equiv [\mathbf{W}^t \mathbf{N}^{-1} \mathbf{W}]^{-1} \mathbf{W}^t \mathbf{N}^{-1}
\mathbf{y}\, .
\end{equation}

For the map-making problem, this solution is known to be minimum variance and un-biased.
Instead of considering the true power spectrum $\mathbf{x}$ as an independent parameter
in each multipole, we treated the power as a piecewise constant simply parameterised by its height
$x_i$ in a suitable number of bands in $\ell$-space.
Furthermore, since our band-power estimator $\mathbf{\hat{x}}$ is a linear combination of the
input data, it is possible to obtain the window function relative to each band-power taking the
same linear combination of the input matrix $\mathbf{W}$. From these band-power window functions
we derived the horizontal bars and the mean point taken to be respectively the 20, 80 and
50\% quantile, respectively, of the absolute value of the band-power window function.

\begin{figure}[here]
\begin{center}
\resizebox{12cm}{!} {\includegraphics{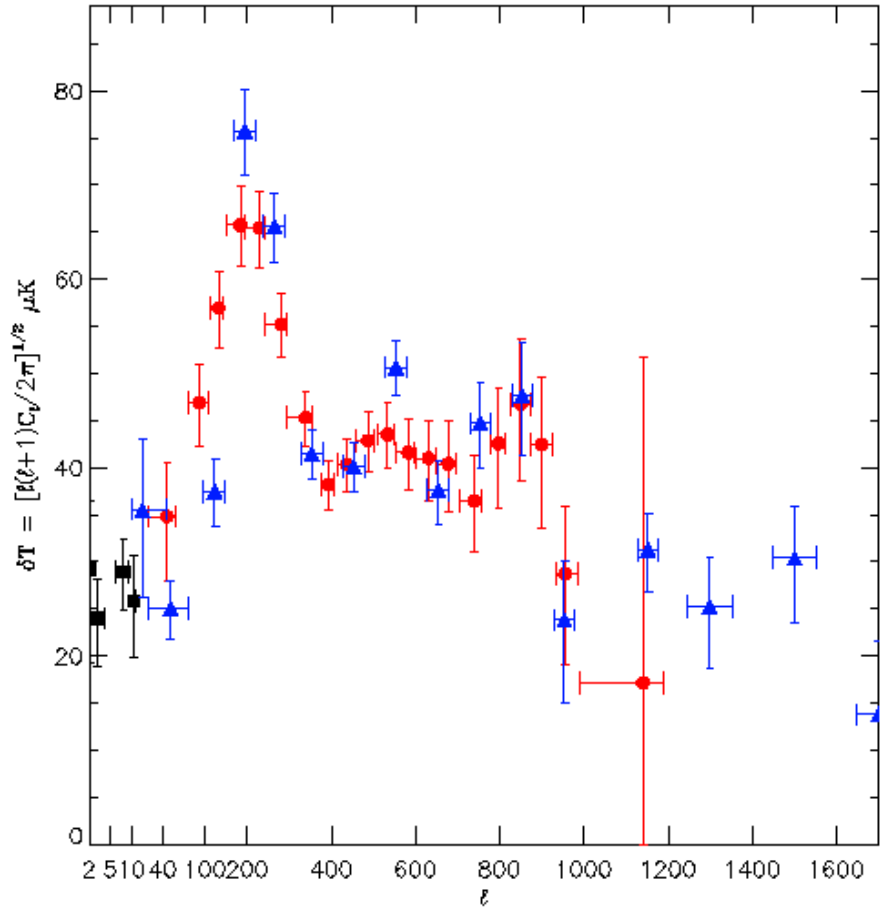}}
\end{center}
\caption{Anisotropy detections combined in a limited number of intervals.
Black data points: $COBE$-DMR results;
Red data points: balloon-borne (mostly bolometers, $\nu >100$\,GHz) experiments;
Blue data points: ground-based (mostly radiometers, $\nu <100$\,GHz) experiments.
Ground-based data extends up to higher $\ell$'s thanks to the recent
results from the CBI experiment.}
\label{fig:summary_plot_b}
\end{figure}

Fig.~\ref{fig:summary_plot_b} shows the results for the CMB
detections divided in ground-based, balloon-borne and space
($COBE$-DMR) experiments. Both ground and balloon results show,
independently, clear evidence of the acoustic pattern of the
power spectrum and in particular of the first two peaks, located
approximately at $\ell \sim 200$ and $\ell \sim 500$,
respectively. A third peak near $\ell \sim 800$ is also suggested
by both data sets. This general agreement is a noteworthy result,
and provides strong indication that the systematic effects and
residual foreground errors that affect the current set of
observations, while representing one of the main sources of
uncertainty, do not dominate the basic features observed in the
power spectrum. On the other hand, it is also clear that much
room exists to improve the current precision in measurements of
the angular power spectrum.

\section{ONGOING AND PLANNED EXPERIMENTS}
\label{subsec:ongoing}

Further observing campaigns are now on-going for several of the
programs described in Sect.~\ref{experiments}, such as Boomerang,
Maxima, CBI, DASI. In addition a number of new ambitious
sub-orbital projects are currently being planned for the near
future, either as continuation of existing collaborations or as
new enterprises. The forthcoming results of the $MAP$ mission, to
be released within one year, are expected to over-ride many of
the current anisotropy results, and will certainly make it much
harder for traditional ground-based and balloon-borne experiments
to be competitive. The {\sc Planck} maps, expected in the second
half of this decade, will definitively cover the whole sky with
exquisite precision and spectral coverage, pushing out of date
most of the current experimental approaches. On the other hand,
we do not believe that {\sc Planck} will represent an end for CMB
observations, but rather the starting point of a new phase (see
Sect.~\ref{future}). In fact, some of the experiments currently
being designed include features that may anticipate future
developments; in particular increasing emphasis is given to very
high resolution and to polarisation capabilities.

\subsection{Temperature anisotropy projects}
\label{subsubsec:temperature_anisotropy_projects}


Two projects in the high ($>100$~GHz) and low ($<100$~GHz)
frequency regimes, Archeops and BEAST, are being carried out as
precursors of the {\sc Planck} LFI and HFI instruments,
respectively. Archeops is a balloon-borne bolometric instrument
with angular resolution $\sim 8'$, eventually intended to cover
$\sim 25\%$ of the sky at 143, 217, 353 and 545~GHz. The
bolometers cooled to 0.1~K can provide sensitivity to CMB
fluctuations of 100~$\mu$K in as many as $9 \times 10^4$ pixels
with FWHM$\sim 20'$. Results from a test flight from Trapani
(Sicily) to Spain in July 1999 are described by \cite{benoit02},
while data from a second flight are now being analysed. The BEAST
instrument is operated in its present configuration in Ka and Q
band, using LFI-like focal plane and components \cite{natoli01}.
The program, carried out by a collaboration involving UCSB, JPL,
Italian groups in Milano, Bologna and Roma and INPE-Brazil, has
been flown twice and, in addition,
now large amounts of data covering $\sim 3000$ deg$^2$ 
with $20'$ have been collected from the White
Mountain High Altitude Station, California (3800~m) and the
analysis is being completed as we write. Future plans include the
addition of a W band polarisation channel and to fly the
experiment as a LDB balloon. Eventually it may be possible to fly
the payload as a ultra long duration ($>100$ days) flight, a
possibility technically feasible.


When approaching excellent instrument sensitivities, the main
source of concern tends to become the control of systematic
errors rather than sheer sensitivity. This is a crucial issue in
view of the future development of precision CMB imaging. For
example, to create an observing environment maximally free from
local stray radiation a novel instrument concept was proposed for
the TopHat experiment \cite{kowitt95}, a bolometer-based long
duration balloon project in which the instrument is located on
top of the balloon rather than hanging below it. With this
arrangement, the entire sky above the instrument is free from
supporting structures which could scatter radiation into the
sidelobes of the optics.


Most ground based projects currently active or planned are based
on interferometer instruments. Among them is the Very Small Array
(VSA), now taking data from Tenerife\footnote{Very recently,
early results from the VSA have been presented \cite{vsa1}
showing evidence of the first two acoustic peaks in good
agreement with Boomerang, Maxima and DASI. These VSA results are
included in the data sets of Fig.~\ref{fig:summary_plot_a} and
Fig.~\ref{fig:summary_plot_b}}. The project is a collaboration
between MRAO, NRAL and IAC. The VSA design includes an array of
14 horn-reflector antennas in a T-shaped configuration, and it
represents a second-generation after the CAT interferometer. The
VSA aims at obtaining detailed maps with angular sensitivity
between $10'$ and $2^\circ$ with $\sim 5$~$\mu$K noise in 4
frequency bands in the 28--37~GHz range. Another interferometer
experiment called AMiBA (Array for Microwave Background
Anisotropy), is expected to be operative by 2003
\cite{kesteven02, lo00}. This will be a 19 elements, W-band array
by the Academia Sinica, Institute of Astronomy and Astrophysics,
and National University of Taiwan, devoted to CMB polarisation
and S-Z effect observation.


Antarctic winter observations may also have a long future. The
Arcminute Cosmology Bolometer Array Receiver (ACBAR) is a 16
element array of bolometers cooled at 0.23~K designed for
observation of small scale anisotropy and Sunyaev-Zeldovich
effect. The experiment, run by the Center for Astrophysical
Research in Antarctica (CARA), was deployed at the South Pole in
December 2000. Preliminary results in the range $200 < \ell <
2200$ based on data taken in the austral winter of 2001 were
presented \cite{kuo02} including serendipitous discoveries of
high redshift clusters via their S-Z effect.


To gain the angular resolution needed to probe fine structures,
existing radio-telescopes may be employed if properly adapted. It
has been proposed \cite{parijskij97} to use the Ratan-600, the
large Russian radio telescope, coupled with high sensitivity
receivers for CMB anisotropy and polarisation measurements. The
so-called ``Cosmological Gene'' main frequency would be 32~GHz,
with a set of frequencies in the 1-22~GHz range for foreground
subtraction. Very high angular resolution ($\sim 1'$) is in
principle obtainable provided that the optics quality and
atmospheric conditions are adequate.

\subsection{Polarisation anisotropy projects}
 \label{subsubsec:polarisation_anisotropy}

An increasing number of projects are designed to attempt
polarisation detection. Early work \cite{caderni78, nanos79,
lubin81, partridge88} has confirmed the low level of polarisation
expected in the CMB signal. Despite the increasing experimental
effort \cite{wollack97, netter97, hedman01, subrahmanyan01,
keating01}, up to now only upper limits have been obtained,
constraining CMB polarisation at the 10~$\mu$K 
level\footnote{A claim of detection of polarisation anisotropy
using the DASI interferometer has appeared \cite{kovac02} 
just before sending this manuscript for printing}. A complete
discussion of CMB polarisation experiments is out of the scope of
this work. However, it is clear that the trend for most CMB
anisotropy projects is to include polarisation capability.


At large angular scales ($\ell \leq 200$) theory predicts a CMB
polarised component as weak as sub-$\mu$K level (see
Fig.~\ref{pola}), far lower than the sensitivities currently
achieved even by the most advanced instruments. An attempt to
reach a statistical detection at large scales was carried out
with the POLAR correlation radiometer \cite{keating01} using
HEMTs in Ka band cooled to 15~K. Sensitivities of few $\mu$K in
$\sim 30$ pixels could be obtained in several months of observing
time. Analysis of the $\sim 750$ hours of data lead to upper
limits $\sim 10$~$\mu$K. A space experiment to search for
polarisation at $>7^\circ$ scales is the SPORT project
\cite{cortiglioni00, carretti01a, carretti02} proposed for
implementation on the International Space Station. The instrument
is based on HEMT radiometers cooled to $\sim 90$~K and operating
at 4 frequencies in the 20-90~GHz range. A nearly full sky map
can be obtained with a sensitivity of $\sim 5$~$\mu$K per
$7^\circ$ resolution element, which is adequate for a valuable
measurement of the Galactic polarised emission. The balloon-borne
counterpart of SPORT, called BarSport, has a multi-feed focal
plane of polarimeters in W band at the focus of an on-axis
telescope reaching sub-degree angular resolution in small sky
patches \cite{zannoni02}.


A number of ground-based and balloon borne experiments exploiting
the novel technique of polarisation sensitive bolometers (PSBs; see
Sect~\ref{subsec:technology}) are being planned. A new
circum-Antarctic LDF flight (B2K) is being prepared for
Boomerang, with the aim of detecting CMB polarisation
\cite{masi02} using PSBs at frequencies 150, 245 and 345~GHz. In
addition to polarisation capability, the instrument has improved
attitude reconstruction (a main limiting factor in the first
Boomerang flight) and calibration (goal precision better than
1\%). QUEST (Q and U Extragalactic Sub-mm Telescope) is another
CMB polarimetric instrument being designed and built at Cardiff
and Caltech \cite{piccirillo02}. The QUEST on-axis Cassegrain
optics will produce $1.5^\circ$ images with $\sim 5'$ resolution
with multi-element arrays of polarisation sensitive bolometers at
100, 150, and 220~GHz. A ground-based bolometric receiver with
polarisation capability, Polatron, is being developed for use at
the Owens Valley 5.5 meter radio telescope. The scanning strategy
and experimental design will be similar to the SuZIE experiment,
while the single-mode feed structure and spider-web bolometers
will be similar to those of Boomerang. The experiment is designed
to run autonomously to cover long integration times. For this
reason, the cryogenic chain  (with a 4~K stage provided by a
mechanical cryocooler, and a 0.25~K stage by a multi-stage
sorption cooler) will consume no liquid cryogens.

\subsection{Towards precision CMB imaging}
\label{subsec:space_missions}

Soon after the $COBE$-DMR detection \cite{smoot92} it become clear
that a reasonable extrapolation into the future of
millimetre-wave technology could support high-sensitivity,
sub-degree surveys of large portions of the sky. At the same
time, the unique scientific potential of extended high resolution
observations became widely recognised. While several ground and
balloon experiments were being prepared, a few groups proposed
space projects to fully accomplish the pioneering work of $COBE$
(e.g. \cite{mandolesi94, bersanelli95, bersanelli96, janssen95,
danese96, lange97, tauber97}). After detailed assessment studies
and competitive peer reviews, in 1996 two CMB missions were
selected for implementation: NASA's Microwave Anisotropy Probe
($MAP$) and ESA's {\sc Planck} Surveyor (formerly known as
COBRAS/SAMBA). The $MAP$ mission was launched in June 2001, it is
currently observing and the first data are expected in early 2003.
$MAP$ will provide a major step forward in the power spectrum
determination with respect to all the present sub-orbital
experiments. The {\sc Planck} mission is designed to {\it fully}
extract the cosmological information contained in CMB anisotropy
by setting angular resolution, spectral coverage and sensitivity
such that the power spectrum reconstruction will be limited by
unavoidable cosmic variance and astrophysical foregrounds. Both
missions are also sensitive to linear polarisation at
proportional levels. Table~\ref{tab:MAP_PLANCK_comparison}
summarises the main performances of $MAP$ and {\sc Planck}.

\begin{center}
\begin{threeparttable}
\caption{Main instrumental performances of the MAP and {\sc
Planck} missions.}\label{tab:MAP_PLANCK_comparison}
\begin{tabular*}{12.35cm}{|l|l|c|c|c|}
\cline{3-5} \cline{3-5}
        \multicolumn{2}{c|}{ }
&           $MAP$        &   \multicolumn{2}{c|}{{\sc Planck}}     \\
\cline{4-5}
        \multicolumn{2}{c|}{ }
&                      &   {\it LFI}           &  {\it HFI}      \\
\hline \hline

\multicolumn{2}{|l|}{Angular resolution} &     14$'$--56$'$     &     33$'$--10$'$      &   9.2$'$--5$'$       \\
\hline {Avg. $\Delta T /T$ per pixel}\tnote{a} &
Intensity          & 4.0
$\times 10^{-5}$  & 6.6$\times 10^{-6}$   &  2.0$\times 10^{-6}$ \\
\cline{2-5} {(1 year mission)}\tnote{b}             &
Polaris.\tnote{c}
&    5.6$\times 10^{-5}$                  & 9.3$\times 10^{-6}$   & 4.2$\times 10^{-6}$  \\
\hline \multicolumn{2}{|l|}{Baseline mission
lifetime}              &  2
years\tnote{d}    & 14 months             & 14 months            \\
\hline \multicolumn{2}{|l|}{Spectral
coverage}                      &  22--90
GHz          & 30--100 GHz           & 100--857 GHz         \\
\hline \multicolumn{2}{|l|}{Detector
technology}                    &
HEMT                & HEMT                  & Bolometers           \\
\hline \multicolumn{2}{|l|}{Detector
temperature}                   &  $\sim 90$ K        & 20 K                  & 0.1 K                \\
\hline
\multicolumn{2}{|l|}{Cooling}                                &
Passive             & Active                & Active               \\

\hline \hline
\end{tabular*}
\begin{tablenotes}
    \small
    \item[a] The sensitivity in thermodynamic temperature and
             refers to the W-band detectors (94 GHz for MAP and 100 GHz
for
             {\sc Planck}) on a 10$'$ pixel. A pixel is a square whose
side
             is the FWHM extent of the beam.
    \item[b] These figures are calculated for the average integration
time per pixel.
             In general the integration time will be inhomogeneously
distributed
             on the sky and will be much higher in certain regions of
it.
    \item[c] Refers to E-mode polarisation (Stokes parameters U and Q).
    \item[d] Recently the MAP team has requested an extention to 4 years of the
mission lifetime.
\end{tablenotes}
\end{threeparttable}
\end{center}

\begin{figure}[here]
\centerline{
    \psfig{figure=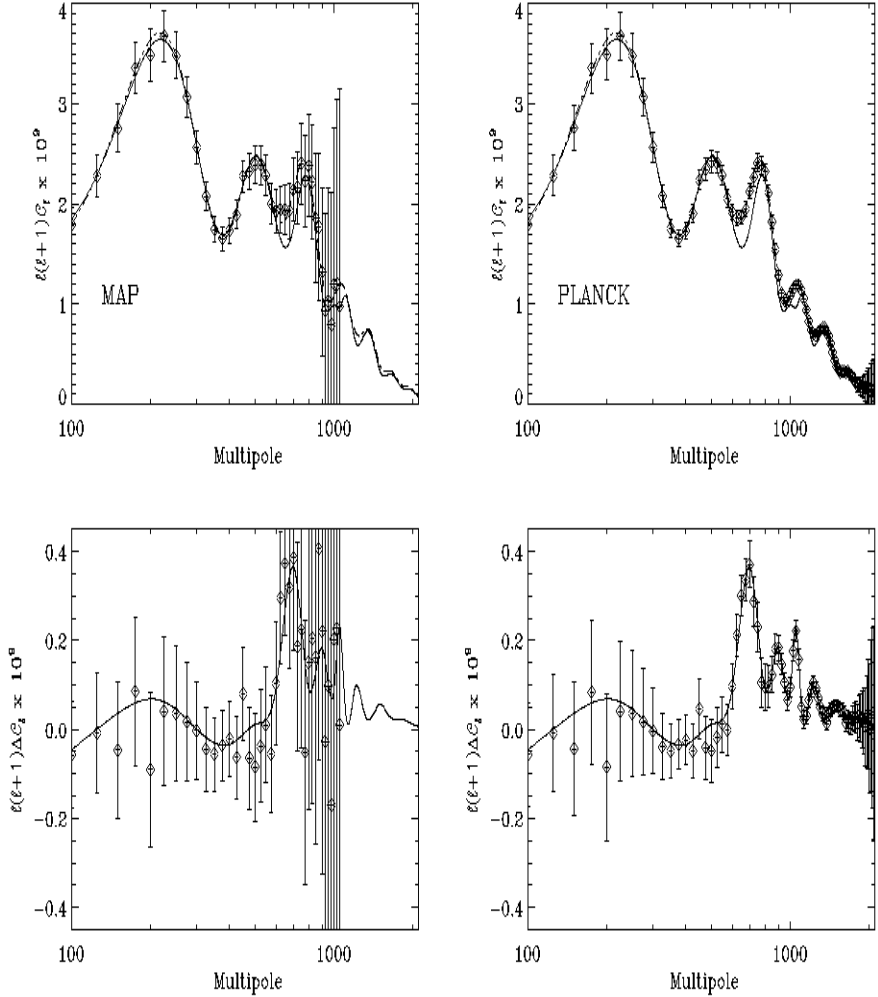,width=12cm,height=14cm}
} \caption{Simulated power spectrum measurements by $MAP$ (left) and {\sc Planck}
(right). The upper panels show two nearly degenerate CDM models
(solid and dotted curves) with significantly different
parameters. Although barionic density, $\Omega_b$, and dark
matter density, $\Omega_{\rm CDM}$, differ by as much as 24\% and
5\% respectively, the two curves are very similar. The residuals
are ploted in the lower panels. The error bars are simulated
power spectra reconstructions according to the $MAP$ and {\sc
Planck} instrument specifications. Current anisotropy data (see
Fig.~\ref{fig:summary_plot_b}) are completely unable to
discriminate the two models. $MAP$ may tentatively distinguish
them, while {\sc Planck} can precisely discriminate between the two
scenarios.} \label{mapplanck}
\end{figure}

The technical challenge posed by these missions is twofold. First,
it is necessary to reach very high sensitivity in order to obtain
the wanted signal to noise ratio per pixel over the full sky.
Given the key requirement of high angular resolution (up to $14'$
for $MAP$ and $5'$ for {\sc Planck}), the requirement on
instrument sensitivity is great for a reasonable mission lifetime,
$\tau_{\rm mission} \approx 1$--4 yrs. From
Eq.~(\ref{eq:sensitivity}) the noise per 1 second integration
time ($\mu$K$\times$Hz$^{-1/2}$), is $\Delta T_{1\,{\rm sec}} =
k_r {T_{\rm sys}+ T_{\rm sky} \over \sqrt{\Delta \nu} }$.
Therefore to get an ultimate noise per pixel $\delta T_{\rm
pix}$, $\Delta T_{1\, {\rm sec}}$ must satisfy $\Delta T_{1\,
{\rm sec}} < \delta T_{\rm pix} \sqrt{n_{\rm R} {\tau_{\rm
mission} \over N_{\rm pix}} }$. This calls for arrays of
detectors ($n_{\rm R} \sim 10$) with state-of-the-art technology
and instruments operating at cryogenic temperatures. The $MAP$
radiometers are operated on a passively cooled payload at
$\sim$90~K, while {\sc Planck} will make full use of cryogenic
technology, with radiometers cooled at 20~K and bolometric
detectors cooled at 0.1~K.

The second challenge is to make sure that the data accuracy is
indeed limited by instrument sensitivity. The more ambitious the
sensitivity goal, the more stringent the required rejection of
systematic effects, which for $MAP$ and {\sc Planck} must be
controlled at $\mu$K level. The space environment, coupled with
an appropriate choice of orbit and scanning scheme, provides a
unique chance to avoid most of the environmental limitation of
sub-orbital experiments such as atmospheric emission, thermal
instability, stray light from the earth, sun and moon, and allows
full sky coverage. Both $MAP$ and {\sc Planck} selected Lissajous
orbits around the Lagrangian point L2 of the sun-earth system at
1.5 million km from earth. This vantage point offers an
exceptionally stable environment for CMB observations (as well as
for other missions, such as Herschel or NGST), since the
instrument can always maintain an unobstructed view of deep
space. The satellites passive thermal design combined with a
constant sun angle results in low thermal drift rates
($\sim$mK/hr) unattainable from any platform near earth.

Both $MAP$ and {\sc Planck} will exploit the CMB dipole as main
calibration source, now known to an accuracy of 20~$\mu$K \cite{bennett96}. In
addition, the seasonal variation of the dipole amplitude due to
satellite motion will further improve absolute calibration. At
frequencies $<300$~GHz photometric calibration better than 0.5\%
can be obtained \cite{bersanelli96}. The highest frequencies of
{\sc Planck} will rely on measurements of the emission of the
Milky Way, calibrated to 1\% by $COBE$/FIRAS (of course {\sc
Planck} will be able to use the $MAP$ data for its own
calibration). Also, both missions will measure the main beams
in-flight using multiple observations of Jupiter, Saturn and Mars.

As shown in Sect.~\ref{experiments}, remarkable progress has been
achieved by sub-orbital experiments since the first proposal of
$MAP$ and {\sc Planck}, as it was anticipated at
that time. However, it is clear that only unobstructed, deep,
full-sky surveys can fully exploit the opportunity for {\it
precision cosmology} offered by CMB observations. In fact, the
present positive detections of acoustic peaks in the power
spectrum strongly supports optimism in our ability to reach
exquisite accuracy in parameter estimation from deep CMB
anisotropy observations. Fig.~\ref{mapplanck}
illustrates the capability of the $MAP$ and {\sc Planck} missions to
accurately reconstruct the power spectrum. In particular, the plots show
how the high percision measurements by {\sc Planck} will allow to
discriminate between nearly degenerate sets of $C_\ell$s.

\subsubsection{\underline{$MAP$}}
 \label{subsubsec:MAP}

The Microwave Anisotropy Probe satellite is expected to
reconstruct the angular power spectrum up to $\ell \sim 1000$,
limited by cosmic variance for $\ell < 500$. The project was
proposed to NASA as a MidEx mission in June 1995, and one year
later hardware manufacturing started based on off-the-shelf
components and space qualified technology. In analogy with
$COBE$-DMR, $MAP$ uses differential radiometers in the 23--94~GHz
range that measure instantaneous temperature differences between
two points in the sky. A pair of back-to-back Gregorian
telescopes (1.4~m $\times$ 1.6~m primary, and 0.9~m $\times$
1.0~m sub-reflector) direct two symmetric beams in the sky
separated by $\sim 141^\circ$. The telescopes are coupled to two
symmetric, 10-element arrays of corrugated feed horns connected
to differential receivers. Each focal plane array includes 1 feed
at 22 and 30~GHz, 2 feeds at 40 and 60~GHz and 4 feeds at 90~GHz.
The feed centres occupy a region $\sim 20$~cm in diameter, which
corresponds to $\sim 4.5^\circ$ projected in the sky. The angular
resolution ranges between $\sim 50'$ and $\sim 14'$, depending on
frequency, with typical ellipticity of the beam shapes $\sim
10\%$. The expected sensitivity is $\sim 35$~$\mu$K per
$0.3^\circ$ pixel at each frequency in a nominal lifetime of two
years.

\begin{figure}[here]
\includegraphics{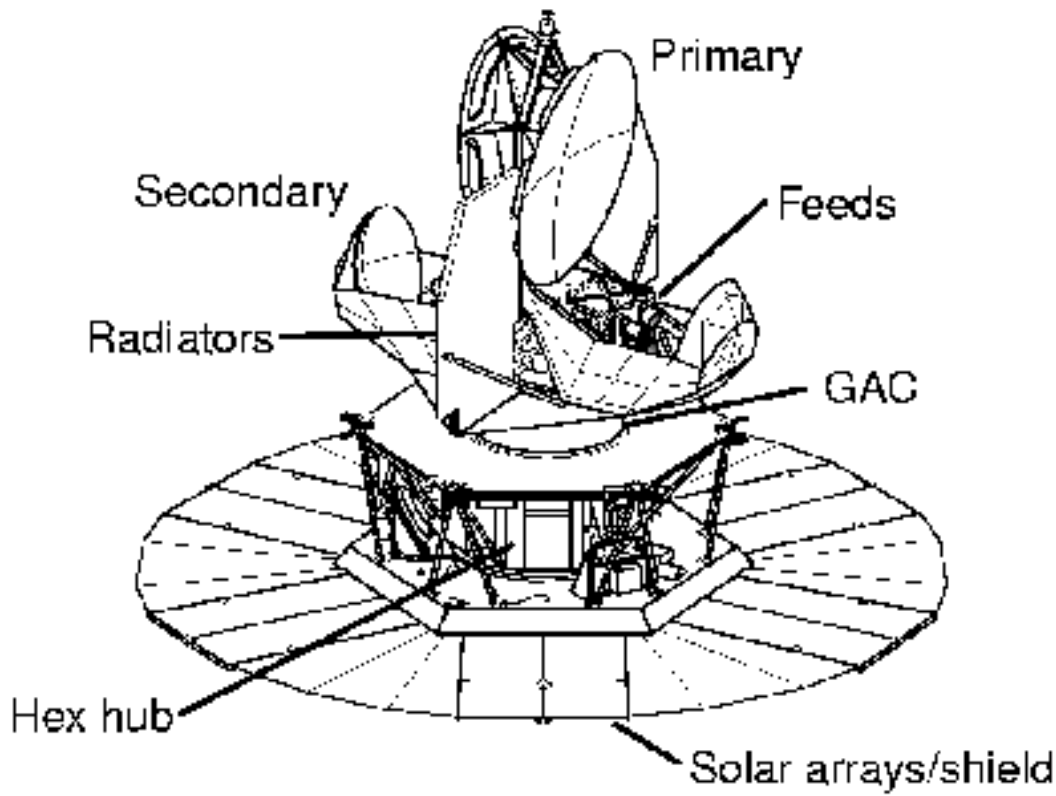}
\caption{An overview of the $MAP$ satellite, showing the
double-telescope system, the large radiators, and the satellite
service module with deployable solar panels on the bottom. One of
the two focal arrays of feed horns is also partially visible on
the right. The overall height is 3.6~m, the mass is 830~Kg and
the diameter of the deployed solar panels reaches 5.1~m. The
solar arrays supply a 400~Watts power to the spacecraft and
instruments. The hexagonal structure (``Hex hub'') above the
solar panel array supports the power supply, instrument
electronics, and other satellite subsystems. The thermally
insulating ``gamma alumina cylinder'' (GAC) supports a 190~K
gradient between the warm and cold portion of the satellite.
(With permission, courtesy of the $MAP$ Science Team).}
\label{fig:MAP}
\end{figure}

Passive cooling to $\sim 90$~K of the optics and front-end portion
of the receivers is ensured by the continuous view of cold space
by two large (5.6~m$^2$) radiators, and by separating the
front-end and back-end sections of the radiometers. No active
cooling is used. Ground tests of the front-end InP HEMT amplifiers
\cite{pospieszalski92, pospieszalski94} achieve noise
temperatures of 25-100~K at 80~K \cite{pospieszalski00}, and are
phase matched over the 17\% nominal bandwidth.

After each feed horn, an orthomode transducer separates the two
linear polarisations which are processed independently by two
differential radiometer chains, thus allowing the system to be
sensitive to the linearly polarised component. The sensitivity to
the $Q$ and $U$ Stokes parameters is degraded by $\sim \sqrt{2}$
with respect to temperature anisotropy. The receiver scheme is a
so-called ``pseudo-correlation'' receiver \cite{jarosik00}, in
which two hybrid tees and two square-law detectors replace the
more traditional multiplier of a typical correlation receiver. A
similar design solution was adopted for the {\sc Planck}-LFI
radiometers, where one of the inputs is constituted by an
internal cryogenic load. The advantage of these solutions is that
they minimise systematic effects due to fluctuations and
instabilities of the gain and noise temperature of HEMT
amplifiers, typically exhibiting 1/$f$ noise spectrum
\cite{seiffert02}. In addition, phase switching (at kHz rates) is
used to reject low frequency noise from the detector diodes and
residual 1/$f$ noise from the back-end.

Large efforts have been made to minimise pixel-to-pixel correlations
and many systematic uncertainties that are avoided by design. Possible artifacts in the map-making
process are controlled at sub-$\mu$K level. Because
of the long time spent in a very stable orbit, and using the observation redundancy
of the scan strategy, numerous consistency and systematic checks are possible.

The spacecraft reached L2 using a trajectory with 4 lunar phasing
loops followed by a transfer cruise. To satisfy communications
requirements and to avoid eclipses, during the Lissajous orbit,
the spacecraft-earth vector remains between $1^\circ$ and
$10^\circ$ off the sun-earth vector. The instrument is
continuously maintained in the shadow for passive cooling and
avoidance of straylight signals. The scan strategy also allows to
maintain a constant angle between the sun and the plane of the
solar panels for thermal and power stability. The satellite spins
at 0.464~rpm ($\sim$2.2~min/spin), and precesses every hour
around a $22.5^\circ$ cone around the sun-satellite line, which
is always within $0.1^\circ$ of the sun-earth line. The scan
strategy was designed to rapidly scan a large fraction of the sky
(30\% every day) and to observe each sky pixel through many
azimuth angles and at different time scales. The spin and
precession periods are out of phase, so that their combined
motion causes the observing beams to fill an annulus centred on
the local solar vector. As $MAP$ orbits the sun, the annular scan
pattern covers the full sky every 6 months, which can be used for
long-term stability checks. Algorithms for data analysis and
processing of $MAP$-like data have been presented by Wright et al
\cite{wright95}, Hinshaw \cite{hinshaw01}, Hivon et al
\cite{hivon02}.

The multifrequency instrument is designed to disentangle
synchrotron and free-free Galactic emission and radio sources. A
recent analysis of simulated $MAP$ data at 90~GHz with
$0.3^\circ$ FWHM \cite{park01} showed that foreground
contamination on the correlation function is small compared to
cosmic variance. However, Galactic emission at $|b| > 20^\circ$,
significantly affects the topology of CMB anisotropy and
non-Gaussianity analysis. It was also showed that IRAS and DIRBE
far-infrared extragalactic sources have little effect on the CMB
anisotropy while radio sources raise the amplitude of the
correlation function considerably on scales $<30'$. $MAP$ will
highly improve over the current measurements of diffuse
synchrotron and free-free emissions and polarisation, and the K
and Ka bands will be able to verify the possible contribution
from spinning dust \cite{draine99}.

In addition to power spectrum measurements, the $MAP$ data can be
used for testing Gaussianity and polarisation, in particular for
polarisation-temperature cross-correlation. The S-Z effect should
be observable in 5 to 10 clusters and possibly also in 
a few bright extragalactic radio sources.

\subsubsection{\underline{{\sc Planck} Surveyor}}
\label{subsubsec:planck}


Following $COBE$-DMR and $MAP$, {\sc Planck} represents the third
generation space mission dedicated to CMB anisotropies. The
mission \cite{bersanelli96a, tauber98} is designed to reach the
fundamental limits imposed by cosmic variance and astrophysical
foregrounds across the entire power spectrum. The observations
will produce 10 full sky maps at frequencies spanning over the CMB
blackbody spectrum in the range 30-850~GHz, with angular
resolution from $33'$ to $5'$, and with a typical sensitivity per
resolution element $\Delta T/T \simeq 2 \times 10^{-6}$. This
unique combination of mapping capabilities will constitute
an unprecedented advancement in the field. Multipoles up to $\ell
\simeq 2500 \div 3000$ can be measured with an accuracy  $\sim
1\%$, surpassing any existing or planned experiment. The {\sc
Planck} instruments are designed to be sensitive to polarisation
as well, thus allowing to break parameters degeneracy at a level
much more powerful than in any previous data sets \cite{bond97,
efstathiou02}. The data will be used to test models for the origin
of primordial perturbations, to constrain the global
properties (topology, rotation, shear, etc.), and to test the gaussianity
of the CMB distribution \cite{martinez00}. {\sc Planck} will
detect the S-Z effect towards thousands of clusters of galaxies
\cite{aghanim97, haehnelt96, dasilva00}, allowing an independent
determination of the Hubble parameter and providing information
on the intercluster medium complementary to those from X-rays.

Though {\sc Planck} is a CMB mission, its multifrequency,
full-sky survey will also represent a major breakthrough for a
number of astrophysical studies. The data will provide complete
samples of thousands of extragalactic sources, selected in a
frequency range essentially unexplored at these sensitivities,
and new classes of sources may be revealed. Moreover, the {\sc
Planck} maps will deliver a rich database for studies  of Galactic
evolution, the interstellar medium, and discrete Galactic
sources. The enormous amount of data envisaged for {\sc Planck}
and complex data processing required \cite{pasian99} represent a
challenging task.


To meet its ambitious objectives, {\sc Planck} employs frontier
technology in millimetre-wave detectors and cryogenic equipment.
A unique feature of {\sc Planck} is its capability of combining
in a single experiment both leading (and complementary)
technologies in CMB observations: coherent receivers and
bolometric detectors. An off-axis shaped aplanatic telescope
\cite{villa98, mandolesi00} with a primary of physical size $1.9
\times 1.5$~m is coupled to two focal plane instruments, combined
in a sophisticated thermo-mechanical cryogenic design. The
frequency range $30-100$~GHz is covered by the Low Frequency
Instrument (LFI), an array of coherent, differential radiometers
based on cryogenic InP HEMT amplifiers and operated at $\sim
20$~K \cite{mandolesi00, mandolesi98, bersanelli00}. To minimise
power dissipation in the focal plane the radiometers are split
into two subassemblies connected by a set of waveguides. The
radiometer design uses a pseudo-correlation scheme to suppress
1/$f$ noise induced by amplifier gain and noise temperature
fluctuations. While the $MAP$ radiometers measure temperature
differences between two widely separated regions of the sky, the
LFI radiometers measure differences between the sky and a stable
internal cryogenic reference load cooled at $\sim$~4~K, taking
advantage of the pre-cooling stage of the bolometric instrument.
In addition, in LFI the effects of the residual input offset
($<$2~K in nominal conditions) is compensated by introducing a
{\it gain modulation factor} which balances the output in the
on-board signal processing \cite{mennella02, bersanelli95}. As
demonstrated by advanced LFI prototype units, this receiver scheme
greatly improves the stability of the measured signal
\cite{meinhold98, tuovinen00}. A detailed analytical study of the
impact of non-idealities in the radiometer on its performances, in
particular 1/$f$ noise effects, has been carried out
\cite{seiffert02}.

Current LFI prototypes establish world-record performances in the
30-100~GHz range for noise, bandwidth and low power consumption.
The amplifiers at 30 and 44~GHz are incorporated into a microwave
integrated circuit (MIC). At these frequencies the parasitics and
uncertainties introduced by the bond wires in a MIC amplifier are
controllable, particularly given the relatively low number of
channels (total of 10), and the additional tuning flexibility
facilitates optimisation for low noise. The LFI amplifiers have
demonstrated noise temperatures $\sim 7.5$~K at 30~GHz with 20\%
bandwidth. At 70 and 100~GHz there will be 12 and 24 channels,
respectively. Amplifiers at these frequencies will use MMICs
(monolothic microwave integrated circuits), which incorporate all
circuit elements and the HEMT transistors on a single InP chip.
The amplifiers can thus be mass-produced in a controlled process,
a real advantage given the large number of amplifers. Cryogenic
MMIC amplifiers have been demonstrated at 75--115~GHz which
exhibit $<35$~K over the required bandwidth. The LFI will thus
fully exploit both MIC and MMIC technologies at their best. The
LFI is ``naturally'' polarisation-sensitive in all of its
channels, with a sensitivity to the Q and U Stokes parameters
$\sim \sqrt{2}$ lower than for temperature anisotropy. In
particular, in the LFI 100~GHz channel the polarised foreground
component is expected to be well below the cosmological polarised
signal, thus providing an ideal window for CMB polarisation
measurements.

\begin{figure}[here]
\resizebox{14.cm}{!} {\includegraphics{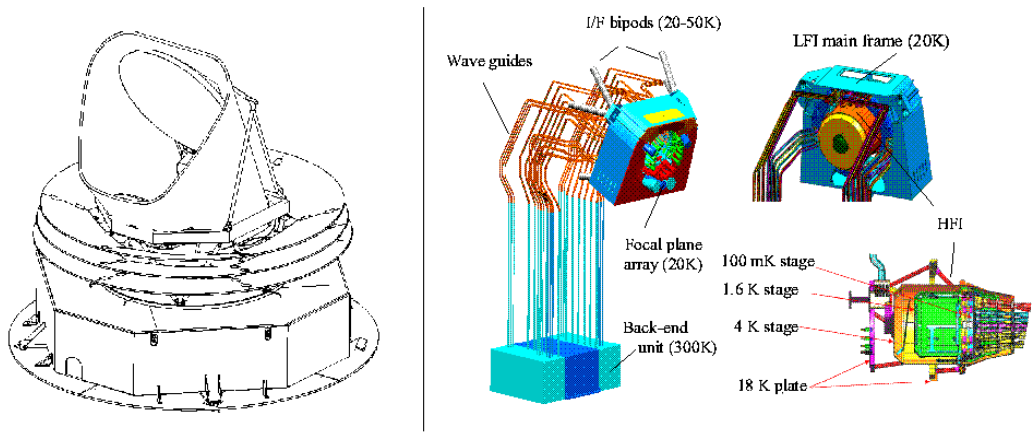}}
\caption{Left: overview of the {\sc Planck} satellite, showing
the primary reflector, the baffle and the three thermal shields
(``V-grooves'') used to thermally decouple the cold ($\sim$50 K)
telescope enclosure from the warm ($\sim$300 K) service module. The
payload design is driven by thermal and radiation (stray-light)
requirements. Right: The LFI radiometer array assembly (left)
with details of the front-end main frame (upper right) and the HFI 
instrument (lower right). The two instruments are integrated so that the 
LFI and HFI horns are located
at the focus of the {\sc Planck} telescope. See text for further details
on the two instruments.
}
\label{fig:Planck}
\end{figure}


The spectral region above 100~GHz is covered by the High
Frequency Instrument (HFI), consisting of an array of 48
spider-web bolometers operated at 0.1~K, and distributed in 6
frequency bands from 100 to 850~GHz \cite{lamarre97, puget98}.
Cooling the bolometers to 0.1~K in space is a key driver in the
design of the instrument and spacecraft. After the 50-60~K
passive cooling and 18~K stage, the HFI cryo-chain uses a
mechanical cooler to provide a 4~K stage to HFI, which is also
used by LFI to cool the reference loads. The 4~K cooler is
designed to minimise microphonic effects, and it operates at a
frequency near 40~Hz adjustable in-flight to minimise resonances.
Finally, a $^3$He - $^4$He closed-cycle dilution cooler provides
0.1~K temperature to the bolometers. The focal plane is
constituted by an array of back-to-back horns at 4~K with
spectral filters placed at $\sim$1.6~K, and a third horn
re-images radiation onto the bolometric detector \cite{church96}.
For the HFI 100, 143 and 217~GHz channels, single-mode
propagation is used. For these channels, as well as for LFI, the
feed horns use double-profiled corrugated design to optimise telescope
illumination and edge taper. At higher frequencies the angular
resolution is achieved without diffraction-limited conditions,
and multi-moded horns are being employed.

Extraordinary sensitivities ($NEP \sim 1 \times
10^{-17}$~W$\times$Hz$^{-1/2}$) are achievable with HFI, and have
been demonstrated by prototype measurements. Thermal stability
requirements are proportionally stringent, at sub-$\mu$K in the
0.1~K stage environment. The use of ``spider-web'' bolometer
technology \cite{bock96, lange96}, with silicon nitride
micromesh, can provide near-background limited performance 
(see Sect.\ref{subsec:technology}). Originally designed as
a temperature sensitive instrument only, HFI has been modified to
include polarisation sensitivity in a subset of the channels
(140--350~GHz) \cite{delabrouille02}. Eight of the 12 HFI
detectors at 143, 217 and 353~GHz use PSBs allowing sensitive
polarisation measurements at these frequencies, a factor of 2
less sensitive than for temperature anisotropy. The HFI PSBs are
built so that two absorbers made with parallel wires coupled to
orthogonal polarisation modes are located in the same cavity and
share the same feed horn, filter and optical path.


The {\sc Planck} payload thermal design is largely driven the
extreme thermal requirements (both absolute levels and stability)
of the instruments. The spacecraft design allows efficient
passive cooling of the payload module (see
Fig.~\ref{fig:Planck}), as necessary to reach passively
$\sim$50~K in the telescope enclosure. To thermally decouple the
300~K service module from the cold focal environment, three
intermediate conical V-groove radiators are implemented at
temperatures of approximately 140~K, 80~K, and 50~K.

Cooling of both instruments to 20~K is achieved with a
closed-cycle hydrogen sorption cryocooler \cite{wade00,
bhandari00, bhandari01}. The cooler operates by thermally cycling
a set of compressors filled with metal hydride to absorb and
desorb hydrogen gas, which is used as the working fluid in a
Joule-Thomson refrigerator. The compressor assembly is attached
to a radiator at 280~K in the warm spacecraft, and the hydrogen
flow-lines are passively pre-cooled at 50-60~K by the V-groove
radiators. The required pressure variations can be obtained by
varying by a factor $\sim 2$ the temperature of the compressors.
In the complete system six identical compressors are used, each
provided with a gas-gap heat switch to optimise their thermal
performance. An additional sorbent bed is used to damp pressure
fluctuations of the low pressure gas. This principle of operation
ensures complete freedom from vibration effects, a unique
property of this kind of coolers which is very beneficial to {\sc
Planck}.


{\sc Planck} requires control of systematic errors at $\mu$K
level. The thermal stability at L2 and the thermal design of the
payload are such that in general the time scale of thermal drifts
will be long compared to the 60 s spin period. Detailed
analytical and numerical studies of the main possible sources of
systematic effects and their impact on the observations were
carried out, including sidelobe pickup of the Galaxy and solar
system bodies \cite{buri01}, distorted beam shapes
\cite{burigana98, burigana02, arnau02}, effects induced by
temperature instability \cite{mennella02}, residual non-white
noise components \cite{maino99, maino02a}, non-idealities in the
instruments \cite{seiffert02, hanany98, delabrouille02a},
spacecraft pointing errors and nutation \cite{vanleeuwen02} and
calibration accuracy \cite{bersanelli95, piat01, cappellini02}.
Similarly, the issue of foreground contamination has to be treated
at an unprecedented level of detail, and several studies on
temperature and polarisation foreground components \cite{hobson98, cayon00,
dezotti99a, dezotti99b, vielva01, hobson99, prunet00,
tegmark00, knox99, bouchet99} have been motivated by the
perspective of {\sc Planck}'s survey.


Although {\sc Planck}'s primary goal is temperature anisotropy,
polarisation has taken on increasing importance, even if it does
not drive the mission requirements. It is expected that {\sc
Planck} will be able to measure with good accuracy the angular
power spectrum of the E-component of CMB polarisation
\cite{huwh97}, with the consequent improved estimation of
cosmological parameters \cite{prunet00, bouchet99a}. Based on
recent estimates of polarised foreground fluctuations as a
function of frequency \cite{bacci01, dezotti99b, prunet01}, the
minimum fluctuation level should be reached at $\sim 100$~GHz,
where the LFI channel has polarisation capability. Foreground
polarised fluctuations at 143~GHz are expected to be 1.5 to 2.5
higher, depending on angular scale and location in the sky. The
combination of {\sc Planck}'s polarised channels will help the
separation of polarised foregrounds.


{\sc Planck} is scheduled for an Ariane 5 launch in 2007 together
with Herschel (previously called FIRST) from the European
spaceport in Kourou (French Guiana). The two satellites will
separate soon after launch, and will proceed independently to
different orbits around L2. {\sc Planck} will carry out (at least)
two complete surveys, each requiring about 7 months of observing
time. While the details of the scanning strategy are still to be
finalised, the nominal scanning law consists of hourly manoeuvre
of amplitude $2.5'$ along the ecliptic plane so as to maintain
the spin axis aligned with the sun-satellite direction. The
scanning law will be optimised taking advantage of the degree of
freedom of moving the spin axis within a $10^\circ$ cone centred
on the spacecraft-sun line.


While both LFI and HFI, have unprecedented capabilities, it is
their combination that gives to {\sc Planck} the imaging power,
the redundancy and the control of systematic effects and
foreground emissions needed to achieve the extraordinary
scientific goals of the mission. In particular, because
systematic effects would in general produce different responses
in the two instruments, their frequency overlap at 100~GHz, near
the minimum of foreground contamination and with similar angular
resolution $\sim 10'$, provides {\sc Planck} with a unique tool
to ensure that the final maps are limited only by the instrument
sensitivity and unavoidable astrophysical foregrounds.

\section{PERSPECTIVES AND CONCLUSIONS}

\subsection{The Future}
\label{future}

The {\sc Planck} mission is expected to bring the long-lasting
effort of measurements of the temperaure power spectrum to its
completion. {\sc Planck}'s precision in the determination of the
$C_\ell$ may be analogous to that achieved by FIRAS in the
measurement of the frequency spectrum \cite{fixsen97, mather94}.
After {\sc Planck}, it is reasonable to expect a decrease of
activity in traditional anisotropy experments, similar to the one
we have witnessed in spectrum projects after
FIRAS\footnote{Actually, a few high precision experiments have
been performed in recent years \cite{staggs96a, staggs96b} and
some new plans are being made \cite{kogut02}}. However, we argue
that a lot of exciting CMB observations should happen after {\sc
Planck}, and indeed in some areas {\sc Planck} will act as a
pioneering mission rather than as a conclusive one. Two main
research directions can be anticipated: precision measurements of
CMB polarisation; and deep imaging at sub-arcminute-scales.

{\it Polarised anisotropy --} Though extraordinary, the
cosmological information obtainable with temperature anisotropy
alone is far less than what could be achieved with a future
high-precision full-sky observation of the CMB polarisation.
Linear polarisation in the microwave background arises from
Thomson scattering of anisotropic radiation at last scattering
\cite{rees68}, with an expected amplitude 1\% to 10\% of the
temperature anisotropy. The polarisation depends sensitively on
the fluctuations on the LSS, and thus encodes a wealth of
cosmological information, some of which is complementary to the
temperature anisotropies. Current upper limits (see
\cite{hedman01} and references therein) are at a level 10-15
$\mu$K, but the expected signal ($\sim 5$~$\mu$K r.m.s., peaking at
multipoles $\ell \sim 1000$) is now in reach of experiments
underway: it is likely that statistical detection at sub-degree
scales will be achieved before $MAP$ and {\sc Planck} data become
available (see Section~\ref{subsec:ongoing})
\footnote{A claim of detection of polarisation anisotropy
using the DASI interferometer has appeared \cite{kovac02} 
just before sending this manuscript for printing}. Even the remarkable
sensitivity of {\sc Planck} to $Q$ and $U$ (10--20~$\mu$K per
$10'$ pixel) will be far from fully extracting the information
encoded in the cosmic polarised signal. {\sc Planck} polarisation
data can provide fundamental tests on structure formation from
initial adiabatic perturbations and break parameter degeneracy,
e.g. between tensor mode amplitude and reionisation optical depth
\cite{efstathiou02}. However, much will be left for further
experiments to measure with increasing precision the signature of
a background of gravitational waves (tensor modes) as anticipated
by inflationary models \cite{selzal97, kamionkowsky97}. A high
resolution, high sensitivity polarisation map can be used to
discriminate between different inflation models \cite{caldwell96}
and, in particular, to determine the inflationary energy scale. A
broad range of inflation models fit the present data, with a flat
geometry ($\Omega_0 \approx 1$) and nearly scale invariant
primordial spectrum ($n_S \approx 1$).
Polarisation maps will be needed to determine the inflation energy
scale, thus probing ultra-high energy physics to levels beyond
what can be obtained with any conceivable terrestrial particle
accelerator. Even more subtle, but not less interesting, will be
to observe the effects of weak gravitational lensing through the
distortion of the CMB polarisation on small scales
\cite{zaldarriaga97}.

Significant progress can be anticipated in CMB polarisation
experiments from ground and balloon, but a space mission will be
needed in the end to avoid atmospheric effects and to map the
entire sky -- which is needed to extend the measurements to low
$\ell$'s and to optimise polarised foreground separation
\cite{bacci01, kogut00}. 
In fact fluctuations in the synchrotron
diffuse component and from extragalactic sources have poorly known 
and possibly significant impact on the polarised signal. 
The sensitivity requirements for a
full-sky, high-precision polarisation survey are extremely
demanding. The expected polarisation amplitude induced by
gravitational waves is $\sim 0.1\%$ to 1\% of temperature $\Delta
T/T$: a high signal-to-noise imaging requires a noise per pixel
$\sim$0.05~$\mu$K, i.e. about 300 times better than {\sc Planck}.
This is beyond the foreseeable future, but an intermediate step
with sensitivity 0.5 $\mu$K per $10'$ resolution element can be
approached extrapolating existing technology. Multi-frequency
observations, possibly with matching FWHM, would be needed for
foreground separation. It is difficult to anticipate which
systematic effects could represent limiting factors at sub-$\mu$K
levels; it is likely that thermal stability at the cryogenic
temperatures needed by the ultra-high sensitivity detectors
(either radiometers or bolometers) will be a major challenge;
also, a multi-channel off-axis instrument, pushed at extreme
sensitivities, would place critical requirements for ultra-low
cross-polarisation.

{\it Fine scale CMB -- } On angular scales $\sim 1'$ and below
the CMB is influenced by interaction with intervening ionised
material (see Sect.~\ref{subsec:secondary_anisotropies}). The CMB
passed through the cosmic ``dark ages'', before star and quasar
formation, from which direct observation is extremely difficult
to obtain. The study of arcmin scale features in the CMB may well
turn out to be the only, or one of the most powerful, technique
to gather observational evidence from the very early processes of
structure formation. Using the CMB as a high redshift backlight,
deep arcminute CMB imaging can probe the early history of galaxy
clusters and their gradual acceleration as they fell in their
potential wells. In addition, CMB fine scale observations can
provide images of the largest structures in the universe as they
started to dissipate the heat of their gravitational collapse.
{\sc Planck} will open up this field, but its $>5'$ angular
resolution is not sufficient to fully cover this promising area
of CMB studies. Furthermore, accurate S-Z measurements can be
made to very high redshifts \cite{stebbins97}, all the way back
to the cluster formation era \cite{aghanim97}. A future sky
survey of clusters velocities based on their S-Z features
throughout the entire Hubble volume could, in principle, map the
evolution of velocity fields over much of the history of the
universe.

The S-Z effect from clusters is by far the dominant secondary
source (see Sect.~\ref{subsec:secondary_anisotropies}). Once this
will be well identified and mapped, sky regions free from S-Z and
other local foregrounds could be searched at $\sim 1'$ scales for
fainter secondary signatures, such as those from gravitational
collapse of large scale ($\sim 100$\,Mpc) structures, bulk motion
of plasma (the ``Ostriker-Vishniac effect''), effects of the
evolution of gravitational potentials on CMB photons (the
``integrated Sachs-Wolfe effect'' and the ``Rees-Sciama effect''),
lensing-induced signatures from clusters \cite{seljak00},
signature of local ionisation events \cite{platania02}, and
details of the ionisation history of the universe.

It is easy to expect that the $MAP$ and (especially) {\sc Planck}
surveys will trigger new observations in selected sky areas
searching for detailed, physically interesting features. New
instruments and observing strategies will be developed for the
purpose. At wavelengths $\lambda \approx 1~$mm, a $\lesssim 1'$
resolution translates in a telescope aperture typically $D \sim
10$~m. The brightest clusters give a S-Z thermal component $\sim
1$~mK, while the kinetic effects is expected to be $\sim 10$
times lower. Signatures such as filaments from in-falling
clusters are expected at $\sim$10~$\mu$K level. Current
bolometric or HEMT-based instruments (either filled aperture
arrays or interferometers) are able to approach sub-$\mu$K
sensitivity in very localised areas. High precision fine-scale
CMB imaging would call for very wide frequency coverage, with
ancillary monitoring at low frequency ($<5$~GHz) and high
frequency ($>5$~THz) to safely remove unrelated foregrounds as
well as  backgrounds (such as primary CMB anisotropy, a source of
confusion in this context!). These observations can be expected
to progress for several years with ground-based instrumentation.
Eventually, an arcminute-scale, full-sky survey to map the
evolution of the large scale velocity field will only be possible
with a space programme, currently out of reach.

\subsection{Conclusions}
\label{sec:conclusions}

Observations of the CMB anisotropy can be considered at present as
the most powerful method to obtain high quality data on the early
universe. We have reviewed the great effort that is being carried
out by sub-orbital projects after the $COBE$-DMR discovery. In
spite of the major experimental challanges, a coherent picture is
emerging. Both balloon-borne and ground-based experiments
independently demonstrate the acoustic pattern of the angular
power spectrum showing the first three peaks, and both provide a
determination $\sim 10\%$ accuracy of the angular scale of first
peak consistent with a flat geometry, $\Omega_0 \approx 1$. This
is a truly remarkable achievement. However, it is only a
foretaste of the precise (percent-level) determination of several
cosmological parameters attainable by the forthcoming generations
of experiments, culminating with the {\sc Planck} space mission.
Beyond {\sc Planck}, it is likely that CMB observations will
concentrate on precise polarisation measurements and deep
sub-arcmin imaging searching for secondary anisotropies.

Precision measurements of the CMB temperature and polarisation
fluctuations are able to shed light on very high energy phenomena
occurring in the primordial cosmic environment. Cosmologists need
to look more and more towards particle physics to interpret the
physical processes probed by CMB data. On the other hand,
exploiting the natural laboratory of the early universe to study
ultra-high energy scales is of great interest for particle
physics. The enormous data sets expected from the planned and
future CMB missions are likely to attract the interest of a wider
scientific community.

\begin{acknowledgments}

ACKNOWLEDGMENTS. The completion of this review has been greatly
helped, either directly or indirectly, by the work of many
people, in particular by the {\sc Planck} Science Team and by the
{\sc Planck}-LFI consortium. We
are indebited to Bruce Partridge and G. De Zotti for a careful 
reading of this paper. We also warmly thank Max Tegmark for useful
discussion and R. Davies, P. De Bernardis,
G. Smoot, and the $MAP$ Science Team for figure permission.

\end{acknowledgments}

\end{document}